\documentclass[prd,twocolumn,nofootinbib,superscriptaddress,amsmath,amssymb]{revtex4-1}
\usepackage{mathtools}
\usepackage{graphics}
\usepackage{graphicx}
\usepackage{dcolumn}
\usepackage{bm}
\usepackage{dsfont} 
\usepackage{amsmath,amssymb}
\usepackage{hyperref}
\usepackage{tabularx}
\usepackage{epsf,epsfig}
\usepackage[normalem]{ulem}
\usepackage[usenames]{color}
\usepackage{multirow}
\usepackage{makecell}
\usepackage{diagbox}
\usepackage[abs]{overpic}
\allowdisplaybreaks
\hypersetup{
    colorlinks=true,
    linkcolor=blue,
    filecolor=magenta,      
    urlcolor=blue,
    citecolor=blue
}
\definecolor{red(ncs)}{rgb}{0.77, 0.01, 0.2}

\newcommand{\sci}[2]{#1 \times 10^{#2}}

\usepackage{array}
\newcolumntype{C}[1]{>{\centering\let\newline\\\arraybackslash\hspace{0pt}}m{#1}}

\newcolumntype{C}[1]{>{\centering\arraybackslash}m{#1}}

\begin{document}

\title{Equation-of-state insensitive relations after GW170817}

\author{Zack Carson}
\affiliation{%
 Department of Physics, University of Virginia, Charlottesville, Virginia 22904, USA
}%

\author{Katerina Chatziioannou}
\affiliation{%
 Center for Computational Astrophysics, Flatiron Institute, 162 5th Ave, New York, NY 10010, USA
}%

\author{Carl-Johan Haster}
\affiliation{%
 LIGO Laboratory and Kavli Institute for Astrophysics and Space Research, Massachusetts Institute of Technology, Cambridge, Massachusetts 02139, USA
}

\author{Kent Yagi}
\affiliation{%
 Department of Physics, University of Virginia, Charlottesville, Virginia 22904, USA
}%

\author{Nicol\'as Yunes}
\affiliation{%
eXtreme Gravity Institute, Department of Physics, Montana State University, Bozeman, MT 59717, USA.
}%

\date{\today}


\begin{abstract}
The thermodynamic relation between pressure and density (i.e.~the equation of state) of cold supranuclear matter is critical in describing neutron stars, yet it remains one of the largest uncertainties in nuclear physics. 
The extraction of tidal deformabilities from the gravitational waves emitted in the coalescence of neutron star binaries, such as GW170817, is a promising tool to probe this thermodynamic relation.
Equation-of-state insensitive relations between symmetric and antisymmetric combinations of individual tidal deformabilities, the so-called  ``binary Love relations", have proven important to infer the radius of neutron stars, and thus constrain the equation of state, from such gravitational waves. 
A similar set of relations between the moment of inertia, the tidal deformability, the quadrupole moment, and the compactness of neutron stars, the so-called ``I-Love-Q" and ``C-Love" relations, allow for future tests of General Relativity in the extreme gravity regime. 
But even the most insensitive of such relations still presents some degree of equation-of-state variability that could introduce systematic uncertainties in parameter extraction and in model selection. 
We here reduce this variability by more than $50\%$ by imposing a prior on the allowed set of equations of state, derived from the posteriors generated from the analysis of GW170817.  
The resulting increase in insensitivity reduces systematic uncertainties in the extraction of the tidal deformability from future gravitational wave observations, although statistical uncertainties currently dominate the error budget, and will continue to do so until the era of Voyager-class detectors.  
\end{abstract}

\maketitle


\section{Introduction}
\label{sec:intro}

The thermodynamic relation between pressure and density in cold, supranuclear matter, the so-called equation of state (EoS), remains an unsolved problem in nuclear physics. The EoS is critical to our understanding of neutron stars (NSs) because it determines many NS observables, such as their mass, radius, moment of inertia ($I$), quadrupole moment ($Q$) and tidal deformability (or Love number). Unfortunately, terrestrial experiments can only probe the EoS to around nuclear saturation density ($\rho_{\text{sat}} \approx 2.5 \times 10^{14} \text{ g/cm}^3$)~\cite{Li:HeavyIon,Tsang:SymmetryEnergy,Centelles:NeutronSkin,Li:CrossSections,Chen:SymEnergy}. Although some temperature-dependent heavy-ion collision experiments can probe higher densities~\cite{Danielewicz:2002pu}, astrophysical observations of NSs remain ideal for constraining the EoS of cold and ultra dense, nuclear matter.

Independent measurements of NS observables can be used to constrain the nuclear EoS. For example, electromagnetic observations of the mass and radius of certain NSs have been used to place confidence limits in the mass-radius plane, and thus constrain the EoS~\cite{guver,ozel-baym-guver,steiner-lattimer-brown,Lattimer2014,Ozel:2016oaf}. These observations, however, may potentially suffer from large systematic errors~\cite{Miller:2016pom, Miller2013} due to uncertainties in the astrophysical modeling of X-ray bursts. The gravitational waves (GWs) emitted in the coalescence of NS binaries may be a cleaner probe of nuclear physics. During the early inspiral, the orbital separation is large enough that the tidal fields are negligible; but as the orbital separation decreases due to GW emission, tidal forces grow and the NSs respond by developing deformations determined by their nuclear EoS. 
These deformations source additional multipole radiation as well as affect the orbital trajectory of the binary, thus altering the GWs emitted, encoding within the latter the NS EoS~\cite{hinderer-love,Flanagan2008}.

The GWs emitted by binary NSs in the late inspiral must then depend on the \emph{tidal deformabilities} $\Lambda_1$ and $\Lambda_2$\footnote{Throughout this work we use the convention that binary quantities are indexed as $1,2$ where 1 always correspond to the more massive object.}, which control the linear response of the star's quadrupole deformation to the (electric-type, quadrupole) tidal field of the companion (to leading order in a post-Newtonian expansion~\cite{Blanchet:2013haa})~\cite{Flanagan2008,Vines:2011ud}. These parameters, however, enter the GW waveform model at the same post-Newtonian order, making them degenerate, and thus, very difficult to estimate independently with current GW data~\cite{Wade:tidalCorrections}. Instead, one can extract certain combinations of the tidal deformabilities, such as a certain mass-weighted tidal deformability $\tilde{\Lambda}$~\cite{Favata:2013rwa,Wade:tidalCorrections}, or one can extract the (mass-independent) coefficients $(\lambda_{0},\lambda_{1},\ldots)$ of a Taylor expansion of the tidal deformabilities about some fiducial mass~\cite{Messenger:2011gi,delPozzo:TaylorTidal,Yagi:binLove}. Current detectors are not sensitive enough to accurately measure any of these coefficients, but future detectors will, and the information from multiple events could then be combined, since the Taylor expansion coefficients should be common to all events. 

Lacking enough sensitivity in current GW observations, one is forced to only estimate the mass-weighted tidal deformability, but this prevents the independent extraction of $\Lambda_{1}$ and $\Lambda_{2}$. Yagi and Yunes~\cite{Yagi:2015pkc,Yagi:binLove} proposed a solution to this problem by finding ``approximately universal'' or ``EoS-insensitive'' relations between the symmetric and anti-symmetric combinations of the tidal deformabilities $\Lambda_{s,a}=\frac{1}{2}(\Lambda_1 \pm \Lambda_2)$, the so-called ``binary Love relations." These relations can be used to analytically express $\Lambda_{s}$ in terms of $\Lambda_{a}$ (or vice-versa), making the mass-weighted tidal deformability a function of only $\Lambda_{a}$. A measurement of the mass-weighted tidal deformability then implies a measurement of $\Lambda_{a}$, and through the use of the binary Love relations, also a measurement of $\Lambda_{s}$, which then allows for the inference of the individual tidal deformabilities $\Lambda_{1}$ and $\Lambda_{2}$~\cite{Yagi:2015pkc,Yagi:binLove}. With those at hand, one can further use EoS-insensitive relations between the tidal deformabilities and the compactness, the so-called ``C-Love relations''~\cite{Yagi:2013bca,Yagi:ILQ,Maselli:2013mva,Yagi:2016bkt}, to infer the radii of the NSs, and thus, to place two constraints in the mass-radius plane, one for each star in the binary. This idea was recently implemented for GW170817~\cite{TheLIGOScientific:2017qsa}, allowing EoS-independent constraints on the mass-radius curve using GW data~\cite{Katerina:residuals,LIGO:posterior}.     

EoS-insensitive relations can in fact be used for more than just measuring the nuclear EoS. For years, the theoretical physics community considered the possibility of using measurements of NS properties, such as the mass, the radius and the moment of inertia, to constrain deviations from General Relativity in the strong-field regime. Certain modified theories of gravity, such as scalar tensor theories with spontaneous scalarization~\cite{Damour:1996ke}, Einstein-\AE ther and Ho\v rava gravity~\cite{Eling:2007xh,Yagi:2013ava,Yagi:2013qpa}, dynamical Chern-Simons gravity~\cite{Yunes:2009ch,Yagi:2013mbt,Gupta:2017vsl}, beyond Horndesky theories~\cite{Babichev:2016jom,Sakstein:2016oel}, modify such NS observables, but unfortunately, these modifications are typically degenerate with the nuclear EoS~\cite{Pani:2014jra,Minamitsuji:2016hkk,Maselli:2016gxk}. Yagi and Yunes proposed to solve this problem by finding EoS-insensitive relations between the moment of inertia, the tidal deformability (or Love number) and the quadrupole moment, the so-called ``I-Love-Q'' relations~\cite{Yagi:2013bca,Yagi:ILQ}. Given a measurement of the Love number for a given NS, for example through GW observations, the I-Love-Q relations can be used to infer the moment of inertia or the quadrupole moment. A second independent measurement of either of these two quantities, for example through binary pulsar observations~\cite{Lattimer:2004nj} or observations with the Neutron star Interior Composition ExploreR (NICER)~\cite{Ozel:2015ykl}, then allows an EoS-insensitive test of General Relativity in the strong field regime~\cite{Yagi:2013bca,Yagi:ILQ,Gupta:2017vsl,Doneva:2017jop}. 

The implementation of the EoS-insensitive relations in data analysis has to somehow contend with the fact that these relations are in fact not exactly universal, but rather present different (albeit small) levels of EoS variability. In the case of the binary Love and C-Love relations to infer the radii of NSs with GW170817, the problem is solved by marginalizing over the EoS variability~\cite{Katerina:residuals}. Presumably, this same procedure can be applied in the future when carrying out tests of General Relativity with the I-Love-Q relations, using a combination of binary pulsar, NICER and GW data. But this marginalization procedure may not always be as important; as constraints in the mass-radius plane become more stringent with future GW observations, the allowed space of EoSs will shrink, which in turn must naturally decrease the degree of EoS variability in all EoS-insensitive relations. This is the main focus of this paper.


\subsection{Executive Summary}

We begin by investigating the increase in EoS insensitivity due to the constraints placed by GW170817 on the allowed space of EoSs. 
We use the posterior probability distribution on the pressure-density plane~\cite{Carney:2018sdv,LIGO:posterior} obtained from GW170817~\cite{TheLIGOScientific:2017qsa} to inform the set of EoSs that is compatible with this observation. We then generate two large samples of spectral EoSs~\cite{Lindblom:2018rfr}, one in which the EoSs are directly sampled from the posterior probability distribution (the ``constrained" sample) and another in which this constraint not enforced (the ``unconstrained" sample).
We repeat the analysis done by Yagi and Yunes~\cite{Yagi:binLove,Yagi:ILQ} on both sets of EoSs and find that the EoS-insensitive relations present less EoS variability with the constrained set. 
In particular, the EoS-insensitivity increases by a factor of $\sim 60$\% in the binary Love relations (for stars with mass ratio larger than $0.75$), by a factor of $\sim 70\%$ in the C-Love relations, and by factors of $\sim 50$\% in the I-Love-Q relations (see Table~\ref{tab:maxVar} and Fig.~\ref{fig:ILQ} for more details). 

With this study at hand, we then carry out additional related studies on EoS-insensitive relations that go beyond the work in~\cite{Yagi:binLove,Yagi:ILQ}. First, we investigate the relation between the NS radius and its tidal deformability, the R-Love relations, for both sets of EoSs, as these are critical in order to place constraints on the radius from a measurement of the Love number.  We use the C-Love relations to construct the R-Love relations and find that the maximum EoS variability drops from $\sim 880$ m in the unconstrained case to $\sim 360$ m in the constrained case (see Sec.~\ref{sec:clove} for more details). Second, we study the EoS-universality of hybrid stars, which experience strong first-order phase transitions from hadronic to quark matter in the core~\cite{Paschalidis2018,Most:2018eaw,Burgio:2018yix,Montana:2018bkb}. We find that the I-Love-Q and C-Love relations remain EoS-insensitive for these hybrid stars, although the EoS variability increases slightly (see Sec.~\ref{sec:ilq-hyb} and~\ref{sec:clove-hyb} for more details). However, we also find that the binary Love relations are not EoS-insensitive for a mixed binary with a (massive) hybrid star and a (low-mass) hadronic star, due to the large separation in mass-weighted tidal deformability between the constituent stars (see Sec.~\ref{sec:binary} for more details).

Last but not least, we study the importance of using the improved EoS-insensitive relations in future GW observations. The use of EoS-insensitive relations introduces systematic uncertainties in parameter estimation because of the intrinsic non-zero EoS-variability in these relations, which one must marginalize over. These uncertainties are currently irrelevant because statistical uncertainties in parameter estimation are much larger with current detectors. But as the detector sensitivity is improved, the signal-to-noise ratio and the number of events that will be detected will increase, therefore decreasing the statistical uncertainties below the systematic one due to EoS variability. We carry out Fisher analyses to estimate when the statistical uncertainties become comparable to the systematic uncertainties due to EoS variability and find that this occurs for Voyager-class detectors. 

\begin{figure}
\begin{center} 
\includegraphics[width=\columnwidth]{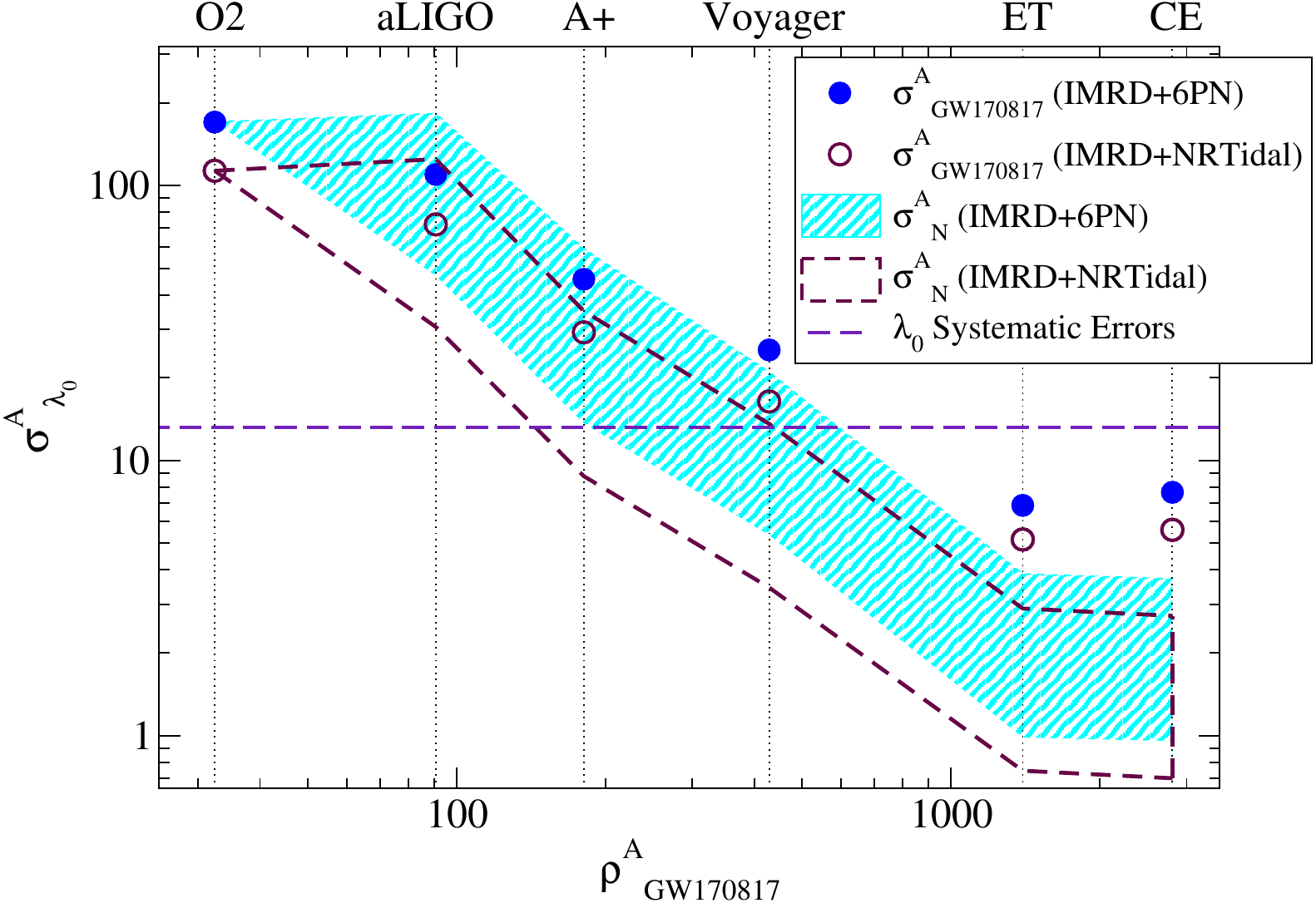}
\end{center}
\caption{(Color online) Fisher-estimated statistical uncertainties on the extraction of $\lambda_{0}$ with interferometer (O2, aLIGO, A\texttt{+}, Voyager, CE, ET-D) as a function of the signal-to-noise-ratio expected in each of these instruments, given a single GW170817 detection (circles). The statistical uncertainties with ET are lower than with CE in spite of a lower signal-to-noise ratio because the former is more sensitive above 300 Hz, where tidal effects matter the most (see Sec.~\ref{sec:futureObservations} for further discussion). We also plot the combined statistical uncertainty given $N$ observations consistent with the NS binary merger rate for a 1 year observation (regions), with the top and bottom edges of the regions corresponding to pessimistic and optimistic merger rates. These statistical uncertainties should be compared to the systematic uncertainty on the extraction of $\lambda_{0}$ due to EoS-variability (horizontal dashed line). The statistical and systematic uncertainties cross for Voyager-class detectors. We confirm this conclusion by repeating the statistical analysis with two different waveform models (PhenomD~\cite{PhenomDI,PhenomDII} plus NRTidal corrections~\cite{Dietrich:2017aum,Samajdar:NRTidal} and PhenomD~\cite{PhenomDI,PhenomDII} plus 6PN tidal corrections~\cite{Vines:2011ud,Wade:tidalCorrections}).}
\label{fig:stackedFisher}
\end{figure}

Figure~\ref{fig:stackedFisher} shows this result in more detail. We present the (Fisher-estimated) statistical uncertainty in the measurement of $\lambda_{0}$ with various detectors (LIGO O2~\cite{aLIGO}, Advanced LIGO at design sensitivity (aLIGO)~\cite{aLIGO}, LIGO A\texttt{+} (A\texttt{+})\cite{Ap_Voyager_CE}, Voyager~\cite{Ap_Voyager_CE}, the Einstein Telescope (ET)~\cite{ET}, and the Cosmic Explorer (CE)~\cite{Ap_Voyager_CE}) for an event similar to GW170817. The x-axis shows the signal-to-noise ratio for a GW170817 event detected with each of these detectors. We also present the combined statistical uncertainty $\sigma^A_N$ after $N$ detections with each of these instruments, with the top and bottom of the region representing the most optimistic and pessimistic expectation for the number of detections expected from the binary NS merger rate~\cite{Abbott2017} (see also Table~\ref{tab:variances}). These statistical uncertainties should be compared to the systematic uncertainty in $\lambda_{0}$ due to EoS-variability improved with the constrained set. We find that the statistical and the systematic uncertainties cross for Voyager-class detectors.

The remainder of this paper presents the details of the results summarized above and it is organized as follows. 
We begin with complementary background and theory material in Sec.~\ref{sec:theory}.
We continue in Sec.~\ref{sec:universal} by finding new and improved binary Love, I-Love-Q, and C-Love relations, and considering how well hybrid star EoSs agree with the EoS-insensitive relations.
We next examine these improved relations and question whether or not they are useful for future interferometers in Sec.~\ref{sec:observations}.
We conclude in Sec.~\ref{sec:conclusion} by discussing our results and mentioning avenues of future work.
Throughout this paper, we have adopted geometric units of $G=1=c$, unless otherwise stated.


\section{Background and theory}\label{sec:theory}

In this section we review how the EoS can be represented analytically
through a spectral decomposition, and how the observation of GW170817
constrains the space of possible EoSs. We then proceed to discuss 
how one computes the tidal deformabilities and the EoS-insensitive relations. 

\subsection{Spectral representations of NS EoSs}
\label{sec:eos}

The structure of a NS and its tidal interactions in a binary system rely heavily on the underlying state function (or equation of state - EoS) describing the relationship between the pressure ($p$) and energy density ($\epsilon$) of nuclear matter.
Given that all currently proposed EoSs utilize certain approximations~\cite{Oertel:Review,Baym:Review}, one method to study a wide range of physically realizable EoSs is to parameterize them such that any realistic EoS can be represented with a small number of parameters.
Spectral representations~\cite{Lindblom:2010bb,Lindblom:2012zi,Lindblom:2013kra,Lindblom:2018rfr,Abbott:2018exr} parameterize EoSs by performing spectral expansions on the adiabatic index $\Gamma(p)$\footnote{Another way of parameterizing EoSs is through a piecewise polytropic formulation~\cite{Read2009,Lackey:2014fwa,Carney:2018sdv}.}:
\begin{equation}
\Gamma(x) = \exp{\sum_k^{N}\gamma_k x^k},
\end{equation}
where $x \equiv \log{(p/p_0)}$ for a minimum pressure $p_0$.
The equation of state is then determined by an integration of the differential equation:
\begin{equation}
\frac{d \epsilon(p)}{dp}=\frac{\epsilon(p)+p}{p \Gamma(p)}.
\end{equation}
Using this formalism, any valid EoS can be approximated through the choice of $N$ spectral coefficients $\gamma_k$, and we here choose $N=4$, tabulated for several common EoSs in Table 1 of~\cite{Lindblom:2018rfr}.

\begin{figure*}
\begin{center} 
\includegraphics[width=.97\columnwidth]{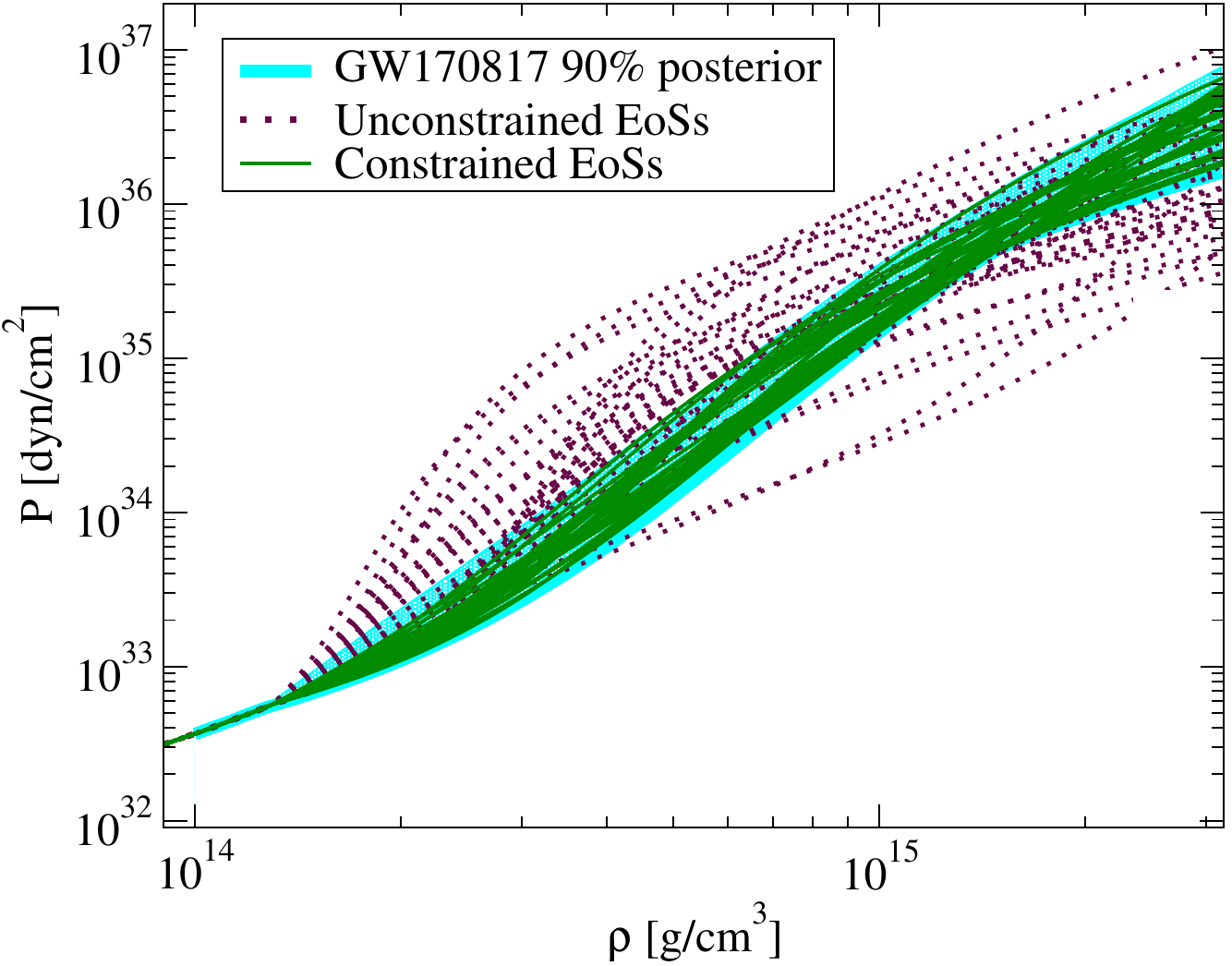}
\includegraphics[width=\columnwidth]{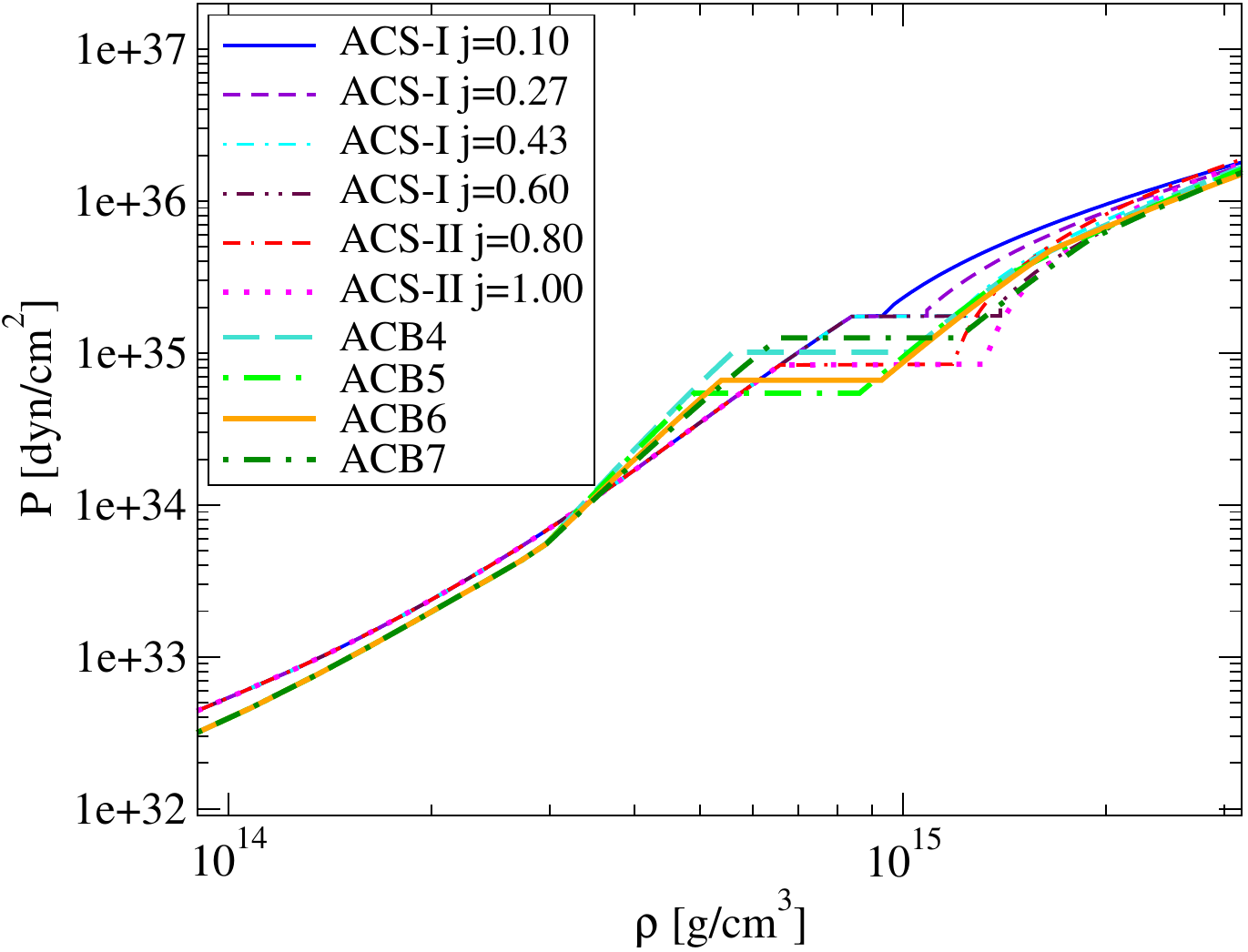}
\end{center}
\caption{(Color Online) Left: Small representative samples of the unconstrained (dotted) and constrained (solid) sets, together with the 90\% marginalized posterior distribution from the observation of GW170817 (cyan shaded region)~\cite{LIGO:posterior}. There is significantly less variability in the constrained set of EoSs due to the requirement that they be consistent with the GW170817 observation. Right: EoSs for ACS and ACB hybrid stars~\cite{Paschalidis2018}, each transitioning from a hadronic branch (corresponding to a pure hadronic-matter NS) into a quark-matter branch (quark-matter inner core surrounded by hadronic matter) at various transition pressures $P_{\text{tr}}$.
}
\label{fig:eos}
\end{figure*} 

We here wish to consider EoS-insensitive relations using two sets of EoSs: a ``constrained set'' consistent with the observation of GW170817 and an ``unconstrained set'' that does not impose this prior. In \emph{both} cases, we model the EoS with a $C^{0}$-piecewise function that equals the low-density crust EoS of SLy\footnote{SLy represents the unified equation of state based on the effective Skyrme Lyon nucleon-nucleon interactions, developed by Douchin and Haensel, 2001~\cite{Douchin:2001sv}.}~\cite{Douchin:2001sv} below half nuclear saturation density $\rho_{\text{stitch}}=1.3 \times 10^{14} \text{ g/cm}^3$~\cite{Read2009}, and equals the spectral decomposition described above outside the crust. For the latter, we restrict the spectral coefficients to the ranges $\gamma_0 \in \lbrack 0.2,2 \rbrack$, $\gamma_1 \in \lbrack -1.6,1.7 \rbrack$, $\gamma_2 \in \lbrack -0.6,0.6 \rbrack$, $\gamma_3 \in \lbrack -0.02,0.02 \rbrack$, and the adiabatic index is further restricted to $\Gamma \in \lbrack 0.6,4.5 \rbrack$~\cite{Lindblom:parameters}. Moreover, we impose the following two restrictions:  (i) causality within 10\%, i.e. that the speed of sound of the fluid be less than the speed of light to 10\% (following the analysis of Ref.~\cite{LIGO:posterior}), and (ii) a high maximum mass, i.e.~that the resulting EoS supports NSs with masses at least as high as $1.97 \text{ M}_{\odot}$, consistent with astrophysical observations~\cite{1.97NS,2.01NS,Zhao:massiveNS}. 

The unconstrained set is then defined by drawing random samples in the spectral coefficients within their allowed prior ranges, and then eliminating any EoS that either leads to an adiabatic index $\Gamma$ outside the allowed range, breaks the causality restriction, or breaks the maximum mass restriction. 
To obtain the constrained set of EoSs we analyze the publicly available data for GW170817 from the Gravitational Wave Open Science Center~\cite{GWOSC,Vallisneri:2014vxa}. 
We use the publicly available software library {\tt LALInference}~\cite{Veitch:2014wba,lalinference_o2} to sample the EoS posterior, similar to Ref~\cite{LIGO:posterior}.
Our analysis uses the same settings and prior choice as those of~\cite{LIGO:posterior,Carney:2018sdv}. 
In both cases, each set consists of 100 members, with a subset of these shown in the left panel of Fig.~\ref{fig:eos}.

In addition, we investigate 10 transitional quark-hadron matter stars, which undergo strong first-order phase transitions at a pressure $P_{\text{tr}}$, leading to the hadronic branch departing into a quark-matter branch at a given transitional mass~\cite{Paschalidis2018,Alford:2017qgh,1971SvA....15..347S,Zdunik:2012dj,Alford:2013aca}. In particular, we focus on the ACS and ACB models (corresponding to speed of sound, and piecewise polytropic representations of the EoS beyond the phase transition, respectively) described in~\cite{Paschalidis2018}, and also shown in the right panel of Fig.~\ref{fig:eos}. These result in two distinct types of NSs, based on their mass: (i) massive ($M \geq \text{ M}_{\text{tr}}$) hybrid stars which have quark-matter inner cores and nuclear matter elsewhere (henceforth, we denote such stars as hybrid stars (HSs)), and (ii) low-mass ($M \leq \text{ M}_{\text{tr}}$) hadronic stars with no internal transition to quark matter (henceforth, we denote these stars as simply NSs).

\subsection{NS Tidal deformability}\label{tidal}

The GWs emitted in the coalescence of binary NSs, such as GW170817, provide valuable insight into their internal structure. As the stars spiral into each other, they tidally deform due to their companion's tidal field. The tidal field can be decomposed into multipole moments, with the leading-order field being even parity and quadrupolar, and thus leading to an even parity and quadrupolar deformation in response. The constant of proportionality that determines how quadrupolarly deformed a star becomes in the presence of a quadrupolar tidal field is called the \emph{tidal deformability} and denoted $\lambda$. The magnitude of $\lambda$ depends strongly on the underlying structure, as well as the fluid nature of the NS~\cite{PhysRevD.98.084010}. Therefore, this parameter has the largest impact on the GW phase, and thus, it is encoded in GW observations of binary NS coalescence.  

Consider a NS of mass $M$ in the presence of the (spatial) tidal tensor $\varepsilon_{ij}$ of its companion. In response to this tidal field, the NS will deform away from sphericity and acquire a (spatial) quadrupole moment tensor $Q_{ij}$. The tidal deformability $\lambda$ quantifies this linear response via~\cite{Flanagan2008,hinderer-love,Yagi:ILQ}:
\begin{equation}
Q_{ij}=-\lambda \; \varepsilon_{ij}.
\end{equation}
One can extract $Q_{ij}$ and $\varepsilon_{ij}$ -- and thus $\lambda$, or its dimensionless form $\Lambda \equiv \lambda/M^5$ -- via a double asymptotic expansion of the gravitational potential in a buffer zone, defined by ${\cal{L}} \gg r \gg R$:
\begin{align}
\Phi=- \frac{1+g_{tt}}{2}&=-\frac{M}{r} - \frac{3}{2}\frac{Q_{ij}}{r^3} \Bigg(\frac{x^i}{r} \frac{x^j}{r}-\frac{1}{3}\delta_{ij} \Bigg) + \mathcal{O} \Bigg( \frac{{\cal{L}}^4}{r^4} \Bigg)
\nonumber \\ &+ \frac{1}{2} \varepsilon_{ij} x^i x^j + \mathcal{O} \Bigg( \frac{r^3}{R^3} \Bigg),
\end{align}
where $R$ is the radius of the star and ${\cal{L}}$ is the length scale associated with the radius of curvature induced by the companion. 

The calculation of the tidal deformability therefore requires the calculation of the metric component $g_{tt}$. One typically starts by first constructing a spherically symmetric, non-spinning background solution, whose stellar radius and mass are determined from $p(R)=0$ and $M=(1-g_{rr}(R)^{-1})R/2$. One then perturbatively introduces a tidal deformation and solves the perturbed Einstein equations in the NS interior, and matches this solution to an exterior solution at the surface, modulo two integration constants. The requirement that the metric be at least $C^{1}$ fixes the ratio between the two constants of integration, yielding the tidal deformability, as discussed in more detail in~\cite{hinderer-love}. 

Similar to tidal perturbations, one can consider rotational perturbations. At first order in rotation, the stellar moment of inertia can be extracted by looking at the asymptotic behavior of the $g_{t \phi}$ metric component. At second order in rotation, the quadrupole moment can be extracted by studying the asymptotic behavior of the $g_{t t}$ metric component. 

Consider now a binary NS system, as that which produced the GW170817 event, where each star individually experiences the tidal field of the other star.
Thus, each star possesses tidal deformabilities $\Lambda_1$ and $\Lambda_2$ that enter the GW phase and encode information about each star's EoS.
Due to strong correlations between these parameters, individual extraction is very difficult with current interferometer sensitivity limitations.
However, the waveform templates can be strategically reparameterized to instead include independent linear combinations of $\Lambda_1$ and $\Lambda_2$ that mitigate correlations. One common reparameterization is through the introduction of the mass-weighted tidal deformability $\tilde{\Lambda}=\tilde{\Lambda}(\Lambda_1,\Lambda_2)$ and the parameter $\delta \tilde{\Lambda}=\delta \tilde{\Lambda}(\Lambda_1,\Lambda_2)$~\cite{Favata:2013rwa,Wade:tidalCorrections}. Since these parameters enter the GW phase at 5 and 6PN orders respectively\footnote{A term of $N$PN order is proportional to $v^{2N}$ relative to the leading-order term in the expression, where $v$ represents the relative velocity of binary constituents.}, they partially break the degeneracies between $\Lambda_{1}$ and $\Lambda_{2}$.

\subsection{EoS-insensitive relations}\label{sec:eosInsensitive}
Current GW interferometry is not yet sensitive enough to accurately extract both tidal parameters $\tilde{\Lambda}$ and $\delta\tilde{\Lambda}$.
In a search to remedy this, Yagi and Yunes~\cite{Yagi:binLove} found that symmetric and antisymmetric combinations of the tidal deformabilities
\begin{equation}
\Lambda_s \equiv \frac{\Lambda_2 + \Lambda_1}{2}, \hspace{6mm} \Lambda_a \equiv \frac{\Lambda_2 - \Lambda_1}{2},
\end{equation}
display EoS-insensitive properties to a high degree, showing EoS variations of at most 20\% for binaries with masses less than $1.7 \text{ M}_{\odot}$ and using a representative sample of 11 EoSs. 
These ``binary Love relations" allow one to analytically break degeneracies between the tidal parameters: one can substitute $\Lambda_{a}=\Lambda_{a}(\Lambda_{s})$ in the GW model, thus completely eliminating $\Lambda_{a}$ from the parameter list (or vice-versa). This is important for two reasons: (i) the new model allows for the more accurate extraction of $\Lambda_{s}$ (or $\Lambda_{a}$ if $\Lambda_{s}$ is eliminated), and (ii) the relations allow for the inference of $\Lambda_{a}$ given a measurement of $\Lambda_{s}$ (or vice-versa), and from this for the inference of the individual tidal deformabilities $\Lambda_{1}$ and $\Lambda_{2}$. A simple Fisher analysis has shown that the binary Love relations improve parameter estimation of $\tilde{\Lambda}$ by up to an order of magnitude~\cite{Yagi:2015pkc,Yagi:binLove}.

Similar EoS-insensitive relations have been found between individual NS observables: the moment of inertia ($I$), the tidal deformability (Love), the quadrupole moment ($Q$), and the compactness ($C$), known as the ``I-Love-Q" and ``C-Love" relations~\cite{Yagi:2013bca,Yagi:ILQ, Maselli:2013mva}.
These relations are EoS-insensitive to better than 1\% and 6\% respectively, and they have been important applications in both GW astrophysics~\cite{Kumar:2019xgp} and experimental relativity~\cite{Yagi:2013bca,Yagi:ILQ,Gupta:2017vsl,Doneva:2017jop}. For example, these relations and the measurement of the tidal deformabilities allow for the inference of several other stellar properties, such as the moment of inertia, the  compactness, the spin ($\chi$), and the radius of NSs~\cite{Kumar:2019xgp}.
Analyses such as that of~\cite{Kumar:2019xgp} could benefit from the improvement of such EoS-insensitive relations, which we derive in this paper. 


\begin{table*}[htb]
\centering
\begin{tabular}{ c  c  | c c c c c c c c c c} 
 \hline
 \hline
 $y$ & $x$ & $\alpha$ & $K_{yx}$ & $a_1$ & $a_2$ & $a_3$ & $a_{4}$ & $a_{5}$ & $b_1$ & $b_2$ & $b_3$ \\
 \hline
  $\bar{I}$ & $\Lambda$ & (--) & (--) & $1.493$ & $0.06410$ & $0.02085$ & $-5.018 \times 10^{-4}$ & $3.16 \times 10^{-7}$ & (--) & (--) & (--)  \\
 $\bar{Q}$ & $\Lambda$ & (--) & (--) & $0.2093$ & $0.07404$ & $0.05382$ & $-5.018 \times 10^{-3}$ & $1.576 \times 10^{-4}$ & (--)  &(--)  & (--) \\ 
  $\bar{I}$ & $\bar{Q}$ & (--) & (--) & $1.383$ & $0.5931$ & $-0.02161$ & $0.04190$ & $-2.968 \times 10^{-3}$ &(--)  &(--) & (--) \\
 \hline 
 $\bar{I}$ & $\Lambda$ & $2/5$ & $0.5313$ & $1.287$ & $0.09888$ & $-2.300$ & (--) & (--) & $-1.347$ & $0.3857$ & $-0.02870$\\
 $\bar{Q}$ & $\Lambda$ & $1/5$ & $3.555$ & $-2.122$ & $2.72$ & $-1.491$ & (--) & (--) & $0.8644$ & $-0.1428$ & $-1.397$\\
 $\bar{I}$ & $\bar{Q}$ & $2$ & $0.008921$ & $10.59$ & $-37.46$ & $43.18$ &(--)  &(--)  & $-2.361$ & $1.967$ & $-0.5678$\\
 $C$ & $\Lambda$ & $-1/5$ & $0.2496$ & $-919.6$ & $330.3$ & $-857.2$ & (--) & (--) & $-383.5$ & $192.5$ & $-811.1$\\
\hline
\hline
\end{tabular}
\caption{
Fit parameters for the I-Love-Q and C-Love relations using the constrained set and the fitting functions in Eq.~\eqref{eq:ILQfit} (top) and in Eq.~\eqref{eq:ILQfitNew} (bottom).
}\label{tab:ILQfitNew}
\end{table*}

\section{EoS-insensitive relations}
\label{sec:universal}

In this section, we repeat and improve the analyses of~\cite{Yagi:binLove,Yagi:ILQ} through the use of the two sets of EoSs (the constrained and the unconstrained set) and through new fitting functions that properly limit Newtonian results.   


\subsection{I-Love-Q relations}
\label{sec:ilq}

Here we present our results on the I-Love-Q universality, comparing when possible to Fig.~1 of~\cite{Yagi:ILQ}.
In particular, we consider two distinct classes of NSs: nuclear matter EoSs and hybrid quark-hadron star EoSs as described in Sec.~\ref{sec:theory}.
We begin in Sec.~\ref{sec:ilq-nuc} by fitting the new I-Love-Q relations using the constrained set of EoSs.
This is followed in Sec.~\ref{sec:ilq-hyb} by an analysis and discussion of how well the hybrid star EoSs agree with the improved binary Love relations. 

\subsubsection{Nuclear matter stars}
\label{sec:ilq-nuc}

Following~\cite{Yagi:ILQ}, we first fit the data for each EoS-insensitive relation to the following function:
\begin{equation}\label{eq:ILQfit}
\ln{y}=a_{1}+a_{2} \ln{x} + a_{3} (\ln{x})^2 + a_{4} (\ln{x})^3 + a_{5} (\ln{x})^4,
\end{equation}
where $y$ and $x$ correspond to NS observables $\bar{I}$, $\bar{Q}$, and $\Lambda$, and the updated coefficients are given in the top of Table~\ref{tab:ILQfitNew}. This fitting function, however, does not limit properly to the Newtonian results~\cite{Yagi:ILQ}:
\begin{equation}\label{eq:Newtonian}
\bar{I}^{\text{N}} = K_{\bar{I}\Lambda}\Lambda^{2/5}, \hspace{3mm} \bar{Q}^{\text{N}} = K_{\bar{Q}\Lambda}\Lambda^{1/5}, \hspace{3mm} \bar{I}^{\text{N}} = K_{\bar{I}\bar{Q}}\bar{Q}^{2},
\end{equation}
where $\Lambda^{-1/5} \sim C$ when $C \ll 1$. One can thus improve the fitting function to 
\begin{equation}\label{eq:ILQfitNew}
y=K_{yx} x^{\alpha} \frac{1+\sum_{i=1}^3 a_i x^{-i/5}}{1+\sum_{i=1}^3 b_i x^{-i/5}},
\end{equation}
where $\alpha$ is either $2/5$, $1/5$, or $2$ for the $\bar{I}-\Lambda$, $\bar{Q}-\Lambda$, and $\bar{I}-\bar{Q}$ relations respectively, and where the new fitting coefficients are presented in the bottom of Table~\ref{tab:ILQfitNew}. While the two fits result in similar $R^2$ values\footnote{$R^2$ is the coefficient of determination, defined as $\sum_i(f_i-\bar{y})^2/\sum_i(y_i-\bar{y})^2$, where $\bar{y}$ is the mean data value, and $f_i$, $y_i$ are the modeled and actual data values} of $\sim 0.999995$, the fit in Eq.~\eqref{eq:ILQfitNew} has the advantage that it properly limits to the Newtonian result as $\Lambda \gg 1$~\cite{Yagi:binLove}.

\begin{figure*}[htb]
\begin{center} 
\includegraphics[width=.32\textwidth]{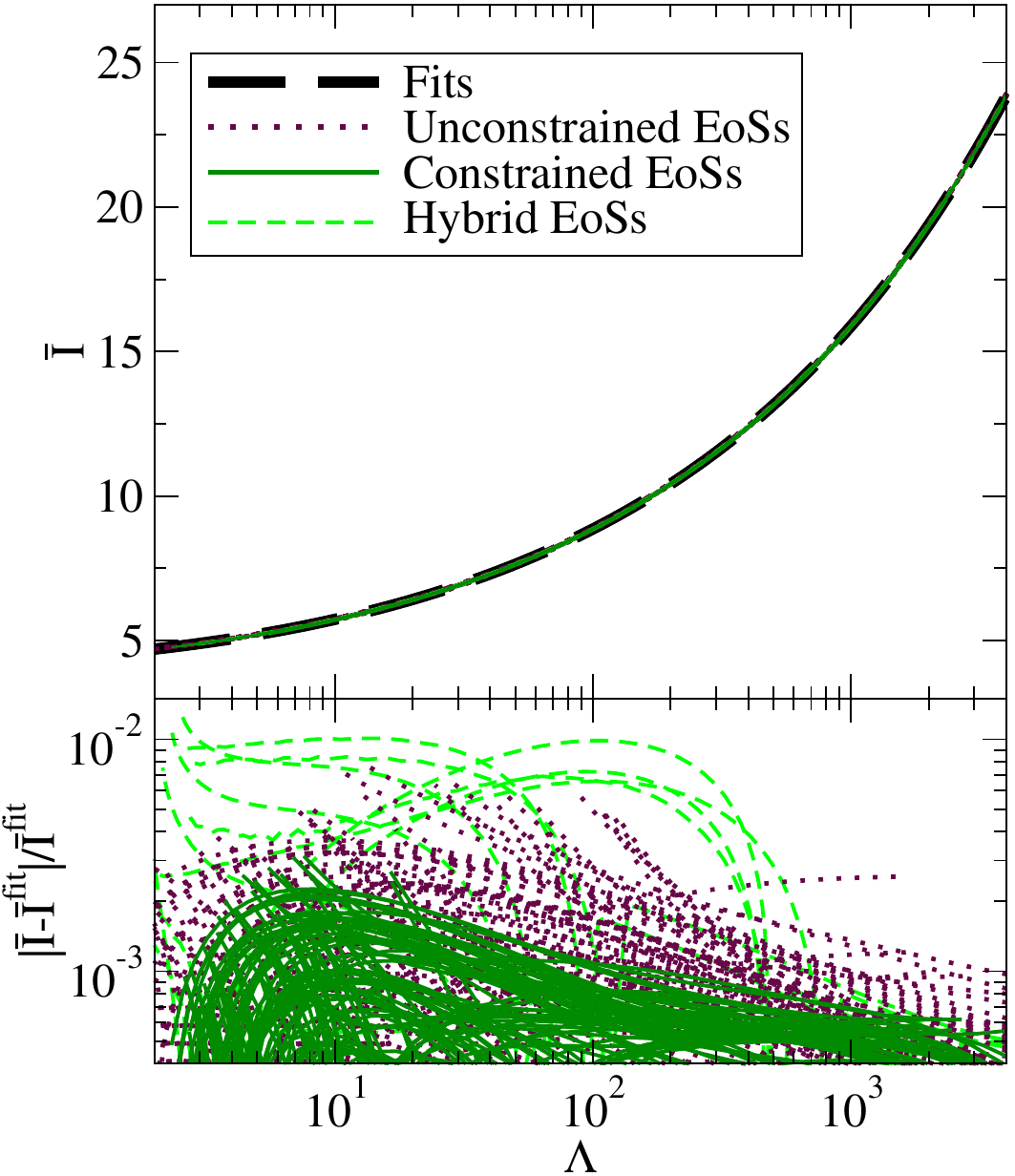}
\includegraphics[width=.32\textwidth]{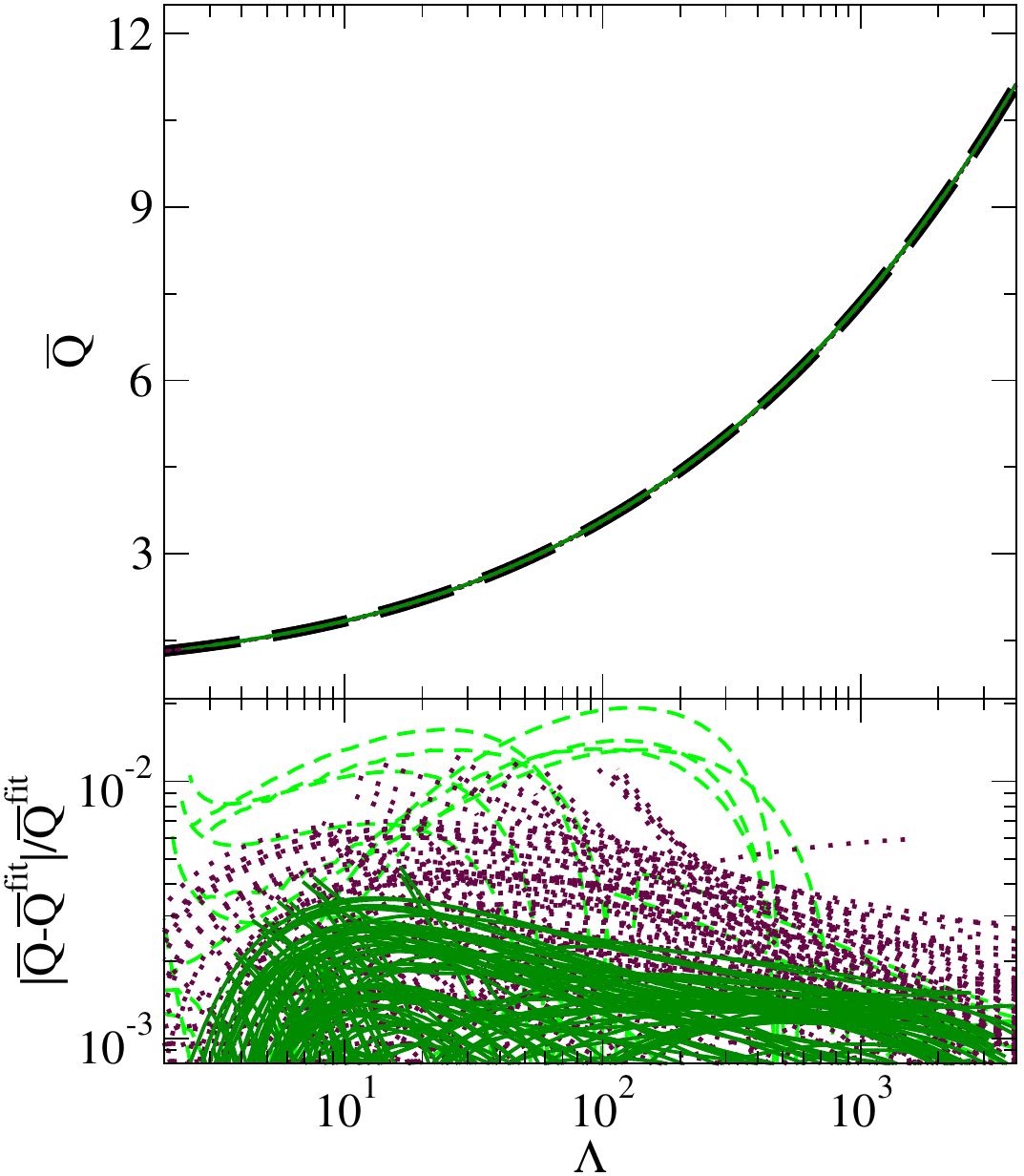}
\includegraphics[width=.32\textwidth]{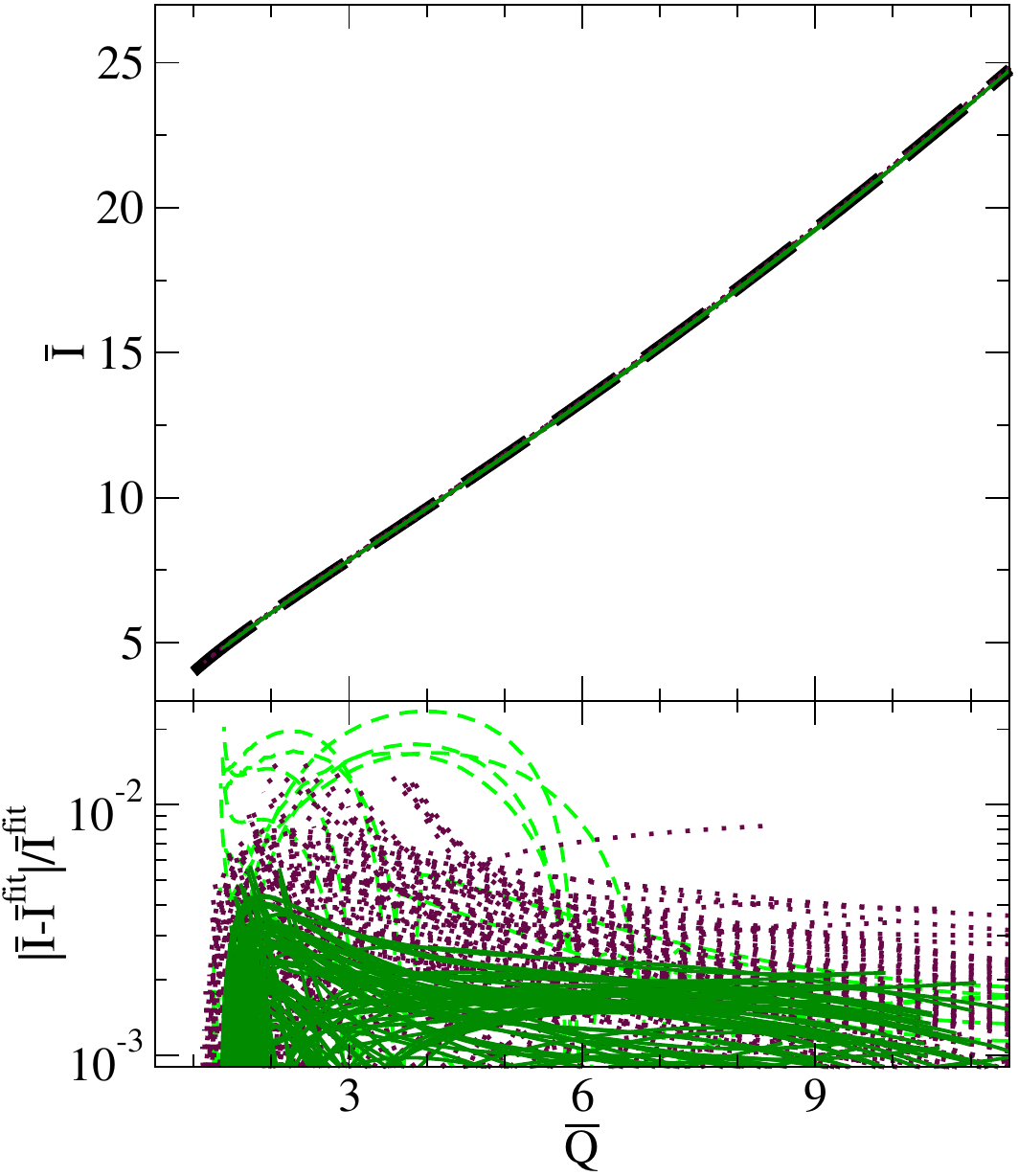}
\end{center}
\caption{
(Color Online) Individual I-Love-Q relations $\bar{I}-\Lambda$ (left), $\bar{Q}-\Lambda$ (center), and $\bar{I}-\bar{Q}$ (right), shown for both the constrained EoSs (solid green) and unconstrained EoSs (dotted maroon).
In these figures, the black dashed lines correspond to the fits given by Eq.~\eqref{eq:ILQfitNew}.
The fractional difference from the fits, shown in the bottom panels, is greatly suppressed for the constrained case, compared to both the unconstrained case, and results from previous works~\cite{Yagi:2013bca,Yagi:ILQ}.
The maximal EoS variation from the fits for the unconstrained and constrained sets of EoSs are compared in Table~\ref{tab:maxVar}.
Additionally shown in this figure is the fractional difference from the nuclear matter fits for the 10 hybrid star EoSs (dashed green).
}
\label{fig:ILQ}
\end{figure*}

Figure~\ref{fig:ILQ} shows the improved I-Love-Q relations between the dimensionless moment of inertia $\bar{I} \equiv I/M^3$, the dimensionless quadrupole moment $\bar{Q} \equiv Q/M^3$, and the dimensionless tidal deformability $\Lambda$. The fits are done using the new fitting function in Eq.~\eqref{eq:ILQfitNew} and over either the constrained set or the unconstrained set separately. The bottom panels show the relative fractional difference between the fit and the data for each set of EoSs. The new fits to the constrained set shows considerably more EoS-insensitivity than the fit to the unconstrained set. These results are summarized in Table~\ref{tab:maxVar}, which tabulates the maximum EoS variation in each fit. Clearly then, the EoS-insensitive relations can be made more universal by restricting the EoSs through the use of observations. Such improvements can be beneficial to future studies in GW astrophysics, such as those of~\cite{Kumar:2019xgp}, and in experimental relativity, such as those of~\cite{Gupta:2017vsl}. 

\begin{table}[htb]
\centering
\begin{tabular}{ c  || c c c } 
 \hline
 \hline
 \textbf{EoS-insensitive} & \multicolumn{3}{c}{\textbf{Maximal EoS Variability}} \\
 \cline{2-4}
 \textbf{Relation} & \multicolumn{1}{c|}{\emph{Previous}} & \multicolumn{1}{c|}{\emph{Unconstrained}} & \emph{Constrained}\\
 \hline
 $\bar{I}-\Lambda$ &  $0.0059$ & $0.0077$ & $0.0031$\\
 $\bar{Q}-\Lambda$ & $0.010$ & $0.013$ & $0.0047$\\
 $\bar{I}-\bar{Q}$ & $0.012$ & $0.015$ & $0.0057$\\
 \hline
 \multirow{2}{*}{$C-\Lambda$} & $0.065$ & $0.072$ & $0.022$\\
 & (--) & ($0.018$) & ($0.0066$)\\
  \hline
 \multirow{2}{*}{$R-\Lambda$} & -- & $0.056$ & $0.022$\\
 & (--) & ($880 \text{ m}$) & ($360 \text{ m}$) \\
 \hline
 $\Lambda_a-\Lambda_s$ & $\sim0.50$ & $0.57$ & $0.21$\\
 $q=0.90$ & (--) & ($190$) & ($37$) \\
 \cline{1-1}
 $\Lambda_a-\Lambda_s$ & $\sim0.20$ & $0.25$ & $0.083$\\
  $q=0.75$ & (--) & ($320$) & ($52$) \\
  \cline{1-1}
 $\Lambda_a-\Lambda_s$ & $\sim0.025$ & $0.038$ & $0.018$\\
  $q=0.50$ & (--) & ($240$) & ($29$) \\
  \cline{1-1}
\hline
\hline
\end{tabular}
\caption{Maximum relative and fractional EoS variation in the I-Love-Q, C-Love, R-Love, and binary Love relations using the unconstrained set, the constrained set and variations reported in previous work~\cite{Yagi:ILQ,Yagi:binLove}. The maximum absolute EoS variation is also reported in the C-Love, R-Love and binary Love cases in parentheses. The maximum variation in the constrained set case is better than a factor of two smaller than the variability in the unconstrained case and in previous work. The maximum EoS variation in the unconstrained set is slightly larger than that found in previous work because the former is built from a large random sampling of EoSs. 
}
\label{tab:maxVar}
\end{table}

\subsubsection{Hybrid quark-hadron stars}
\label{sec:ilq-hyb}

Let us now focus on the I-Love-Q relations of hybrid stars, and their compatibility with their nuclear matter counterparts. For concreteness, we consider three different sets of EoSs in the hybrid case:
\begin{enumerate}
\item the complete set of 100 constrained EoSs combined with the 10 hybrid star EoSs,
\item the complete set of 100 constrained EoSs alone,
\item the complete set of 10 hybrid star EoSs alone.
\end{enumerate}
For each of these cases, we compute the I-Love-Q relations, we fit the data to Eq.~\eqref{eq:ILQfitNew} and we compute the relative fractional difference. 

\begin{table}[htb]
\centering
\begin{tabular}{ c  || c c } 
 \hline
 \hline
 \textbf{Fitting} & \multicolumn{2}{c}{\textbf{Maximal EoS Variability}} \\
 \cline{2-3}
 \textbf{Case} &  \multicolumn{1}{c|}{\emph{Constrained}} & \emph{Hybrid}\\
 \hline
 \emph{Combined} &  \multirow{2}{*}{$0.0044$} & \multirow{2}{*}{$0.014$}\\
 \emph{(Case 1)} & &\\
 \cline{1-1}
 \emph{Constrained only} & \multirow{2}{*}{$0.0031$} & \multirow{2}{*}{$0.017$}\\
  \emph{(Case 2)} & &\\
  \cline{1-1}
 \emph{Hybrid only} & \multirow{2}{*}{$0.0084$} & \multirow{2}{*}{$0.010$}\\
  \emph{(Case 3)} & &\\
  \cline{1-1}
\hline
\hline
\end{tabular}
\caption{
Maximum relative and fractional EoS variation in the I-Love relation, fitting to three different sets of data: using the constrained set plus hybrid EoSs, using the constrained set alone, and using only the hybrid EoSs. In all 3 cases the hybrid EoSs are EoS-insensitive to $\sim1$\%, which is a slight decrease in universality relative to the hadronic only EoSs.
}\label{tab:hybridCompare}
\end{table}

The fractional differences of hybrid stars from the fit for the second case (fit to only the constrained EoSs) is shown with dashed green lines in Fig.~\ref{fig:ILQ}, while the maximum EoS variation for the I-Love relation is shown in Table~\ref{tab:hybridCompare} for both the constrained and hybrid star cases. The hybrid star EoSs obey the I-Love-Q relations in each case to better than $\sim1.7$\%, a variability that is slightly higher than that found when using only nuclear matter EoSs in previous works~\cite{Yagi:2013bca,Yagi:ILQ}, and is consistent with~\cite{Paschalidis2018}. The universality cannot be improved much through the introduction of new fits, only bringing the maximal EoS variation down to $\sim1$\% for the fits constructed with only hybrid star EoSs. From this study, we conclude that hybrid star EoSs \emph{do} obey the traditional nuclear I-Love-Q relations computed with nuclear EoS data, albeit with a slight decrease in universality to $\sim1.7$\%.  
We find this decrease in universality to be somewhat consistent with similar works~\cite{Paschalidis2018,Bandyopadhyay2018,PhysRevD.95.101302,Han:2018mtj}.
In such investigations, departures from universality of up to $\sim1\%-2\%$ were found with various hybrid star EoSs, compatible with our result of $\sim1.7$\%.


\begin{figure*}
\begin{center} 
\includegraphics[width=\columnwidth]{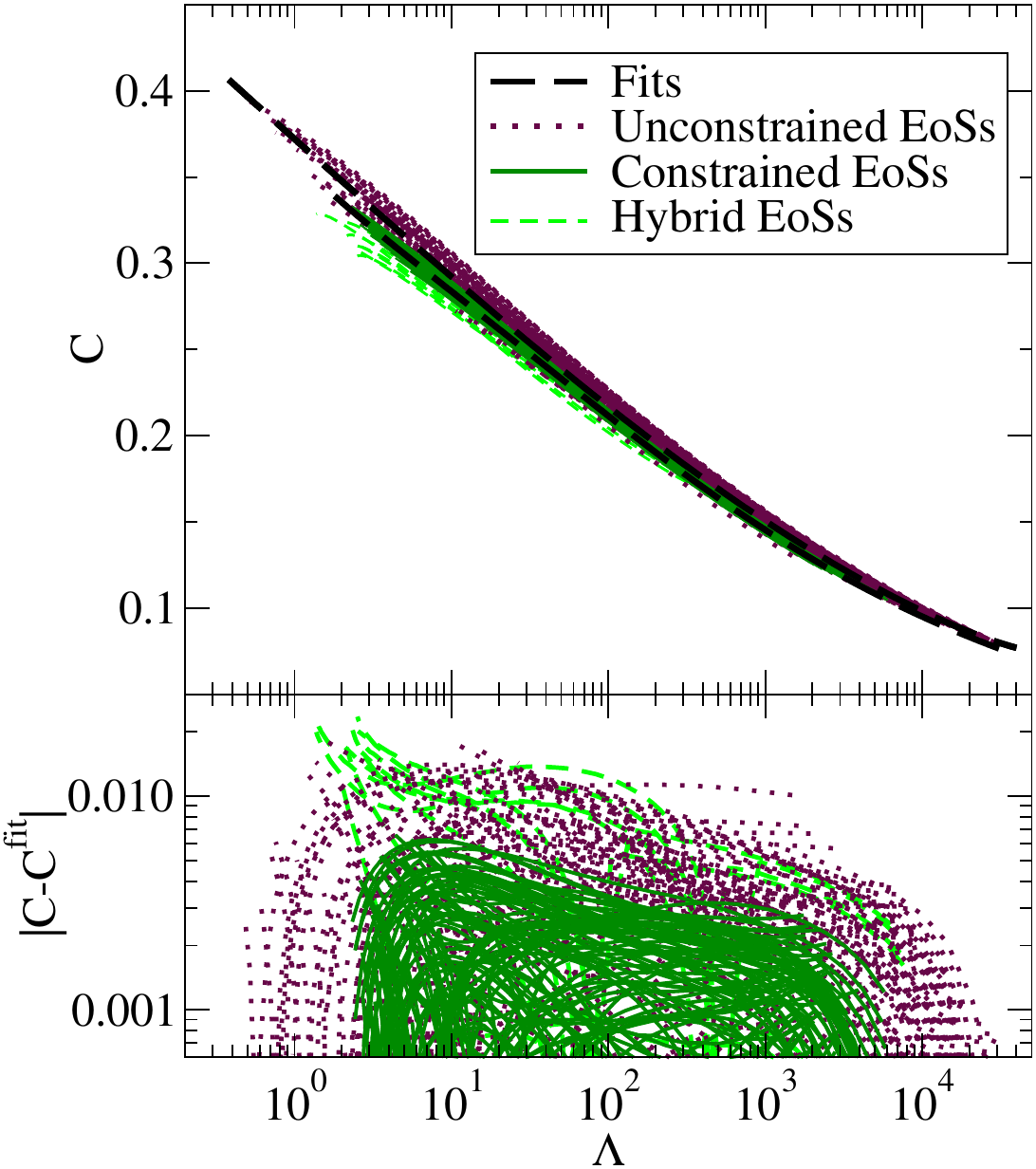}
\includegraphics[width=.96\columnwidth]{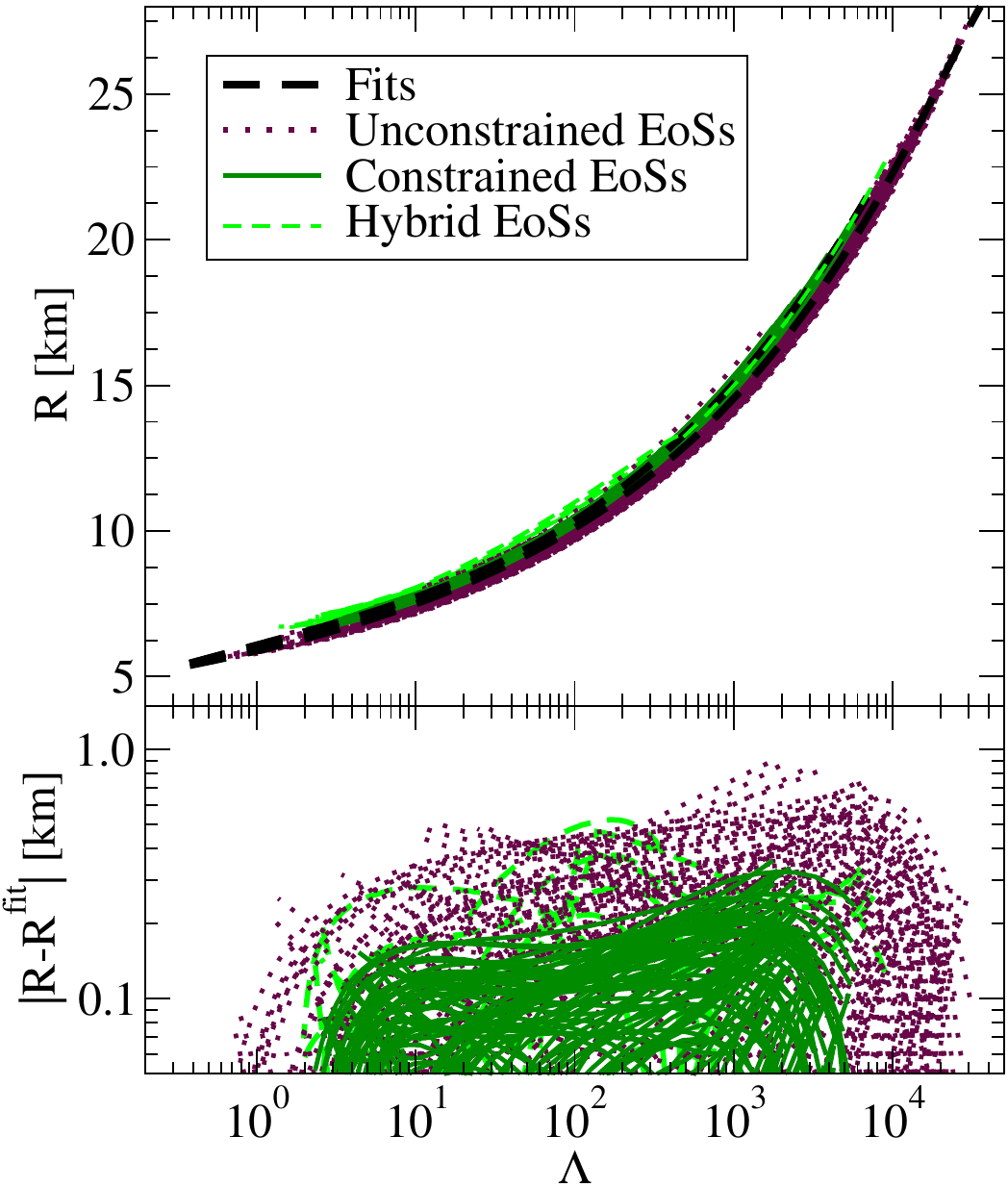}
\end{center}
\caption{
(Color online) Similar to Fig.~\ref{fig:ILQ} but for the C-Love (left) and the R-Love relations (right). In the top panels, we show two different fits, one for the unconstrained set and one for the constrained set of EoSs. The bottom panels show the \emph{absolute differences} (rather than fractional difference as in Fig.~\ref{fig:ILQ}) from the fit. The absolute difference is suppressed in the constrained set case relative to the unconstrained set, and results from previous work~\cite{Maselli:2013mva}. In the left panel, we also present the corresponding relations in the hybrid star cases (dashed green) for comparison, where, although still EoS-insensitive, the degree of universality decreases.
}
\label{fig:clove}
\end{figure*} 

\subsection{C-Love Relations}
\label{sec:clove}

In this subsection we focus on the EoS-insensitive C-Love relations, as introduced in~\cite{Yagi:2013bca,Yagi:ILQ,Maselli:2013mva}, as well as on the R-Love relations, since these play a key role in GW data analysis for the extraction of M-R credible intervals. As in the previous subsection, we consider both nuclear matter EoSs, as well as hybrid quark-hadron star EoSs.

\subsubsection{Nuclear matter stars}
\label{sec:clove-nuc}

Following Ref.~\cite{Maselli:2013mva}, we begin by fitting the data for each set of EoSs to the simple curve
\begin{equation}
C = \sum^2_{k=0} a_k (\ln{\Lambda})^k.
\end{equation}
Doing so, yields $a_0 = 0.3617$, $a_1 = -0.03548$, and $a_2 = 0.0006194$ for the constrained set of EoSs, similar to what was found in~\cite{Maselli:2013mva}. As in the I-Love-Q case, however, the above fitting function does not limit properly to the Newtonian result 
\begin{equation}
C^N=K_{C\Lambda}\Lambda^{-1/5}.\label{eq:cloveFit}
\end{equation}
We thus repeat the fit using Eq.~\eqref{eq:ILQfitNew} as the fitting function, with the new fitting coefficients presented in Table~\ref{tab:ILQfitNew}.

The left panel of Fig.~\ref{fig:clove} shows the C-Love relations for both the constrained and unconstrained sets, along with the corresponding \emph{absolute differences}\footnote{We present absolute differences instead of fractional differences since the former is what matters directly to the GW data analysis.} from the fits (instead of the fractional differences as done back in Fig.~\ref{fig:ILQ}). The fit to the constrained EoSs suppresses the EoS variability compared to the fit to the unconstrained set, as well as that of previous work~\cite{Maselli:2013mva}. The maximal EoS variation is compared between these three cases in Table~\ref{tab:maxVar}.

From the C-Love relations, we can also compute directly the R-Love relation using $R(\Lambda)=M/C(\Lambda)$ for NSs of mass $M$. The right panel of Fig.~\ref{fig:clove} shows the R-Love relations for the constrained and unconstrained sets, with the bottom panel showing the absolute difference of the data and the fits. The C-Love relations allow us to infer the NS radius to better than $\sim 350 \textrm{m} $ in the constrained case, while the error goes up to $1,000 \textrm{m}$ in the unconstrained case, as also tabulated in Table~\ref{tab:maxVar}. The systematic uncertainty in the radius using the constrained fit is thus comparable to the $\sim 140\textrm{m}$ systematic uncertainty in the radius due to the choice of EoS to model the crust~\cite{Gamba:2019kwu}. 
See also Refs.~\cite{Annala:2017llu,Raithel:2018ncd} for related work on the R-Love relations.

\subsubsection{Hybrid quark-hadron stars}
\label{sec:clove-hyb}

Let us now focus on whether the C-Love relations hold for hybrid stars. Much like in Sec.~\ref{sec:ilq-hyb}, we perform 3 separate fits: one to the constrained set plus 10 hybrid EoSs, another to the constrained set only, and a third to the 10 hybrid EoSs only. We then compare the EoS insensitivity in each case.

In the top left panel of Fig.~\ref{fig:clove}, we show the C-Love relations for hybrid stars (dashed green curves), while in the bottom panel, we show the absolute difference of the relations for such stars from the fit constructed only from the constrained set of EoSs (dashed green curves). The C-Love relations for hybrid stars remain EoS-insensitive, but the degree of universality is not as high as in the case of hadronic NSs. Table~\ref{tab:hybridCompareClove} compares the maximal EoS variability for the constrained and hybrid star EoSs fitting to the three data sets described above. As in Sec.~\ref{sec:ilq-hyb}, the maximal EoS variation for hybrid stars fluctuates only slightly ($\sim 4.5\% - 7\%$) in each case. From this, we conclude that hybrid stars do obey the C-Love relation derived for nuclear matter stars, with the caveat that the maximum universality increases to $\sim 7.1\%$.
For completeness, we also show the R-Love relations for hybrid stars in Fig.~\ref{fig:clove}.
The absolute differences from the (constrained EoS) fit are similarly displayed in the bottom panel, showing a maximum EoS variability of $\sim50\text{ m}$, consistent with the uncertainties found in the unconstrained relations.

\begin{table}
\centering
\begin{tabular}{ c  || c c } 
 \hline
 \hline
 \textbf{Fitting} & \multicolumn{2}{c}{\textbf{Maximal EoS Variability}} \\
 \cline{2-3}
 \textbf{Case} &  \multicolumn{1}{c|}{\emph{Constrained}} & \emph{Hybrid}\\
 \hline
 \emph{Combined} &  \multirow{2}{*}{$0.037$} & \multirow{2}{*}{$0.055$}\\
 \emph{(Case 1)} & &\\
 \cline{1-1}
 \emph{Constrained only} & \multirow{2}{*}{$0.022$} & \multirow{2}{*}{$0.072$}\\
  \emph{(Case 2)} & &\\
  \cline{1-1}
 \emph{Hybrid only} & \multirow{2}{*}{$0.058$} & \multirow{2}{*}{$0.045$}\\
  \emph{(Case 3)} & &\\
  \cline{1-1}
\hline
\hline
\end{tabular}
\caption{
Similar to Table~\ref{tab:hybridCompare} but for the C-Love relation.
For all 3 cases, the hybrid EoSs are only universal up to a minimum of $\sim1$\% (fractional difference from the fit), and the constrained EoSs typically outperform the hybrid ones (other than the third case where the hybrid stars show slightly better agreement to the fit). The second case is also shown in Fig.~\ref{fig:clove}.
}\label{tab:hybridCompareClove}
\end{table}


\subsection{Binary love relations}
\label{sec:binary}

\begin{figure}[htb]
\begin{center} 
\includegraphics[width=\columnwidth]{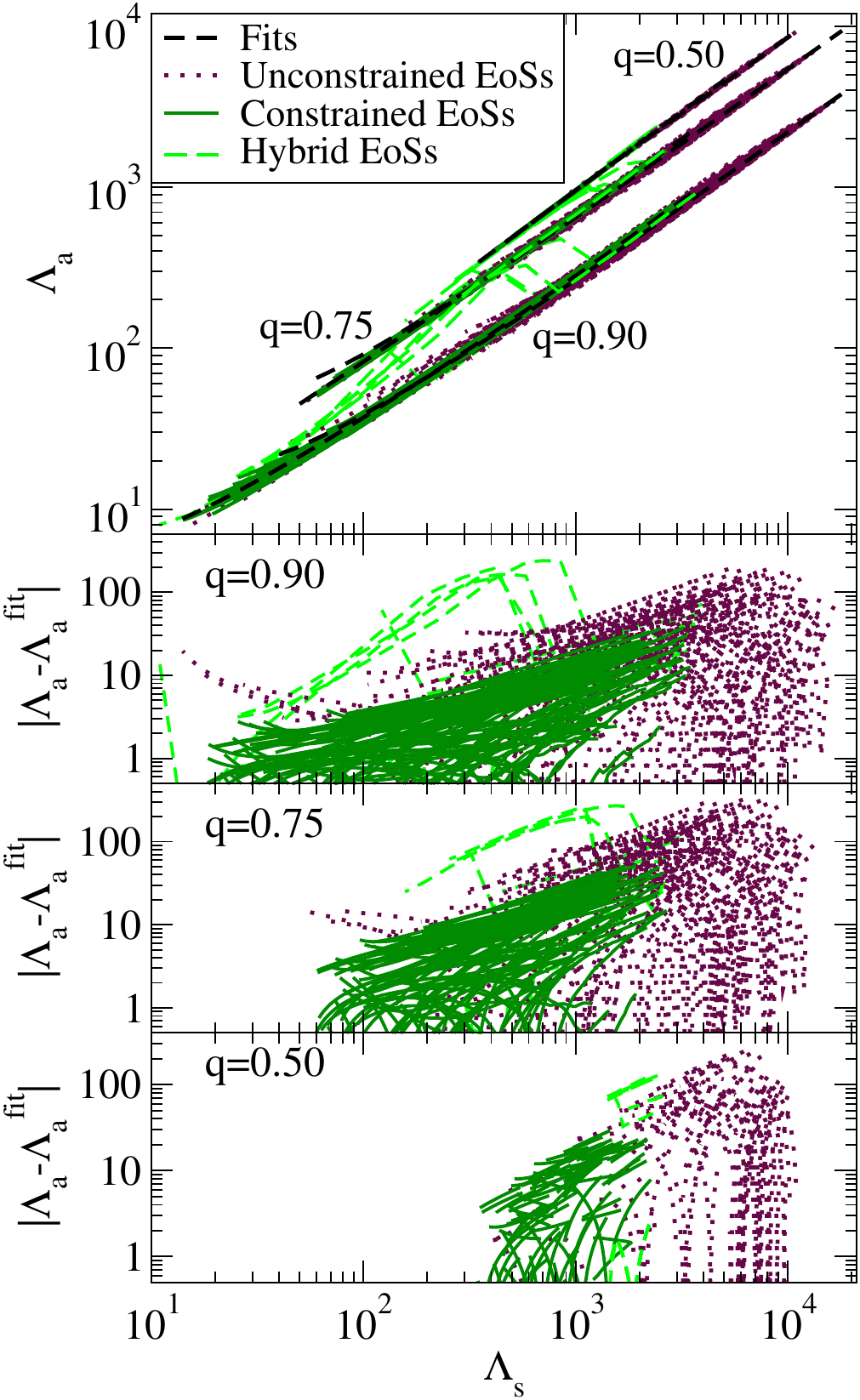}
\end{center}
\caption{(Color online) Binary Love relations using Eq.~\eqref{eq:binLovefit} fitted to the constrained (dotted maroon) and unconstrained sets (solid green), fixing $q=0.90$, 0.75 and 0.50. The bottom panels show the absolute difference (not the relative difference as in Fig.~\ref{fig:ILQ}) of the fit and the data. The fit to the constrained set shows a reduction in EoS variation relative to both the fit to the unconstrained set and previous work~\cite{Yagi:binLove}. For comparison, we also show the binary Love relations for the 10 hybrid star EoSs (dashed green curves), which present much larger EoS variation.
Observe here how both hadronic and hybrid star EoSs alike seemingly observe increasingly small distinction from the fits as the mass ratio $q$ approaches 0.
While the absolute errors remain somewhat consistent with mass ratio, the fractional differences from the fit approach $0$ as $q\rightarrow0$ as the universal relation becomes exact ($\Lambda_s\rightarrow\Lambda_a$).
}
\label{fig:binLove}
\end{figure} 

Let us now consider the binary Love relations. As for other relations, we consider nuclear matter stars and hybrid stars separately.

\subsubsection{Nuclear matter stars}

Following Ref.~\cite{Yagi:binLove}, we begin by fitting the binary Love relations to the constrained and unconstrained sets using the two-dimensional curve:
\begin{equation}\label{eq:binLovefit}
\Lambda_a=F_n(q) \frac{1+ \sum_{i=1}^3 \sum_{j=1}^2 b_{ij}q^j \Lambda_s^{-i/5}}{1 + \sum_{i=1}^3 \sum_{j=1}^2 c_{ij}q^j \Lambda_s^{-i/5}} \Lambda_s^{\alpha},
\end{equation}
where $q\equiv m_{2}/m_{1}$ is the mass ratio with $m_2 \leq m_1$, and $F_n(q)$ is the Newtonian-limiting controlling factor, given by
\begin{equation}\label{eq:control}
F_n(q) \equiv \frac{1-q^{10/(3-n)}}{1+q^{10/(3-n)}}.
\end{equation}
The coefficients of the fit to the constrained set are 
\begin{align}
b_{ij} = \begin{bmatrix}
    -14.40 & 14.45   \\
    31.36 & -32.25   \\
    -22.44 & 20.35   
\end{bmatrix}
\quad
c_{ij} = \begin{bmatrix}
    -15.25 & 15.37   \\
    37.33 & -43.20   \\
    -29.93 & 35.18   
\end{bmatrix},
\end{align}
where we set $n = 0.743$ and $\alpha = 1$ as was done in Ref.~\cite{Yagi:binLove}.
Unlike the individual I-Love-Q relations, the binary Love relations depend also on the mass ratio. 

Figure~\ref{fig:binLove} shows the improved binary Love relation for 3 different mass ratios ($q=0.9$, 0.75 and 0.5) for both the constrained and unconstrained sets of EoSs. Once again we find that the constrained set shows a considerable increase in EoS-insensitivity relative to both the unconstrained set and previous works~\cite{Yagi:binLove}. As before, the fit to the unconstrained set shows similar, yet slightly larger EoS variation than that found in previous work~\cite{Yagi:binLove} due to the random sampling used in our analysis. The absolute error between the fit and the data is approximately independent of the mass ratio $q$, with a subtle maximum at $q=0.75$, which mimics the behavior of the controlling factor $F_n(q)$. As before, maximum EoS variation is tabulated in Table~\ref{tab:maxVar} for each value of mass ratio considered here.

We can extract two additional conclusions from Fig.~\ref{fig:binLove}. First, as the mass ratio becomes more extreme (i.e.~away from unity), the ranges of values that $\Lambda_s$ and $\Lambda_a$ can take decrease. This is because as the separation between $m_1$ and $m_2$ increases, the effective number of possible $\Lambda_1(m_1)$-$\Lambda_2(m_2)$ configurations decrease, and thus, the allowed parameter space shrinks. Second, the fit to the constrained set does not extend to as high values of $\Lambda_s$ as the fit to the unconstrained set does. This is because of how the sets of EoSs were generated in our analysis, with the unconstrained set including some EoSs that are stiffer\footnote{Typically, the ``stiffness" of an EoS is determined by the amount of pressure gained given an increase in density. Stiff EoSs have steep pressures and predict larger maximum mass NSs, while soft EoSs have shallow pressures and predict smaller maximum mass NSs.} and others that are softer than those constrained by the GW170817 event. This results in the unconstrained set  containing a larger range of tidal deformabilities $\Lambda_{1,2}$ (from both above and below) than the constrained set, which ultimately leads to a \emph{larger} range of $\Lambda_s=\frac{1}{2}(\Lambda_1+\Lambda_2)$ values.

\subsubsection{Hybrid quark-hadron stars}

Let us now consider the binary Love relations for hybrid stars. As described in Sec.~\ref{sec:theory}, transitional quark-hadron matter stars undergo strong first-order phase transitions at a transitional pressure $P_{\text{tr}}$, where the hadronic branch departs into a quark-matter branch at the corresponding transitional mass $M_{\text{tr}}$. These transitions result in two distinct types of NSs, based on their observed mass: (i) massive ($M \geq M_{\text{tr}}$) hybrid stars that have quark-matter inner cores and nuclear matter elsewhere (which recall we denote ``HS"); and (ii) low-mass ($M \leq M_{\text{tr}}$) hadronic stars with no internal transition to quark matter (which recall we denote ``NS"). We therefore expect the binary Love relations for hybrid stars to present behavior identical to their purely hadronic counterparts below the transitional mass (or correspondingly above a transitional $\Lambda_{s,\textrm{tr}}$), and very different behavior for higher masses (or correspondingly for smaller $\Lambda_{s,\textrm{tr}}$). 

This behavior is exactly what we observe in Fig.~\ref{fig:binLove} for the binary Love relations of hybrid stars. Indeed, contrary to the case of EoS-insensitive relations for isolated hadronic stars (I-Love-Q, C-Love, and R-Love) which remain moderately EoS-insensitive for all hybrid stars, the binary Love relations depart from their hadronic counterparts below some transitional $\Lambda_{s}$. This is because at low pressures ($\Lambda_s>\Lambda_{s,\textrm{tr}}$) the binary is purely of NS/NS type (with the EoS for both stars a member of the constrained set), but once the critical pressure is reached ($\Lambda_{s}<\Lambda_{s,\textrm{tr}}$), one or both stars transition into the hybrid branch, resulting in large reductions in tidal deformability, as shown in the left panel of Fig.~\ref{fig:clove}. When one star lies on the hadronic-matter branch while the other is on the quark-matter branch, the difference between tidal deformabilities becomes larger than expected for pure hadronic-matter stars. This results in large deviations in the sums and differences of tidal deformabilities $(\Lambda_2 \pm \Lambda_1)$, disrupting the overall universality and generating the large ``bump" in the binary Love relations for hybrid stars\footnote{This discrepancy is not present in the I-Love-Q and C-Love relations due to their single-star nature.} seen in Fig~\ref{fig:binLove}.

From this analysis, we conclude that binaries containing one or more hybrid stars do \emph{not} satisfy the same binary Love relations as binaries that contain only traditional nuclear matter stars. This does not imply, however, that more sophisticated binary Love relations cannot be constructed that will remain EoS-insensitive and able to model both types of binary systems. Such relations, however, would necessarily have to include (at least) one new parameter that determines the transition between the hadronic and the quark branches, such that the ``bump'' in the binary Love relations can be properly modeled. Work along these lines is outside the scope of this paper.


\section{Impact on future observations}
\label{sec:observations}

We have now shown that binary NS merger observations can help improve the EoS-insensitive relations, but the question remains: is it worth it?
Current interferometer sensitivities are not yet high enough to accurately constrain the dominant tidal parameter $\tilde{\Lambda}$.
For example, GW170817 was detected by the second LIGO observing run (``O2")~\cite{aLIGO} and Virgo~\cite{TheVirgo:2014hva}, and was able to constrain $\tilde{\Lambda}$ to a $90\%$ credible interval centered at $\mu_{\tilde{\Lambda}}=395$ and with a width of 325~\cite{Abbott2018} (or $\sigma_{\tilde{\Lambda}} \approx 198$).  This corresponds to statistical uncertainties of ${\cal{O}}(80\%)$, which dominates the error budget compared to the small systematic uncertainties picked up by EoS variation in the EoS-insensitive relations.
This implies that currently, the use of improved EoS-insensitive relations will only make a negligible difference on the extraction of tidal parameters.

In this section, we further explore this question and study when in the future the new set of improved relations will become important as current detectors are improved and new ones are built. In Sec.~\ref{sec:marginalization}, we first estimate the systematic uncertainties introduced by using the improved binary Love relations.
In Sec.~\ref{sec:futureObservations}, we estimate the statistical uncertainties on the extraction of tidal parameters, and compare them to the above-mentioned systematic uncertainties.
This is repeated for 5 future detectors, where multiple detections become important, but to combine constraints from multiple events we cannot use the $\tilde\Lambda$ parameterization, as this depends on the masses of the binary constituents. To remedy this, we re-parameterize the waveform in terms of the coefficients $\lambda_0$ (sometimes called $\Lambda_{1.4}$) and $\lambda_1$, obtained from a Taylor expansion of the dimensionless tidal deformability $\Lambda$ about a ``canonical" reference mass $M_0=1.4\text{ M}_{\odot}$~\cite{delPozzo:TaylorTidal,Yagi:binLove}:
\begin{equation}
\Lambda = \lambda_0 + \lambda_1 \left(1-\frac{M}{M_0}\right) + {\cal{O}}\left[\left(1-\frac{M_{0}}{M}\right)^{2}\right].
\end{equation}
The Taylor coefficients $\lambda_0$ and $\lambda_1$ are mass independent, and thus they are identical in all binary NS observations, and their posteriors may be combined.

\subsection{Error Marginalization}\label{sec:marginalization}

Although the improved binary Love relation $\Lambda_a^{\text{relation}}(\Lambda_s,q)$ shows a high degree of universality, any residual EoS dependence in the relation could introduce a systematic bias and lead to incorrect inferences about the correct EoS. Let us then discuss a method to marginalize over the residual EoS-dependence of the binary Love relations. The residual is here defined as 
\begin{equation}
\label{eq:residual}
r(\Lambda_{s},q) \equiv \Lambda_a^{\text{relation}}(\Lambda_s,q)-\Lambda_a^{\text{true}}\,,
\end{equation}
where $\Lambda_a^{\text{relation}}(\Lambda_s,q)$ is given by the binary Love fit found in Sec.~\ref{sec:binary}, while $\Lambda_a^{\text{true}}$ is the true value predicted by choosing a particular EoS in the constrained set and solving for the tidal deformabilities numerically. 

Following the proposal in~\cite{Katerina:residuals}, one can model the residual EoS-sensitivity by enhancing the binary Love relations through 
\begin{equation}
\Lambda_a^{{\textrm{new relation}}}=\Lambda_a^{\text{relation}}(\Lambda_s,q)+\mathcal{N}(\mu_{r}(\Lambda_s,q),\sigma_{r}(\Lambda_s,q)),
\end{equation}
where $\mathcal{N}(\mu_r,\sigma_r)$ is a normal distribution with mean and variance $\mu_{r}$ and $\sigma_{r}^2$. Let us further assume that the residuals $r$ obey a Gaussian distribution with mean and standard deviation that can be decomposed as
\begin{align}
\mu_{r}(\Lambda_s,q) &=\frac{\mu_{r}(\Lambda_s)+\mu_{r}(q)}{2},\\ 
\sigma_{r}(\Lambda_s,q) &=\sqrt{\sigma_{r}^2(\Lambda_s) + \sigma_{r}^2(q)},
\end{align}
where $\mu_{r}(q)$ and $\mu_{r}(\Lambda_{s})$ are the means marginalized over $\Lambda_{s}$ and $q$ respectively, while $\sigma_{r}(q)$ and $\sigma_{r}(\Lambda_{s})$ are the standard deviations marginalized over $\Lambda_{s}$ and $q$ respectively.

Clearly then, to account for the residual EoS-sensitivity, we must first find the marginalized mean and standard deviation of the residual function, which we accomplish as follows. We first generate $\Lambda_{s}^{\textrm{true}}$ data by sampling through various values of mass ratio $q\in \lbrack0.36,1\rbrack$ and symmetric tidal deformability $\Lambda_{s} \in \lbrack 4,4600 \rbrack$ over all 100 elements of the constrained set of EoSs. For each value of $(q,\Lambda_{s})$ we then compute $\Lambda_{a}^{\textrm{relation}}$ from the binary Love relation, and then compute the residuals through Eq.~\eqref{eq:residual}. We then proceed to marginalize over $q$ by binning the residuals in $\Lambda_{s}$, which returns a distribution of residuals that is only a function of the binned $\Lambda_{s}$, and from which we can compute the mean $\mu_{s}(\Lambda_{s})$ and the standard deviation $\sigma_{r}(\Lambda_{s})$. Repeating this procedure by marginalizing over $\Lambda_{s}$ then allows us to compute the mean $\mu_{s}(q)$ and the standard deviation $\sigma_{r}(q)$. Finally, we fit the means and standard deviations to the functions~\cite{Katerina:residuals}
\begin{align}
\mu_{r}(\Lambda_s) &= \mu_1 \Lambda_s + \mu_2, \label{eq:margFit1}\\ 
\mu_{r}(q) &= \mu_3 q^2 + \mu_4 q + \mu_5, \label{eq:margFit2}\\ 
\sigma_{r}(\Lambda_s) &= \sigma_1 \Lambda_s^{5/2} + \sigma_2 \Lambda_s^{3/2} + \sigma_3 \Lambda_s +  \sigma_4 \Lambda_s^{1/2} + \sigma_5, \label{eq:margFit3}\\ 
\sigma_{r}(q) &= \sigma_6 q^3 + \sigma_7 q^2 + \sigma_8 q + \sigma_9, 
\label{eq:margFit4}
\end{align}
where the fitting parameters $\mu_i$ and $\sigma_i$ are tabulated in Table~\ref{tab:marginalized}. 

\begin{table}
\centering
\addtolength{\tabcolsep}{1pt} 
\begin{tabular}{ c | c || c | c}
\hline 
\noalign{\smallskip}
$\mu_1$ & $3.509 \times 10^{-3}$ & $\sigma_1$ & $-2.074 \times 10^{-7}$\\
$\mu_2$ & $9.351 \times 10^{-1}$ & $\sigma_2$ & $1.492 \times 10^{-3}$\\
$\mu_3$ & $-18.07$ & $\sigma_3$ & $-4.891 \times 10^{-2}$\\
$\mu_4$ & $27.56$ & $\sigma_4$ & $8.207 \times 10^{-1}$\\
$\mu_5$ & $-10.10$ & $\sigma_5$ & $-1.308$\\
 &  & $\sigma_6$ & $-63.76$\\
 &  & $\sigma_7$ & $11.14$\\
 &  & $\sigma_8$ & $75.25$\\
 &  & $\sigma_8$ & $-23.69$\\
 \noalign{\smallskip}
 \hline
\end{tabular}
\caption{
Coefficients to the fits given by Eqs.~(\ref{eq:margFit1})-(\ref{eq:margFit4}) for the relative error on $\Lambda_a$ in the improved binary Love EoS-insensitive relations presented in this paper.
}\label{tab:marginalized}
\addtolength{\tabcolsep}{-1pt}
\end{table}

\begin{figure}
\begin{center} 
\includegraphics[width=\columnwidth]{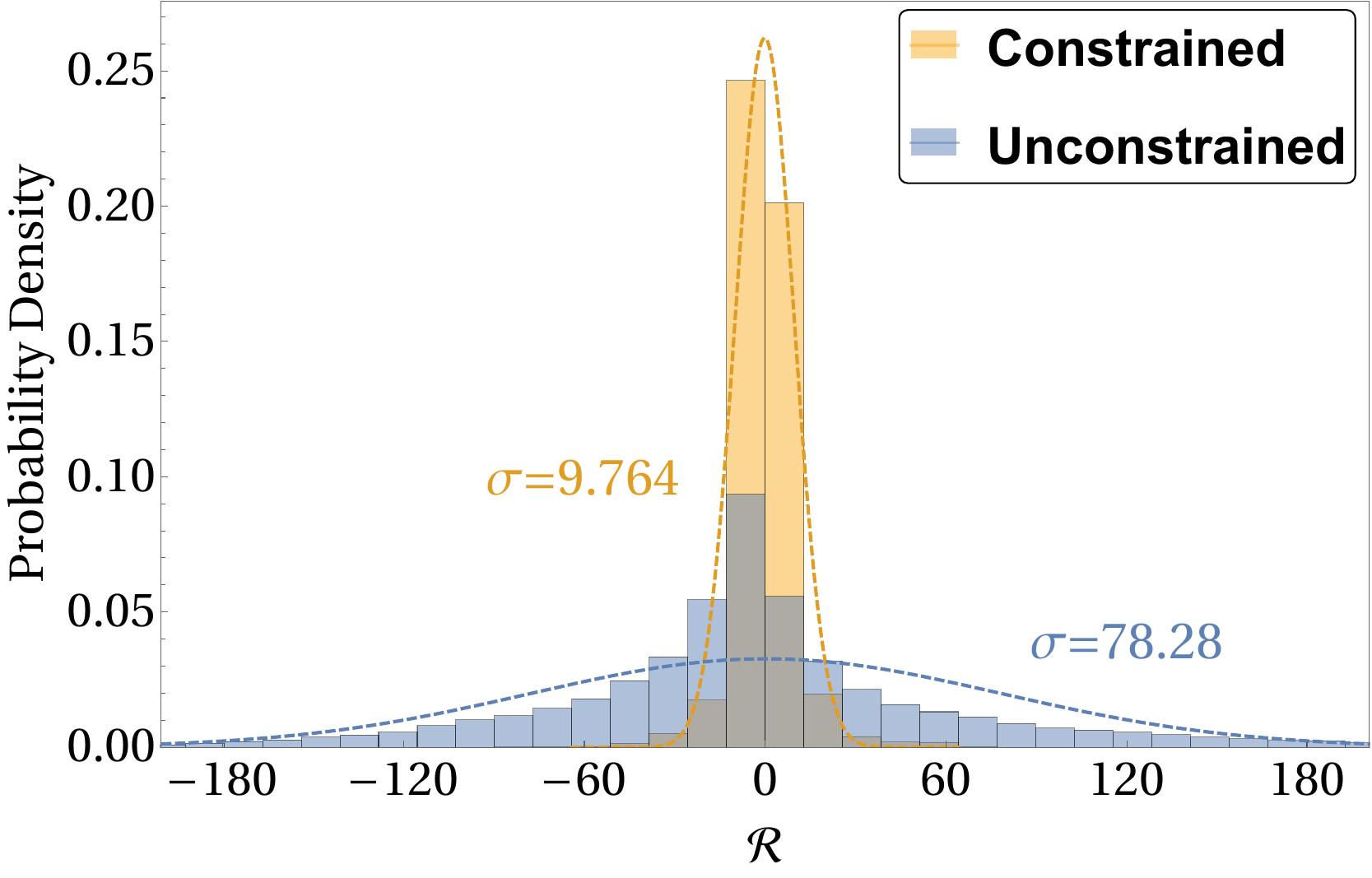}
\end{center}
\caption{
Residuals on $\lambda_0$ computed as $\mathcal{R}=\lambda_0^{\text{relation}}-\lambda_0^{\text{true}}$ (computed by sampling $q\in \lbrack0.36,1\rbrack$ and $\Lambda_{s} \in \lbrack 4,4600 \rbrack$ for each of the 100 EoS samples) for the binary Love relations modeled in Sec.~\ref{sec:binary} for both constrained and unconstrained sets of EoSs.
We fit these residuals with Gaussian distributions centered at $\mu=-0.1530$ and $\mu=0.2710$ with standard deviations of $\sigma=9.764$ and $\sigma=78.28$ for the constrained and unconstrained sets of EoSs, respectively.
These uncertainties correspond roughly to the systematic uncertainties introduced on the parameter extraction of $\lambda_0$ upon the use of binary Love relations.
However, to take into account the systematic uncertainties found in high-error regions of the parameter space, we instead set the systematic uncertainty to be the 90th percentile, $P_{90}=13.19$.
The systematic uncertainties from using the improved (constrained) binary Love relations are negligible compared to the statistical uncertainties accrued on parameter extraction from GW170817, found to be $\sigma_{\lambda_0}=170.1$.
}
\label{fig:residuals}
\end{figure}

In what follows, however, we will be interested in comparing an estimate of the systematic uncertainties in $\lambda_0$ due to the residual EoS-sensitivity in the binary Love relations to the statistical error in the extraction of this parameter. To estimate the former, we first numerically calculate $\lambda_{0}^{\textrm{true}}$ by sampling again in $(q,\Lambda_{s})$ within the same ranges as before and over all 100 elements of the constrained and the unconstrained sets of EoSs. We then use the binary Love relation to calculate $\lambda_0^{\text{relation}}$ over the $(q,\Lambda_{s})$ sampled points, and from this we compute the residual.  
\begin{equation}
{\cal{R}}(\Lambda_{s},q) = \lambda_0^{\text{relation}}(\Lambda_s,q)-\lambda_0^{\text{true}}.
\end{equation}
Figure~\ref{fig:residuals} shows a histogram of ${\cal{R}}$ for both the constrained and unconstrained set of EoSs. The standard deviations of the two histograms, $\sigma=9.764$ and $\sigma=78.28$, show a large decrease in EoS-variability from the unconstrained to the constrained sets of EoSs. From this distribution, we can additionally find the 90\%, 99\%, and 100\% credible interval on $\lambda_0$, namely $P_{90}=13.19$, $P_{99}=42.38$, and $P_{100}=111.32$ for the constrained EoSs.

\begin{figure}
\begin{center} 
\begin{overpic}[width=\columnwidth]{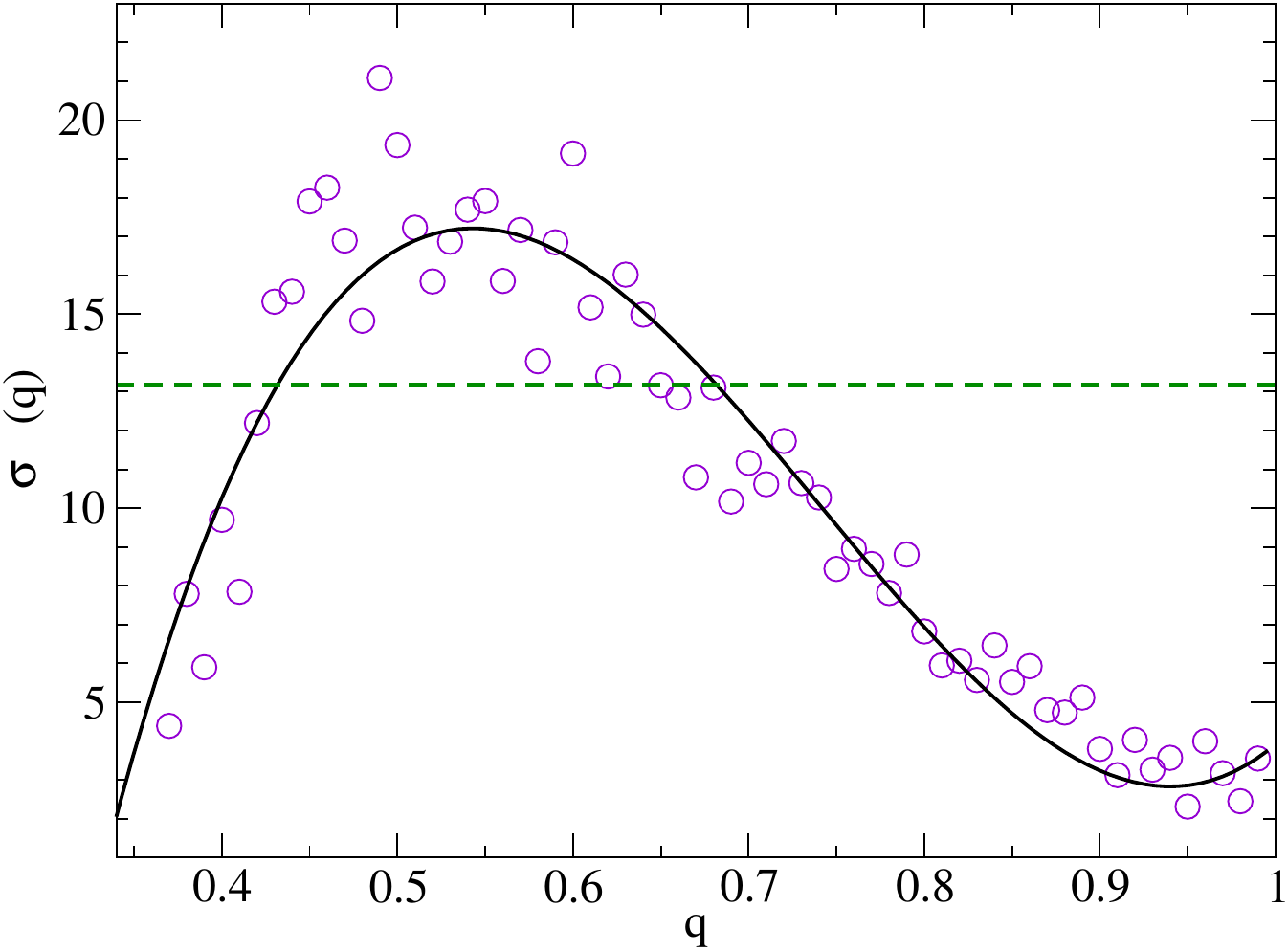}
\put(7,94){\tiny\rotatebox{90}{$\mathcal{R}$}}
\end{overpic}
\begin{overpic}[width=\columnwidth]{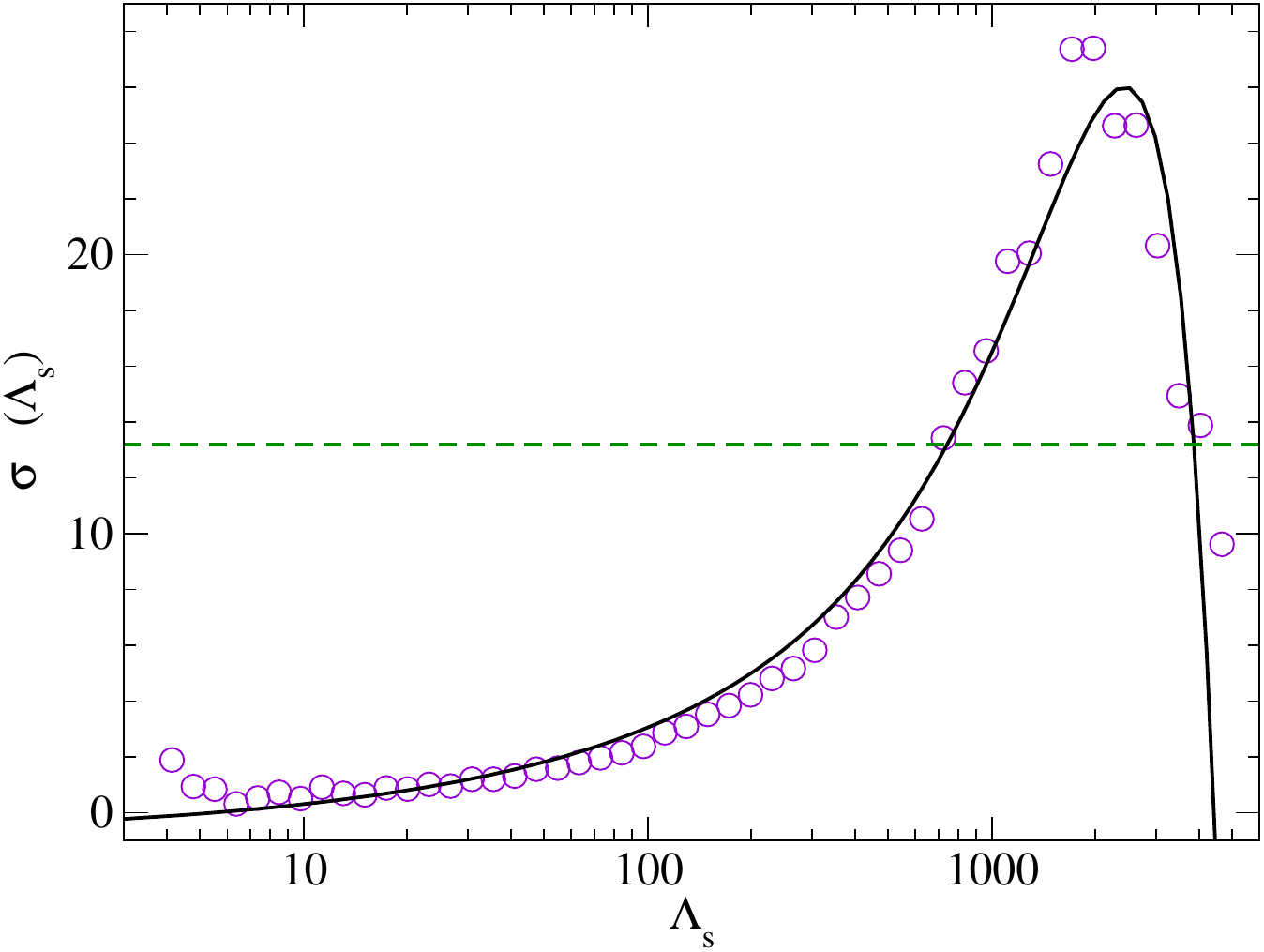}
\put(7,95){\tiny\rotatebox{90}{$\mathcal{R}$}}
\end{overpic}
\end{center}
\caption{Standard deviations of the $\lambda_0$ residuals $\mathcal{R}$ binned in $q$ (top) and $\Lambda_s$ (bottom), highlighting the different error weights across the entire $(q,\Lambda_s)$ parameter space.
In this figure, the violet circles indicate the standard deviation of each bin in $q$-space ($\Lambda_s$-space), while the solid black lines represent the best fit given by Eqs. (\ref{eq:margFit3})-(\ref{eq:margFit4}).
The uncertainty is maximal for $q\sim0.5$ and $\Lambda_s\sim2000$, while it becomes minimal for both low and high values of $q$ and $\Lambda_s$.
Also shown in the figure is the 90th percentile of the un-binned residuals seen in Fig.~\ref{fig:residuals}, taken to be the overall systematic uncertainty introduced by using binary Love relations.
}
\label{fig:qLsResiduals}
\end{figure}

Figure~\ref{fig:qLsResiduals} shows the standard deviations of the $\lambda_0$ residuals binned in $q$ (top panel) and $\Lambda_{s}$ (bottom panel), as was done previously for the $\Lambda_a$ residuals. The EoS-variability in $\lambda_0$ is dominated by the region around $q \sim 0.5$ and $\Lambda_{s} \sim 2000$. Henceforth, in order to take into account these high-error regions of parameter space which may get averaged out, we approximate the EoS-variability in the binary Love relations with the 90\% credible interval (rather than the $1\sigma$ uncertainty), which for the $\lambda_{0}$ parameter is $P_{90}=13.19$ (indicated by the dashed green line in Fig.~\ref{fig:qLsResiduals} and the dashed indigo line in Fig.~\ref{fig:stackedFisher}). We can similarly derive a conservative estimate of the systematic uncertainty due to EoS-variability in the binary Love relations in terms of $\tilde\Lambda_{a}$ to find $12.01$, which is what we use later in Fig.~\ref{fig:singleFisherLt}.
Here we note that these results rely on the assumption of a purely hadronic EoS, as the constrained set of EoSs do not contain any hybrid stars.
This assumption was also made in Ref.~\cite{LIGO:posterior} upon the derivation of the posterior probability distribution.

\subsection{Future Observations}\label{sec:futureObservations}

We now estimate the feasibility of using the improved EoS-insensitive relations in future GW observations of the coalescence of binary NSs. We estimate the statistical accuracy to which parameters can be extracted through a simple Fisher analysis~\cite{Finn:Fisher,Cutler:Fisher}, assuming sufficiently high signal-to-noise ratio and Gaussian noise~\cite{Cutler:Fisher,Berti:Fisher,Poisson:Fisher}. 

Consider then a waveform template $h$ parameterized in terms of
\begin{equation}\label{eq:template}
\theta^a=(\ln{A},\phi_c,t_c,\ln{\mathcal{M}},\ln{\mathcal{\eta}},\chi_s,\chi_a,\lambda_0, \lambda_1),
\end{equation}
where $\eta \equiv m_1 m_2/m^2$ is the symmetric mass ratio with $m_{1,2}$ and $m$ being the individual and total masses, $\mathcal{M}=m \eta^{3/5}$ is the chirp mass, $A \equiv {\mathcal{M}^{5/6}}/({\sqrt{30}\pi^{2/3}D_L})$ is a normalized amplitude factor with $D_L$ the luminosity distance, and $\chi_{s,a}=\frac{1}{2}(\chi_1\pm\chi_2)$ are the symmetric and antisymmetric dimensionless non-precessing spin parameters. Given a sufficiently loud observation and Gaussian noise, the resulting posterior distribution on the estimated parameters is Gaussian with a root-mean-square error given by
\begin{equation}
\Delta \theta^a=\sqrt{\Big( \tilde{\Gamma}^{-1}\Big)^{aa}},
\end{equation}
where the Fisher information matrix $\tilde{\Gamma}$ is
\begin{equation}
\tilde{\Gamma}_{ab} \equiv \Big( \frac{\partial h}{\partial \theta^b} \Big| \frac{\partial h}{\partial \theta^a}\Big) + \frac{1}{\sigma_{\theta^a}^2} \delta_{ab}\,.
\end{equation}
Here $\sigma_{\theta^a}$ are the parameters' prior root-mean-square estimate, and the inner product is defined by:
\begin{equation}
(a|b) \equiv 2 \int^{\infty}_0\frac{\tilde{a}^*\tilde{b}+\tilde{b}^*\tilde{a}}{S_n(f)}df,
\end{equation}
with the overhead tilde and the star representing the Fourier transform and complex conjugation respectively. 

\begin{figure}
\begin{center} 
\includegraphics[width=\columnwidth]{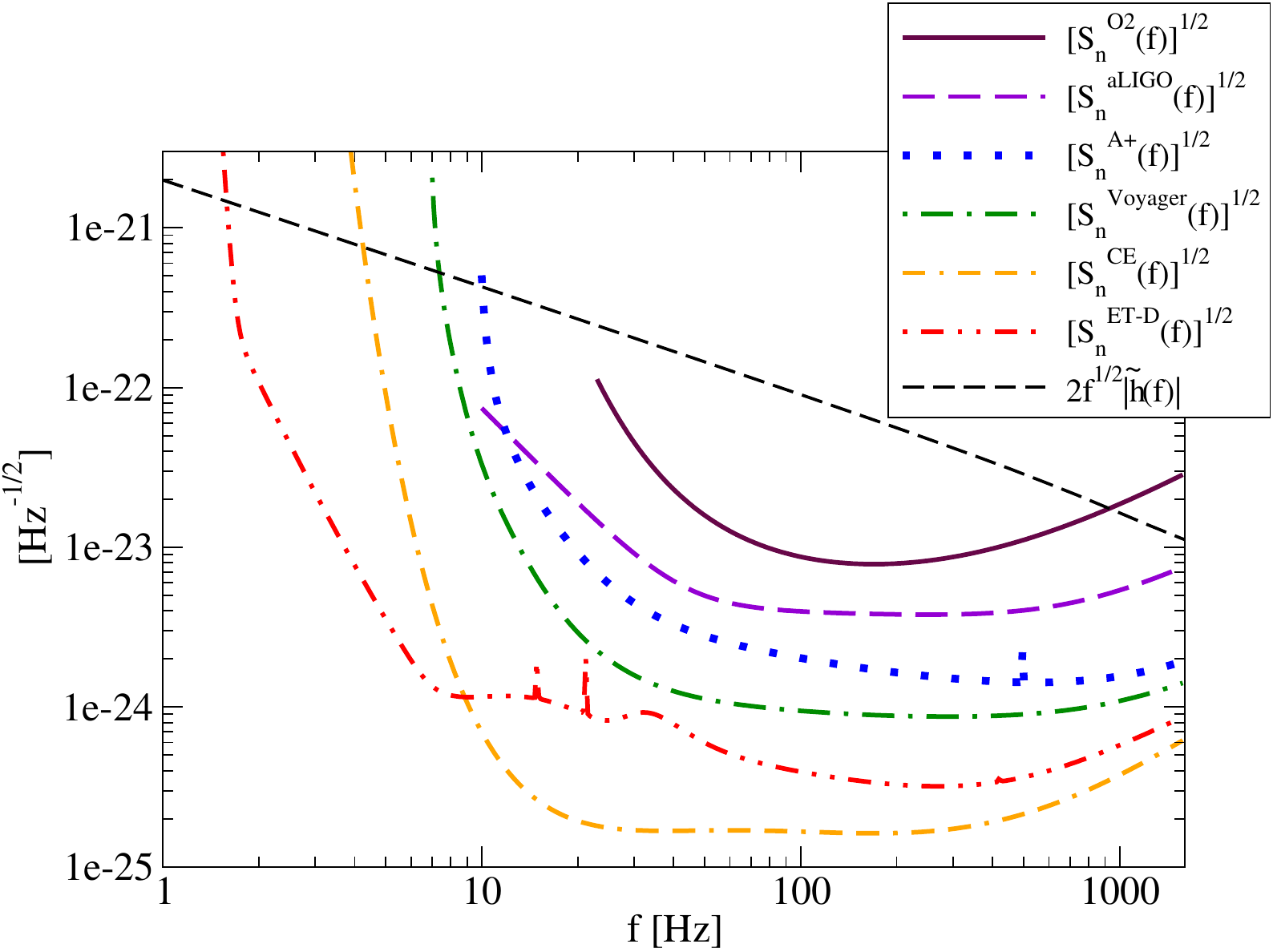}
\end{center}
\caption{(Color online)
Amplitude spectral noise densities $\sqrt{S_n^A(f)}$ plotted for LIGO O2, aLIGO at design sensitivity, A\texttt{+}, Voyager, CE, and ET-D as interpolated from publicly available data.
Spectral noise densities are plotted from $f_{\text{min}}=(23,10,10,7,1,1) \text{ Hz}$, respectively, to $f_{\text{max}}=1649 \text{ Hz}$.
Also shown is $2 \sqrt{f}$ multiplied by the amplitude of the PhenomD~\cite{PhenomDI,PhenomDII} waveform template.
}
\label{fig:sensitivities}
\end{figure}

The inner product, and thus the root-mean-square error, depends on the spectral noise density $S_{n}(f)$ and the waveform template $\tilde{h}$. We consider here 6 different interferometer designs, O2~\cite{aLIGO}, aLIGO at design sensitivity~\cite{aLIGO}, A\texttt{+}~\cite{Ap_Voyager_CE}, Voyager~\cite{Ap_Voyager_CE}, CE~\cite{ET}, ET-D~\cite{Ap_Voyager_CE} shown in Fig.~\ref{fig:sensitivities}, in order to compare the statistical uncertainties accrued on parameter extraction using future upgraded LIGO detectors, as well as third generation detectors. The inner product also depends on the particular waveform model we use for the template. We here consider a ``PhenomD" (IMRD) waveform template~\cite{PhenomDI,PhenomDII} modified by a 6PN tidal correction~\cite{Wade:tidalCorrections} (IMRD + 6PN), as well as a ``PhenomD'' template modified with a NRTidal correction~\cite{Samajdar:NRTidal} (IMRD + NRTidal). Considering two template models will allow us to estimate systematic uncertainties due to mismodeling of the GW signal. 

We begin by testing our Fisher analysis (with an IMRD + 6PN waveform injection) against a simulated event identical to GW170817 with O2 detector sensitivity~\cite{aLIGO}. Because only 1 event was detected, we use $\tilde\Lambda$ and $\delta\tilde\Lambda$ as the tidal parameters for comparison purposes. Further, we scale the luminosity distance such that the signal-to-noise-ratio ($SNR \equiv \rho$) is fixed to $\rho=32.4$, as found in GW170817. We used fiducial template parameter values of $\mathcal{M}=1.22\text{ M}_\odot$ for the chirp mass, $\eta=0.249$ for the symmetric mass ratio, $\tilde\Lambda=395$ (Corresponding to GW170817), and $0$ for the remaining parameters. We also assume low spin priors $|\chi| \leq 0.05$, as well as $\tilde{\Lambda} \leq 3000$ and $|\delta \tilde{\Lambda}| \leq 500$~\cite{Wade:2014vqa}. The resulting $90\%$ credible region of the posterior distribution on $\tilde{\Lambda}$ has a range of $\pm 276.99$, which is in close agreement to that found by LIGO~\cite{TheLIGOScientific:2017qsa,Abbott2018}

Next, we consider events similar to GW170817 detected on upgraded detectors and future detectors, as well as the combined statistical uncertainties of $N_A$ events detected over a 1 year observation with each detector. The latter is calculated by integrating the local binary NS merger rate over redshifts up to the horizon redshift of each detector. 
Similar to before, we used fiducial template parameter values of $\mathcal{M}=1.22\text{ M}_\odot$ for the chirp mass, $\eta=0.249$ for the symmetric mass ratio, $\lambda_0=150$, $\lambda_1=-213$ (corresponding to the GW170817), and $0$ for the remaining parameters.
Also as before, for each Fisher calculation we assume low spin priors $|\chi| \leq 0.05$, as well as $0 \leq \lambda_0 \leq 3207$ and $-4490 \leq \lambda_1 \leq 0$~\cite{delPozzo:TaylorTidal}\footnote{These are converted from the dimensional forms found in~\cite{delPozzo:TaylorTidal} into their corresponding dimensionless forms.}.
The process we use to compute single and combined statistical uncertainties for each detector sensitivity $S_n^A(f)$ is detailed in App.~\ref{app:stackingProcedure}.
From this, we determine if and when the statistical uncertainties associated with the parameter extraction of $\lambda_0$ drop below the systematic EoS variation errors from using the binary Love relations.

\begin{figure}
\begin{center} 
\includegraphics[width=\columnwidth]{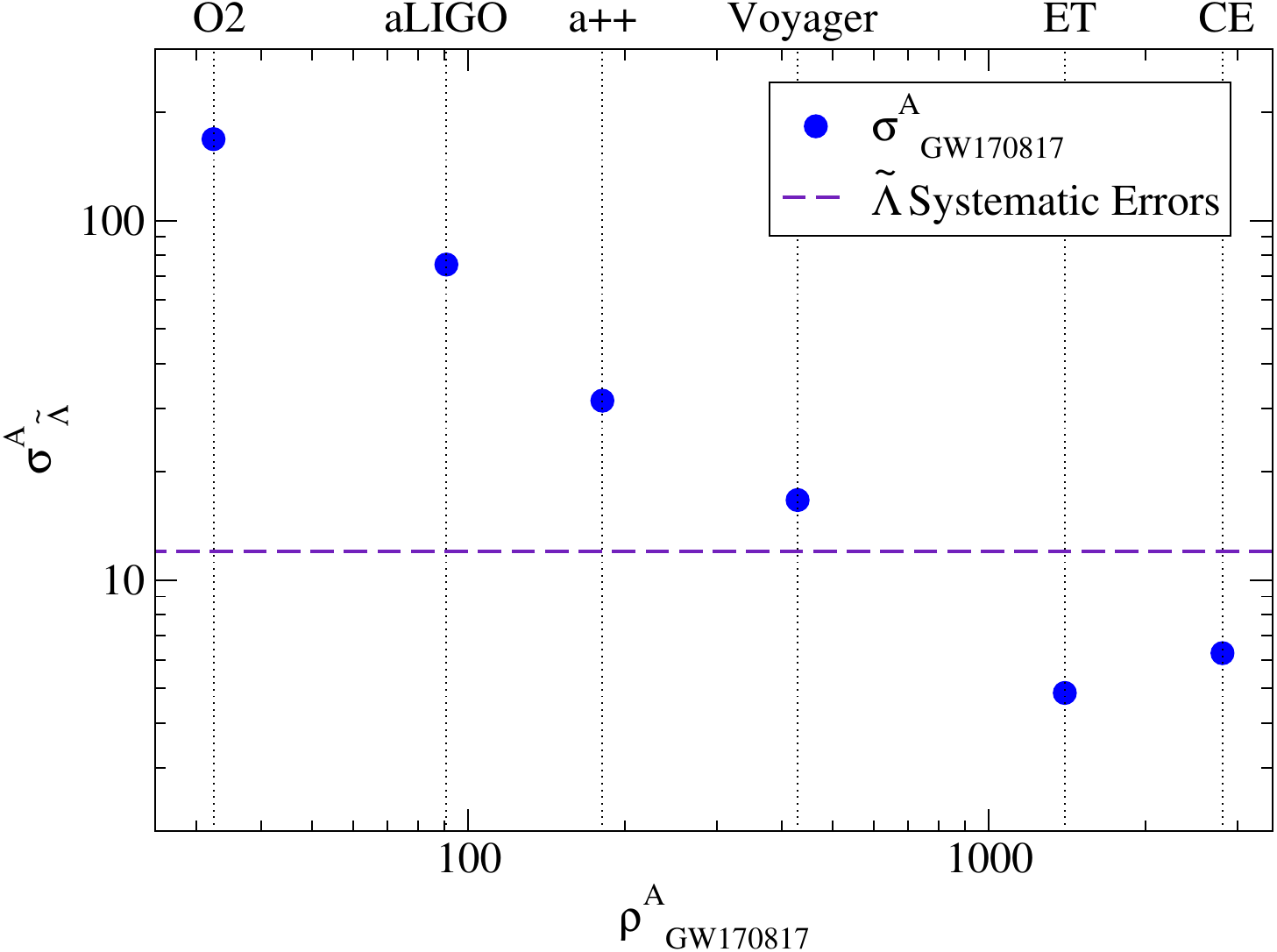}
\end{center}
\caption{(Color Online) Estimated statistical uncertainty $\sigma^A_{\text{GW170817}}$ in the extraction of $\tilde\Lambda$ from a single GW170817-like event as if observed with aLIGO at design sensitivity, A\texttt{+}, Voyager, CE, and ET-D, plotted as a function of the signal-to-noise-ratio $\rho^A_{\text{GW170817}}$ that those detectors would measure for such an event. For comparison, we also plot the systematic uncertainty on $\tilde\Lambda$ due to the use of the binary Love relations. The statistical and the systematic errors become comparable for Voyager-class detectors or better.
}
\label{fig:singleFisherLt}
\end{figure}

Figures~\ref{fig:stackedFisher} and~\ref{fig:singleFisherLt}, and Table~\ref{tab:variances} summarize our results graphically using an IMRD + 6PN template. In Fig.~\ref{fig:singleFisherLt}, one can see the statistical accuracy to which $\tilde{\Lambda}$ can be estimated as a function of $\rho^A_{\text{GW170817}}$ for single GW170817-like events, i.e.~the signal-to-noise ratio that future detector $A$ would have measured for a GW170817 event. The systematic uncertainty in $\tilde{\Lambda}$ (the horizontal dashed line) becomes comparable to the statistical uncertainty for Voyager-class detectors or better. Figure~\ref{fig:stackedFisher} presents similar results but for the parameter $\lambda_{0}$, for which one can combine posteriors and obtained a reduced combined statistical error. This figure also shows that the statistical and systematic uncertainties become comparable for Voyager-class detectors or better. The careful reader will notice that Figs.~\ref{fig:stackedFisher} and~\ref{fig:singleFisherLt} show that the single-event statistical uncertainties using CE is higher than using ET despite the latter having a larger SNR for a GW170817-like event. This is because the 3-detector geometry of ET enables it to have higher sensitivity than CE at frequencies above 300Hz, which is precisely where the tidal deformabilities are encoded, but ET has lower sensitivity at lower frequencies, where a lot of the signal-to-noise ratio accumulates.

\begin{table*}
\centering
\begin{tabular}{|c|@{\extracolsep{4pt}}C{1.7cm}@{\extracolsep{0pt}}|C{1.7cm}|@{\extracolsep{4pt}}C{1.7cm}@{\extracolsep{-2pt}}|C{1.7cm}|@{\extracolsep{-2pt}}C{1.7cm}@{\extracolsep{2pt}}|C{1.7cm}|@{\extracolsep{0pt}}C{1.7cm}@{\extracolsep{0pt}}|C{1.7cm}|}
\cline{1-1}\cline{2-3}\cline{4-9}
    \multicolumn{1}{|c|}{\bfseries Detectors (A)} & \multicolumn{2}{|c|}{\bfseries GW170817} & \multicolumn{6}{|c|}{\bfseries Multiple events} \\
\cline{1-1}\cline{2-3}\cline{4-9}
\noalign{\smallskip}
\cline{2-3}\cline{4-6}\cline{7-9}
\multicolumn{1}{c}{} & \multicolumn{1}{|c|}{} & \multicolumn{1}{c|}{} & \multicolumn{3}{|c|}{} & \multicolumn{3}{|c|}{}
\\[-1em]
\multicolumn{1}{c}{}  &  \multicolumn{1}{|c|}{\multirow{2}{*}{$\rho^A_{\text{GW170817}}$}}  &  \multirow{ 2}{*}{$\sigma^A_{\text{GW170817}}$}  &  \multicolumn{3}{|c|}{$N_A$}  &  \multicolumn{3}{|c|}{$\sigma^A_N$}  \\
\cline{4-6}\cline{7-9}
\multicolumn{1}{c}{}  &  \multicolumn{1}{|c|}{}  &  \multicolumn{1}{c|}{}  &  \multicolumn{1}{|c|}{Low}  &  \multicolumn{1}{c|}{Central} &  \multicolumn{1}{c|}{High}  & \multicolumn{1}{|c|}{Low}  &  \multicolumn{1}{c|}{Central} &  \multicolumn{1}{c|}{High}\\
\cline{2-3}\cline{4-6}\cline{7-9}
\noalign{\smallskip}
\noalign{\smallskip}
\cline{1-1}\cline{2-3}\cline{4-6}\cline{7-9}
&\multicolumn{1}{|c|}{}&&\multicolumn{1}{|c|}{}&&&\multicolumn{1}{|c|}{}&&\\[-1em]
 O2  & \multicolumn{1}{|c|}{$\sci{3.2}{1}$}  & $\sci{1.7}{2}$ &  \multicolumn{1}{|c|}{--} & -- & -- & \multicolumn{1}{|c|}{--} & -- & --\\
\cline{1-1}\cline{2-3}\cline{4-6}\cline{7-9}
&\multicolumn{1}{|c|}{}&&\multicolumn{1}{|c|}{}&&&\multicolumn{1}{|c|}{}&&\\[-1em]
 aLIGO  & \multicolumn{1}{|c|}{$\sci{9.1}{1}$}  & $\sci{1.1}{2}$ &  \multicolumn{1}{|c|}{$\sci{2.0}{1}$} & $\sci{9.8}{1}$ & $\sci{3.0}{2}$ & \multicolumn{1}{|c|}{$\sci{1.8}{2}$} & $\sci{8.3}{1}$ & $\sci{4.7}{1}$\\
\cline{1-1}\cline{2-3}\cline{4-6}\cline{7-9}
&\multicolumn{1}{|c|}{}&&\multicolumn{1}{|c|}{}&&&\multicolumn{1}{|c|}{}&&\\[-1em]
 A\texttt{+}  & \multicolumn{1}{|c|}{$\sci{1.8}{2}$}  & $\sci{4.6}{1}$ &  \multicolumn{1}{|c|}{$\sci{1.6}{2}$} & $\sci{7.9}{2}$ & $\sci{2.4}{3}$ & \multicolumn{1}{|c|}{$\sci{5.9}{1}$} & $\sci{2.5}{1}$ & $\sci{1.4}{1}$\\
\cline{1-1}\cline{2-3}\cline{4-6}\cline{7-9}

\multicolumn{1}{|c|}{} & \multicolumn{1}{|c|}{} & \multicolumn{1}{c|}{} & \multicolumn{1}{|c|}{} & \multicolumn{1}{c|}{} & \multicolumn{1}{c|}{} & \multicolumn{1}{|c|}{} & \multicolumn{1}{c|}{} & \multicolumn{1}{c|}{}
\\[-1em]
&\multicolumn{1}{|c|}{}&&\multicolumn{1}{|c|}{}&&&\multicolumn{1}{|c|}{}&&\\[-1em]
 Voyager  & \multicolumn{1}{|c|}{$\sci{4.3}{2}$}  & $\sci{2.5}{1}$ &  \multicolumn{1}{|c|}{$\sci{2.2}{3}$} & $\sci{1.1}{4}$ & $\sci{3.2}{4}$ & \multicolumn{1}{|c|}{$\sci{2.1}{1}$} & $\sci{9.6}{0}$ & $\sci{5.3}{0}$\\
\cline{1-1}\cline{2-3}\cline{4-6}\cline{7-9}

\multicolumn{1}{|c|}{} & \multicolumn{1}{|c|}{} & \multicolumn{1}{c|}{} & \multicolumn{1}{|c|}{} & \multicolumn{1}{c|}{} & \multicolumn{1}{c|}{} & \multicolumn{1}{|c|}{} & \multicolumn{1}{c|}{} & \multicolumn{1}{c|}{}
\\[-1em]
&\multicolumn{1}{|c|}{}&&\multicolumn{1}{|c|}{}&&&\multicolumn{1}{|c|}{}&&\\[-1em]
 ET-D  & \multicolumn{1}{|c|}{$\sci{1.4}{3}$}  & $\sci{6.9}{0}$ &  \multicolumn{1}{|c|}{$\sci{7.2}{4}$} & $\sci{3.4}{5}$ & $\sci{1.1}{6}$ & \multicolumn{1}{|c|}{$\sci{3.8}{0}$} & $\sci{1.7}{0}$ & $\sci{9.6}{-1}$\\
\cline{1-1}\cline{2-3}\cline{4-6}\cline{7-9}

\multicolumn{1}{|c|}{} & \multicolumn{1}{|c|}{} & \multicolumn{1}{c|}{} & \multicolumn{1}{|c|}{} & \multicolumn{1}{c|}{} & \multicolumn{1}{c|}{} & \multicolumn{1}{|c|}{} & \multicolumn{1}{c|}{} & \multicolumn{1}{c|}{}
\\[-1em]
&\multicolumn{1}{|c|}{}&&\multicolumn{1}{|c|}{}&&&\multicolumn{1}{|c|}{}&&\\[-1em]
 CE  & \multicolumn{1}{|c|}{$\sci{2.8}{3}$}  & $\sci{7.7}{0}$ &  \multicolumn{1}{|c|}{$\sci{3.0}{5}$} & $\sci{1.4}{6}$ & $\sci{4.4}{6}$ & \multicolumn{1}{|c|}{$\sci{3.7}{0}$} & $\sci{1.7}{0}$ & $\sci{9.0}{-1}$\\
\cline{1-1}\cline{2-3}\cline{4-6}\cline{7-9}
\end{tabular}
\caption{
Approximate signal-to-noise ratio $\rho^A_{\text{GW170817}}$ and ($1\sigma$) statistical uncertainty on the extraction of $\lambda_{0}$ had a single event like GW170817 been observed by future interferometer $A$ and had interferometer $A$ observed $N_{A}$ events in a 1 year observation, using aLIGO, A\texttt{+}, Voyager, CE, and ET. The number of events $N_{A}$, and the combined statistical uncertainty depends on the binary NS merger detection rate, and thus we include results assuming an upper, a central and a lower limit on this rate. The statistical uncertainties on $\lambda_0$ becomes comparable with the systematic uncertainty (set to be $P_{90}=13.19$) from using the improved binary Love relations with detectors of Voyager-class or better.
}\label{tab:variances}
\end{table*}

Is the statistical error calculated here robust to mismodeling systematics in the template? We repeat the analysis above using an IMRD + NRTidal template model  and find results consistent with those presented above. Indeed, Fig.~\ref{fig:stackedFisher} shows the individual and combined statistical error on $\lambda_{0}$ when using this template model (maroon circles and dashed region). The accuracy to which $\lambda_{0}$ can be measured with this template model is systematically better than when using the IMRD+6PN model (by roughly a factor of two). This implies that more accurate template models will indeed be required in the third-generation detector era, as previously pointed out in~\cite{Samajdar:NRTidal}. However, our conclusion that for Voyager-class detectors or better the statistical and binary Love systematic uncertainties become comparable seems robust.   


\section{Conclusion and Discussion}
\label{sec:conclusion}

The recent GW observation of a binary NS coalescence, GW170817, placed constraints on the supranuclear matter EoS for NSs. We used this observation to generate a restricted set of spectral EoSs that agree with it, in order to reduce the uncertainties upon the extraction of tidal parameters from future GW events. Previous work by Yagi and Yunes~\cite{Yagi:2013bca,Yagi:ILQ,Yagi:2015pkc,Yagi:binLove} had found EoS-insensitive relations between symmetric and antisymmetric combinations of NS tidal deformabilities, which aid in the extraction of said tidal parameters. We here found that the GW170817-constrained set of EoSs are more EoS-insensitive by a factor of $\sim 60$\% for stars with mass ratios of $0.75$ relative to previous work. Similarly, we find an increase in EoS-insensitivity in the C-Love and I-Love-Q  relations by factors of $\sim 75$\% and $\sim 50$\% respectively. The former further allowed us to improve the R-Love relation leading to uncertainties below $400$ meters in the entire parameter space. We also studied the EoS-insensitive relations of hybrid stars and found that for the most part the relations remain insensitive for isolated stars, albeit with slightly higher EoS variability. The binary Love relations, however, do not satisfy the same EoS-insensitive relations as in the case of purely hadronic star, when the binary contains at least one hybrid star. 

The second half of this paper focused on when the improvements in the EoS-insensitive relations would become necessary in future detectors. Current detectors are not yet sensitive enough that the systematic uncertainties in the EoS-insensitive relations make a large difference. However, we did find that for Voyager-class detectors or better, the systematic uncertainties in the EoS-insensitive relations due to EoS variability become comparable to statistical uncertainties in the estimation of $\tilde{\Lambda}$ and $\lambda_{0}$. We also considered the effect of waveform mismodeling and found that the above conclusion remains robust, but that more accurate waveform models will be necessary to take full advantage of the improved sensitivity of future detectors. 

Future work on this subject could entail an investigation into the improvement of alternative EoS-insensitive relations, such as the multipole Love relations between various $\ell$-th order electric, magnetic, and shape tidal deformabilities, as discussed in~\cite{Yagi:Multipole}. Lackey \emph{et al.}~\cite{Lackey:Surrogate, Lackey:EOB} presented surrogate models of non-spinning effective-one-body waveforms with the use of universal relations. By reducing the number of waveform model parameters, surrogate models aid in the extraction of NS observables from GW detections. The improvement in the multipole Love relations can then be used to increase the accuracy of such surrogate models. Another possible avenue for future research includes a more comprehensive analysis into the intricacies of new hybrid star binary Love relations.


\section*{Acknowledgments}\label{acknowledgments}
KY acknowledges support from NSF Award PHY-1806776. 
K.Y. would like to also acknowledge networking support by the COST Action GWverse CA16104.
N.Y. acknowledges support from  NSF grant PHY-1759615 and NASA grants NNX16AB98G and 80NSSC17M0041.
This research has made use of data, software and/or web tools obtained from the Gravitational Wave Open Science Center (https://www.gw-openscience.org), a service of LIGO Laboratory, the LIGO Scientific Collaboration and the Virgo Collaboration. LIGO is funded by the U.S. National Science Foundation. Virgo is funded by the French Centre National de Recherche Scientifique (CNRS), the Italian Istituto Nazionale della Fisica Nucleare (INFN) and the Dutch Nikhef, with contributions by Polish and Hungarian institutes.
The authors are grateful for computational resources provided by the LIGO Laboratory and supported by National Science Foundation Grants PHY-0757058 and PHY-0823459.  


\appendix

\section{Computation of statistical uncertainties}\label{app:stackingProcedure}

In this appendix, we detail the process used to compute the statistical uncertainties on the extraction of $\lambda_0$ from the gravitational waveform.
This is accomplished with a simple Fisher analysis described in Sec.~\ref{sec:futureObservations}, where we first consider single-events similar to GW170817 as if they were detected on future detectors $A \equiv ($ O2~\cite{aLIGO}, aLIGO~\cite{aLIGO}, A\texttt{+}~\cite{Ap_Voyager_CE}, Voyager~\cite{Ap_Voyager_CE}, ET~\cite{ET}, CE~\cite{Ap_Voyager_CE}$)$.
We conclude by simulating a population of $N_A$ events for each interferometer, approximating the number of events detected on interferometer $A$ over an observing period of one year.
This is used to combine the statistical uncertainties, resulting in an approximation on the overall measurement accuracy of $\lambda_0$, which is compared to the systematic uncertainties computed in Sec.~\ref{sec:marginalization}.
The process used to achieve this is outlined below:

\begin{enumerate}
\item[(i)] Perform a Fisher analysis as outlined in Sec.~\ref{sec:futureObservations} using detector sensitivity $S_n^A(f)$, while restricting the luminosity distance $D_L$ such that an SNR of $\rho^{\text{O2}}_{\text{GW170817}}=32.4$ would be achieved on O2 sensitivity $S_n^{\text{O2}}(f)$.
Here we assume low spin priors $|\chi_{1,2}| \leq 0.05$, as well as $0 \leq \lambda_0 \leq 3207$ and $-4490 \leq \lambda_1 \leq 0$~\cite{delPozzo:TaylorTidal} (these are converted from their corresponding dimensional forms).
This results in an SNR $\rho^A_{\text{GW170817}}$ and a single-event statistical uncertainty $\sigma_\text{GW170817}^A$ accrued in the extraction of $\lambda_0$ on detector $A$.

\item[(ii)] Generate a population of $N_A$ events corresponding to the expected binary NS merger detection rate for detector $A$, following the probability distribution~\cite{Shutz:SNR,Chen:SNR}:
\begin{equation}\label{eq:population}
f(\rho)=3 \rho_{\text{th}}^3/\rho^4
\end{equation}
with a network SNR threshold of $\rho_{\text{th}}=8$.
The number of events $N_A$ is calculated by taking into account the BNS merger rate history throughout all redshift values within detector As' horizon redshift $z_h$, as shown by Eq. (10) of Ref.~\cite{Cutler:BNSmerger}:
\begin{equation}
N_A=\Delta \tau_0 \int\limits^{z_{h}}_0 4 \pi \lbrack  a_0r_1(z)\rbrack^2 \mathcal{R} r(z) \frac{d \tau}{dz} dz.
\end{equation}
Here, $a_0r_1(z)$, $\frac{d\tau}{dz}$, and $r(z)$ for our chosen cosmology are given by:
\begin{equation}
a_0r_1(z) = \frac{1}{H_0}\int\limits^z_0 \frac{dz'}{\sqrt{(1-\Omega_{\Lambda})(1+z')^3+\Omega_{\Lambda}}},
\end{equation}
\begin{equation}
\frac{d\tau}{dz} = \frac{1}{H_0} \frac{1}{1+z}\frac{1}{\sqrt{(1-\Omega_{\Lambda})(1+z')^3+\Omega_{\Lambda}}},
\end{equation}
\begin{equation}
r(z) = \left\{
\begin{array}{ll}
      1+2z & z \leq 1 \\
      \frac{3}{4}(5-z) & 1\leq z\leq 5 \\
      0 & z\geq 5\\ 
\end{array}\,, 
\right.
\end{equation}

where $H_0 = 70 \text{km s}^{-1}\text{Mpc}^{-1}$ is the local Hubble constant and $\Omega_{\Lambda}=0.67$ is the universe's vacuum energy density.
Here we choose an observing period of $\Delta \tau_0 = 1$ year, and calculate the detection rate for the upper, central, and lower limits of the local binary NS coalescence rate density $\mathcal{R}=1540^{+3200}_{-1220} \text{ Gpc}^{-3}\text{yr}^{-1}$~\cite{Abbott2017}, giving the rates $N_A$ shown in the second column of Table~\ref{tab:variances}.

\item[(iii)] Compute the combined population standard deviation $\sigma_{N_A}$, taking into account sources at varying redshifts as was done in Eq. (3) of Ref.~\cite{Takahiro}:
\begin{equation}
\sigma_{N_A}^{-2}=\Delta \tau \int\limits^{z_h}_04 \pi \lbrack a_0 r_1(z)\rbrack^2 \mathcal{R}r(z)\frac{d\tau}{dz}\sigma^A_i(z)^{-2}dz.
\end{equation}
Here, we compute the populations' single-event uncertainties $\sigma_i^A(z)$ via the simple SNR scaling factor: 
\begin{equation}
\frac{\rho_i^A}{\rho_{\text{GW170817}}^A} = \frac{\sigma_{\text{GW170817}}^A}{\sigma_i^A},
\end{equation}
where $\rho_i^A$ is the SNR of simulated event $i$ computed from Eq.~\eqref{eq:population}, and $\rho_{\text{GW170817}}^A$ and $\sigma_{\text{GW170817}}^A$ are the known SNRs and uncertainties of the GW170817 event from step (i).
This results in the combined statistical uncertainties for the lower, central, and upper limits of the local binary NS coalescence rate density $\mathcal{R}$, depicted by the cyan shaded region in Fig.~\ref{fig:stackedFisher}.
\end{enumerate}


\bibliography{Zack}

 \newcommand{\noop}[1]{}
\begin{thebibliography}{102}%
\makeatletter
\providecommand \@ifxundefined [1]{%
 \@ifx{#1\undefined}
}%
\providecommand \@ifnum [1]{%
 \ifnum #1\expandafter \@firstoftwo
 \else \expandafter \@secondoftwo
 \fi
}%
\providecommand \@ifx [1]{%
 \ifx #1\expandafter \@firstoftwo
 \else \expandafter \@secondoftwo
 \fi
}%
\providecommand \natexlab [1]{#1}%
\providecommand \enquote  [1]{``#1''}%
\providecommand \bibnamefont  [1]{#1}%
\providecommand \bibfnamefont [1]{#1}%
\providecommand \citenamefont [1]{#1}%
\providecommand \href@noop [0]{\@secondoftwo}%
\providecommand \href [0]{\begingroup \@sanitize@url \@href}%
\providecommand \@href[1]{\@@startlink{#1}\@@href}%
\providecommand \@@href[1]{\endgroup#1\@@endlink}%
\providecommand \@sanitize@url [0]{\catcode `\\12\catcode `\$12\catcode
  `\&12\catcode `\#12\catcode `\^12\catcode `\_12\catcode `\%12\relax}%
\providecommand \@@startlink[1]{}%
\providecommand \@@endlink[0]{}%
\providecommand \url  [0]{\begingroup\@sanitize@url \@url }%
\providecommand \@url [1]{\endgroup\@href {#1}{\urlprefix }}%
\providecommand \urlprefix  [0]{URL }%
\providecommand \Eprint [0]{\href }%
\providecommand \doibase [0]{http://dx.doi.org/}%
\providecommand \selectlanguage [0]{\@gobble}%
\providecommand \bibinfo  [0]{\@secondoftwo}%
\providecommand \bibfield  [0]{\@secondoftwo}%
\providecommand \translation [1]{[#1]}%
\providecommand \BibitemOpen [0]{}%
\providecommand \bibitemStop [0]{}%
\providecommand \bibitemNoStop [0]{.\EOS\space}%
\providecommand \EOS [0]{\spacefactor3000\relax}%
\providecommand \BibitemShut  [1]{\csname bibitem#1\endcsname}%
\let\auto@bib@innerbib\@empty
\bibitem [{\citenamefont {Li}\ \emph {et~al.}(2008)\citenamefont {Li},
  \citenamefont {Chen},\ and\ \citenamefont {Ko}}]{Li:HeavyIon}%
  \BibitemOpen
  \bibfield  {author} {\bibinfo {author} {\bibfnamefont {B.-A.}\ \bibnamefont
  {Li}}, \bibinfo {author} {\bibfnamefont {L.-W.}\ \bibnamefont {Chen}}, \ and\
  \bibinfo {author} {\bibfnamefont {C.~M.}\ \bibnamefont {Ko}},\ }\href
  {\doibase https://doi.org/10.1016/j.physrep.2008.04.005} {\bibfield
  {journal} {\bibinfo  {journal} {Physics Reports}\ }\textbf {\bibinfo {volume}
  {464}},\ \bibinfo {pages} {113 } (\bibinfo {year} {2008})}\BibitemShut
  {NoStop}%
\bibitem [{\citenamefont {Tsang}\ \emph {et~al.}(2009)\citenamefont {Tsang},
  \citenamefont {Zhang}, \citenamefont {Danielewicz}, \citenamefont {Famiano},
  \citenamefont {Li}, \citenamefont {Lynch},\ and\ \citenamefont
  {Steiner}}]{Tsang:SymmetryEnergy}%
  \BibitemOpen
  \bibfield  {author} {\bibinfo {author} {\bibfnamefont {M.~B.}\ \bibnamefont
  {Tsang}}, \bibinfo {author} {\bibfnamefont {Y.}~\bibnamefont {Zhang}},
  \bibinfo {author} {\bibfnamefont {P.}~\bibnamefont {Danielewicz}}, \bibinfo
  {author} {\bibfnamefont {M.}~\bibnamefont {Famiano}}, \bibinfo {author}
  {\bibfnamefont {Z.}~\bibnamefont {Li}}, \bibinfo {author} {\bibfnamefont
  {W.~G.}\ \bibnamefont {Lynch}}, \ and\ \bibinfo {author} {\bibfnamefont
  {A.~W.}\ \bibnamefont {Steiner}},\ }\href {\doibase
  10.1103/PhysRevLett.102.122701} {\bibfield  {journal} {\bibinfo  {journal}
  {Phys. Rev. Lett.}\ }\textbf {\bibinfo {volume} {102}},\ \bibinfo {pages}
  {122701} (\bibinfo {year} {2009})}\BibitemShut {NoStop}%
\bibitem [{\citenamefont {Centelles}\ \emph {et~al.}(2009)\citenamefont
  {Centelles}, \citenamefont {Roca-Maza}, \citenamefont {Vi\~nas},\ and\
  \citenamefont {Warda}}]{Centelles:NeutronSkin}%
  \BibitemOpen
  \bibfield  {author} {\bibinfo {author} {\bibfnamefont {M.}~\bibnamefont
  {Centelles}}, \bibinfo {author} {\bibfnamefont {X.}~\bibnamefont
  {Roca-Maza}}, \bibinfo {author} {\bibfnamefont {X.}~\bibnamefont {Vi\~nas}},
  \ and\ \bibinfo {author} {\bibfnamefont {M.}~\bibnamefont {Warda}},\ }\href
  {\doibase 10.1103/PhysRevLett.102.122502} {\bibfield  {journal} {\bibinfo
  {journal} {Phys. Rev. Lett.}\ }\textbf {\bibinfo {volume} {102}},\ \bibinfo
  {pages} {122502} (\bibinfo {year} {2009})}\BibitemShut {NoStop}%
\bibitem [{\citenamefont {Li}\ and\ \citenamefont
  {Chen}(2005)}]{Li:CrossSections}%
  \BibitemOpen
  \bibfield  {author} {\bibinfo {author} {\bibfnamefont {B.-A.}\ \bibnamefont
  {Li}}\ and\ \bibinfo {author} {\bibfnamefont {L.-W.}\ \bibnamefont {Chen}},\
  }\href {\doibase 10.1103/PhysRevC.72.064611} {\bibfield  {journal} {\bibinfo
  {journal} {Phys. Rev. C}\ }\textbf {\bibinfo {volume} {72}},\ \bibinfo
  {pages} {064611} (\bibinfo {year} {2005})}\BibitemShut {NoStop}%
\bibitem [{\citenamefont {Chen}\ \emph {et~al.}(2005)\citenamefont {Chen},
  \citenamefont {Ko},\ and\ \citenamefont {Li}}]{Chen:SymEnergy}%
  \BibitemOpen
  \bibfield  {author} {\bibinfo {author} {\bibfnamefont {L.-W.}\ \bibnamefont
  {Chen}}, \bibinfo {author} {\bibfnamefont {C.~M.}\ \bibnamefont {Ko}}, \ and\
  \bibinfo {author} {\bibfnamefont {B.-A.}\ \bibnamefont {Li}},\ }\href
  {\doibase 10.1103/PhysRevLett.94.032701} {\bibfield  {journal} {\bibinfo
  {journal} {Phys. Rev. Lett.}\ }\textbf {\bibinfo {volume} {94}},\ \bibinfo
  {pages} {032701} (\bibinfo {year} {2005})}\BibitemShut {NoStop}%
\bibitem [{\citenamefont {Danielewicz}\ \emph {et~al.}(2002)\citenamefont
  {Danielewicz}, \citenamefont {Lacey},\ and\ \citenamefont
  {Lynch}}]{Danielewicz:2002pu}%
  \BibitemOpen
  \bibfield  {author} {\bibinfo {author} {\bibfnamefont {P.}~\bibnamefont
  {Danielewicz}}, \bibinfo {author} {\bibfnamefont {R.}~\bibnamefont {Lacey}},
  \ and\ \bibinfo {author} {\bibfnamefont {W.~G.}\ \bibnamefont {Lynch}},\
  }\href {\doibase 10.1126/science.1078070} {\bibfield  {journal} {\bibinfo
  {journal} {Science}\ }\textbf {\bibinfo {volume} {298}},\ \bibinfo {pages}
  {1592} (\bibinfo {year} {2002})},\ \Eprint
  {http://arxiv.org/abs/nucl-th/0208016} {arXiv:nucl-th/0208016 [nucl-th]}
  \BibitemShut {NoStop}%
\bibitem [{\citenamefont {Guver}\ and\ \citenamefont {Ozel}(2013)}]{guver}%
  \BibitemOpen
  \bibfield  {author} {\bibinfo {author} {\bibfnamefont {T.}~\bibnamefont
  {Guver}}\ and\ \bibinfo {author} {\bibfnamefont {F.}~\bibnamefont {Ozel}},\
  }\href {\doibase 10.1088/2041-8205/765/1/L1} {\bibfield  {journal} {\bibinfo
  {journal} {Astrophys. J.}\ }\textbf {\bibinfo {volume} {765}},\ \bibinfo
  {pages} {L1} (\bibinfo {year} {2013})},\ \Eprint
  {http://arxiv.org/abs/1301.0831} {arXiv:1301.0831 [astro-ph.HE]} \BibitemShut
  {NoStop}%
\bibitem [{\citenamefont {Ozel}\ \emph {et~al.}(2010)\citenamefont {Ozel},
  \citenamefont {Baym},\ and\ \citenamefont {Guver}}]{ozel-baym-guver}%
  \BibitemOpen
  \bibfield  {author} {\bibinfo {author} {\bibfnamefont {F.}~\bibnamefont
  {Ozel}}, \bibinfo {author} {\bibfnamefont {G.}~\bibnamefont {Baym}}, \ and\
  \bibinfo {author} {\bibfnamefont {T.}~\bibnamefont {Guver}},\ }\href
  {\doibase 10.1103/PhysRevD.82.101301} {\bibfield  {journal} {\bibinfo
  {journal} {Phys.Rev.}\ }\textbf {\bibinfo {volume} {D82}},\ \bibinfo {pages}
  {101301} (\bibinfo {year} {2010})},\ \Eprint {http://arxiv.org/abs/1002.3153}
  {arXiv:1002.3153 [astro-ph.HE]} \BibitemShut {NoStop}%
\bibitem [{\citenamefont {Steiner}\ \emph {et~al.}(2010)\citenamefont
  {Steiner}, \citenamefont {Lattimer},\ and\ \citenamefont
  {Brown}}]{steiner-lattimer-brown}%
  \BibitemOpen
  \bibfield  {author} {\bibinfo {author} {\bibfnamefont {A.~W.}\ \bibnamefont
  {Steiner}}, \bibinfo {author} {\bibfnamefont {J.~M.}\ \bibnamefont
  {Lattimer}}, \ and\ \bibinfo {author} {\bibfnamefont {E.~F.}\ \bibnamefont
  {Brown}},\ }\href {\doibase 10.1088/0004-637X/722/1/33} {\bibfield  {journal}
  {\bibinfo  {journal} {Astrophys.J.}\ }\textbf {\bibinfo {volume} {722}},\
  \bibinfo {pages} {33} (\bibinfo {year} {2010})}\BibitemShut {NoStop}%
\bibitem [{\citenamefont {Lattimer}\ and\ \citenamefont
  {Steiner}(2014)}]{Lattimer2014}%
  \BibitemOpen
  \bibfield  {author} {\bibinfo {author} {\bibfnamefont {J.~M.}\ \bibnamefont
  {Lattimer}}\ and\ \bibinfo {author} {\bibfnamefont {A.~W.}\ \bibnamefont
  {Steiner}},\ }\href {\doibase 10.1140/epja/i2014-14040-y} {\bibfield
  {journal} {\bibinfo  {journal} {The European Physical Journal A}\ }\textbf
  {\bibinfo {volume} {50}} (\bibinfo {year} {2014}),\
  10.1140/epja/i2014-14040-y}\BibitemShut {NoStop}%
\bibitem [{\citenamefont {Ozel}\ and\ \citenamefont
  {Freire}(2016)}]{Ozel:2016oaf}%
  \BibitemOpen
  \bibfield  {author} {\bibinfo {author} {\bibfnamefont {F.}~\bibnamefont
  {Ozel}}\ and\ \bibinfo {author} {\bibfnamefont {P.}~\bibnamefont {Freire}},\
  }\href {\doibase 10.1146/annurev-astro-081915-023322} {\bibfield  {journal}
  {\bibinfo  {journal} {Ann. Rev. Astron. Astrophys.}\ }\textbf {\bibinfo
  {volume} {54}},\ \bibinfo {pages} {401} (\bibinfo {year} {2016})},\ \Eprint
  {http://arxiv.org/abs/1603.02698} {arXiv:1603.02698 [astro-ph.HE]}
  \BibitemShut {NoStop}%
\bibitem [{\citenamefont {Miller}\ and\ \citenamefont
  {Lamb}(2016)}]{Miller:2016pom}%
  \BibitemOpen
  \bibfield  {author} {\bibinfo {author} {\bibfnamefont {M.~C.}\ \bibnamefont
  {Miller}}\ and\ \bibinfo {author} {\bibfnamefont {F.~K.}\ \bibnamefont
  {Lamb}},\ }\href {\doibase 10.1140/epja/i2016-16063-8} {\bibfield  {journal}
  {\bibinfo  {journal} {Eur. Phys. J.}\ }\textbf {\bibinfo {volume} {A52}},\
  \bibinfo {pages} {63} (\bibinfo {year} {2016})},\ \Eprint
  {http://arxiv.org/abs/1604.03894} {arXiv:1604.03894 [astro-ph.HE]}
  \BibitemShut {NoStop}%
\bibitem [{\citenamefont {Miller}()}]{Miller2013}%
  \BibitemOpen
  \bibfield  {author} {\bibinfo {author} {\bibfnamefont {M.~C.}\ \bibnamefont
  {Miller}},\ }\href {https://arxiv.org/abs/1312.0029} {\bibfield  {journal}
  {\bibinfo  {journal} {Arxiv}\ }}\Eprint {http://arxiv.org/abs/1312.0029v1}
  {1312.0029v1} \BibitemShut {NoStop}%
\bibitem [{\citenamefont {Hinderer}(2008)}]{hinderer-love}%
  \BibitemOpen
  \bibfield  {author} {\bibinfo {author} {\bibfnamefont {T.}~\bibnamefont
  {Hinderer}},\ }\href {http://stacks.iop.org/0004-637X/677/i=2/a=1216}
  {\bibfield  {journal} {\bibinfo  {journal} {The Astrophysical Journal}\
  }\textbf {\bibinfo {volume} {677}},\ \bibinfo {pages} {1216} (\bibinfo {year}
  {2008})}\BibitemShut {NoStop}%
\bibitem [{\citenamefont {Flanagan}\ and\ \citenamefont
  {Hinderer}(2008)}]{Flanagan2008}%
  \BibitemOpen
  \bibfield  {author} {\bibinfo {author} {\bibfnamefont {{\'{E}}.~{\'{E}}.}\
  \bibnamefont {Flanagan}}\ and\ \bibinfo {author} {\bibfnamefont
  {T.}~\bibnamefont {Hinderer}},\ }\href {\doibase 10.1103/physrevd.77.021502}
  {\bibfield  {journal} {\bibinfo  {journal} {Physical Review D}\ }\textbf
  {\bibinfo {volume} {77}} (\bibinfo {year} {2008}),\
  10.1103/physrevd.77.021502}\BibitemShut {NoStop}%
\bibitem [{\citenamefont {Blanchet}(2014)}]{Blanchet:2013haa}%
  \BibitemOpen
  \bibfield  {author} {\bibinfo {author} {\bibfnamefont {L.}~\bibnamefont
  {Blanchet}},\ }\href {\doibase 10.12942/lrr-2014-2} {\bibfield  {journal}
  {\bibinfo  {journal} {Living Rev. Rel.}\ }\textbf {\bibinfo {volume} {17}},\
  \bibinfo {pages} {2} (\bibinfo {year} {2014})},\ \Eprint
  {http://arxiv.org/abs/1310.1528} {arXiv:1310.1528 [gr-qc]} \BibitemShut
  {NoStop}%
\bibitem [{\citenamefont {Vines}\ \emph {et~al.}(2011)\citenamefont {Vines},
  \citenamefont {Flanagan},\ and\ \citenamefont {Hinderer}}]{Vines:2011ud}%
  \BibitemOpen
  \bibfield  {author} {\bibinfo {author} {\bibfnamefont {J.}~\bibnamefont
  {Vines}}, \bibinfo {author} {\bibfnamefont {E.~E.}\ \bibnamefont {Flanagan}},
  \ and\ \bibinfo {author} {\bibfnamefont {T.}~\bibnamefont {Hinderer}},\
  }\href {\doibase 10.1103/PhysRevD.83.084051} {\bibfield  {journal} {\bibinfo
  {journal} {Phys. Rev.}\ }\textbf {\bibinfo {volume} {D83}},\ \bibinfo {pages}
  {084051} (\bibinfo {year} {2011})},\ \Eprint {http://arxiv.org/abs/1101.1673}
  {arXiv:1101.1673 [gr-qc]} \BibitemShut {NoStop}%
\bibitem [{\citenamefont {Wade}\ \emph
  {et~al.}(2014{\natexlab{a}})\citenamefont {Wade}, \citenamefont {Creighton},
  \citenamefont {Ochsner}, \citenamefont {Lackey}, \citenamefont {Farr},
  \citenamefont {Littenberg},\ and\ \citenamefont
  {Raymond}}]{Wade:tidalCorrections}%
  \BibitemOpen
  \bibfield  {author} {\bibinfo {author} {\bibfnamefont {L.}~\bibnamefont
  {Wade}}, \bibinfo {author} {\bibfnamefont {J.~D.~E.}\ \bibnamefont
  {Creighton}}, \bibinfo {author} {\bibfnamefont {E.}~\bibnamefont {Ochsner}},
  \bibinfo {author} {\bibfnamefont {B.~D.}\ \bibnamefont {Lackey}}, \bibinfo
  {author} {\bibfnamefont {B.~F.}\ \bibnamefont {Farr}}, \bibinfo {author}
  {\bibfnamefont {T.~B.}\ \bibnamefont {Littenberg}}, \ and\ \bibinfo {author}
  {\bibfnamefont {V.}~\bibnamefont {Raymond}},\ }\href {\doibase
  10.1103/PhysRevD.89.103012} {\bibfield  {journal} {\bibinfo  {journal} {Phys.
  Rev. D}\ }\textbf {\bibinfo {volume} {89}},\ \bibinfo {pages} {103012}
  (\bibinfo {year} {2014}{\natexlab{a}})}\BibitemShut {NoStop}%
\bibitem [{\citenamefont {Favata}(2014)}]{Favata:2013rwa}%
  \BibitemOpen
  \bibfield  {author} {\bibinfo {author} {\bibfnamefont {M.}~\bibnamefont
  {Favata}},\ }\href {\doibase 10.1103/PhysRevLett.112.101101} {\bibfield
  {journal} {\bibinfo  {journal} {Phys. Rev. Lett.}\ }\textbf {\bibinfo
  {volume} {112}},\ \bibinfo {pages} {101101} (\bibinfo {year} {2014})},\
  \Eprint {http://arxiv.org/abs/1310.8288} {arXiv:1310.8288 [gr-qc]}
  \BibitemShut {NoStop}%
\bibitem [{\citenamefont {Messenger}\ and\ \citenamefont
  {Read}(2012)}]{Messenger:2011gi}%
  \BibitemOpen
  \bibfield  {author} {\bibinfo {author} {\bibfnamefont {C.}~\bibnamefont
  {Messenger}}\ and\ \bibinfo {author} {\bibfnamefont {J.}~\bibnamefont
  {Read}},\ }\href {\doibase 10.1103/PhysRevLett.108.091101} {\bibfield
  {journal} {\bibinfo  {journal} {Phys. Rev. Lett.}\ }\textbf {\bibinfo
  {volume} {108}},\ \bibinfo {pages} {091101} (\bibinfo {year} {2012})},\
  \Eprint {http://arxiv.org/abs/1107.5725} {arXiv:1107.5725 [gr-qc]}
  \BibitemShut {NoStop}%
\bibitem [{\citenamefont {Del~Pozzo}\ \emph {et~al.}(2013)\citenamefont
  {Del~Pozzo}, \citenamefont {Li}, \citenamefont {Agathos}, \citenamefont {Van
  Den~Broeck},\ and\ \citenamefont {Vitale}}]{delPozzo:TaylorTidal}%
  \BibitemOpen
  \bibfield  {author} {\bibinfo {author} {\bibfnamefont {W.}~\bibnamefont
  {Del~Pozzo}}, \bibinfo {author} {\bibfnamefont {T.~G.~F.}\ \bibnamefont
  {Li}}, \bibinfo {author} {\bibfnamefont {M.}~\bibnamefont {Agathos}},
  \bibinfo {author} {\bibfnamefont {C.}~\bibnamefont {Van Den~Broeck}}, \ and\
  \bibinfo {author} {\bibfnamefont {S.}~\bibnamefont {Vitale}},\ }\href
  {\doibase 10.1103/PhysRevLett.111.071101} {\bibfield  {journal} {\bibinfo
  {journal} {Phys. Rev. Lett.}\ }\textbf {\bibinfo {volume} {111}},\ \bibinfo
  {pages} {071101} (\bibinfo {year} {2013})},\ \Eprint
  {http://arxiv.org/abs/1307.8338} {arXiv:1307.8338 [gr-qc]} \BibitemShut
  {NoStop}%
\bibitem [{\citenamefont {Yagi}\ and\ \citenamefont
  {Yunes}(2016{\natexlab{a}})}]{Yagi:binLove}%
  \BibitemOpen
  \bibfield  {author} {\bibinfo {author} {\bibfnamefont {K.}~\bibnamefont
  {Yagi}}\ and\ \bibinfo {author} {\bibfnamefont {N.}~\bibnamefont {Yunes}},\
  }\href {\doibase 10.1088/1361-6382/34/1/015006} {\bibfield  {journal}
  {\bibinfo  {journal} {Classical and Quantum Gravity}\ }\textbf {\bibinfo
  {volume} {34}},\ \bibinfo {pages} {015006} (\bibinfo {year}
  {2016}{\natexlab{a}})}\BibitemShut {NoStop}%
\bibitem [{\citenamefont {Yagi}\ and\ \citenamefont
  {Yunes}(2016{\natexlab{b}})}]{Yagi:2015pkc}%
  \BibitemOpen
  \bibfield  {author} {\bibinfo {author} {\bibfnamefont {K.}~\bibnamefont
  {Yagi}}\ and\ \bibinfo {author} {\bibfnamefont {N.}~\bibnamefont {Yunes}},\
  }\href {\doibase 10.1088/0264-9381/33/13/13LT01} {\bibfield  {journal}
  {\bibinfo  {journal} {Class. Quant. Grav.}\ }\textbf {\bibinfo {volume}
  {33}},\ \bibinfo {pages} {13LT01} (\bibinfo {year} {2016}{\natexlab{b}})},\
  \Eprint {http://arxiv.org/abs/1512.02639} {arXiv:1512.02639 [gr-qc]}
  \BibitemShut {NoStop}%
\bibitem [{\citenamefont {Yagi}\ and\ \citenamefont
  {Yunes}(2013{\natexlab{a}})}]{Yagi:2013bca}%
  \BibitemOpen
  \bibfield  {author} {\bibinfo {author} {\bibfnamefont {K.}~\bibnamefont
  {Yagi}}\ and\ \bibinfo {author} {\bibfnamefont {N.}~\bibnamefont {Yunes}},\
  }\href {\doibase 10.1126/science.1236462} {\bibfield  {journal} {\bibinfo
  {journal} {Science}\ }\textbf {\bibinfo {volume} {341}},\ \bibinfo {pages}
  {365} (\bibinfo {year} {2013}{\natexlab{a}})},\ \Eprint
  {http://arxiv.org/abs/1302.4499} {arXiv:1302.4499 [gr-qc]} \BibitemShut
  {NoStop}%
\bibitem [{\citenamefont {Yagi}\ and\ \citenamefont
  {Yunes}(2013{\natexlab{b}})}]{Yagi:ILQ}%
  \BibitemOpen
  \bibfield  {author} {\bibinfo {author} {\bibfnamefont {K.}~\bibnamefont
  {Yagi}}\ and\ \bibinfo {author} {\bibfnamefont {N.}~\bibnamefont {Yunes}},\
  }\href {\doibase 10.1103/PhysRevD.88.023009} {\bibfield  {journal} {\bibinfo
  {journal} {Phys. Rev. D}\ }\textbf {\bibinfo {volume} {88}},\ \bibinfo
  {pages} {023009} (\bibinfo {year} {2013}{\natexlab{b}})}\BibitemShut
  {NoStop}%
\bibitem [{\citenamefont {Maselli}\ \emph {et~al.}(2013)\citenamefont
  {Maselli}, \citenamefont {Cardoso}, \citenamefont {Ferrari}, \citenamefont
  {Gualtieri},\ and\ \citenamefont {Pani}}]{Maselli:2013mva}%
  \BibitemOpen
  \bibfield  {author} {\bibinfo {author} {\bibfnamefont {A.}~\bibnamefont
  {Maselli}}, \bibinfo {author} {\bibfnamefont {V.}~\bibnamefont {Cardoso}},
  \bibinfo {author} {\bibfnamefont {V.}~\bibnamefont {Ferrari}}, \bibinfo
  {author} {\bibfnamefont {L.}~\bibnamefont {Gualtieri}}, \ and\ \bibinfo
  {author} {\bibfnamefont {P.}~\bibnamefont {Pani}},\ }\href {\doibase
  10.1103/PhysRevD.88.023007} {\bibfield  {journal} {\bibinfo  {journal} {Phys.
  Rev.}\ }\textbf {\bibinfo {volume} {D88}},\ \bibinfo {pages} {023007}
  (\bibinfo {year} {2013})},\ \Eprint {http://arxiv.org/abs/1304.2052}
  {arXiv:1304.2052 [gr-qc]} \BibitemShut {NoStop}%
\bibitem [{\citenamefont {Yagi}\ and\ \citenamefont
  {Yunes}(2017)}]{Yagi:2016bkt}%
  \BibitemOpen
  \bibfield  {author} {\bibinfo {author} {\bibfnamefont {K.}~\bibnamefont
  {Yagi}}\ and\ \bibinfo {author} {\bibfnamefont {N.}~\bibnamefont {Yunes}},\
  }\href {\doibase 10.1016/j.physrep.2017.03.002} {\bibfield  {journal}
  {\bibinfo  {journal} {Phys. Rept.}\ }\textbf {\bibinfo {volume} {681}},\
  \bibinfo {pages} {1} (\bibinfo {year} {2017})},\ \Eprint
  {http://arxiv.org/abs/1608.02582} {arXiv:1608.02582 [gr-qc]} \BibitemShut
  {NoStop}%
\bibitem [{\citenamefont {Abbott}\ \emph
  {et~al.}(2017{\natexlab{a}})\citenamefont {Abbott} \emph
  {et~al.}}]{TheLIGOScientific:2017qsa}%
  \BibitemOpen
  \bibfield  {author} {\bibinfo {author} {\bibfnamefont {B.~P.}\ \bibnamefont
  {Abbott}} \emph {et~al.} (\bibinfo {collaboration} {Virgo, LIGO
  Scientific}),\ }\href {\doibase 10.1103/PhysRevLett.119.161101} {\bibfield
  {journal} {\bibinfo  {journal} {Phys. Rev. Lett.}\ }\textbf {\bibinfo
  {volume} {119}},\ \bibinfo {pages} {161101} (\bibinfo {year}
  {2017}{\natexlab{a}})},\ \Eprint {http://arxiv.org/abs/1710.05832}
  {arXiv:1710.05832 [gr-qc]} \BibitemShut {NoStop}%
\bibitem [{\citenamefont {Chatziioannou}\ \emph {et~al.}(2018)\citenamefont
  {Chatziioannou}, \citenamefont {Haster},\ and\ \citenamefont
  {Zimmerman}}]{Katerina:residuals}%
  \BibitemOpen
  \bibfield  {author} {\bibinfo {author} {\bibfnamefont {K.}~\bibnamefont
  {Chatziioannou}}, \bibinfo {author} {\bibfnamefont {C.-J.}\ \bibnamefont
  {Haster}}, \ and\ \bibinfo {author} {\bibfnamefont {A.}~\bibnamefont
  {Zimmerman}},\ }\href {\doibase 10.1103/PhysRevD.97.104036} {\bibfield
  {journal} {\bibinfo  {journal} {Phys. Rev. D}\ }\textbf {\bibinfo {volume}
  {97}},\ \bibinfo {pages} {104036} (\bibinfo {year} {2018})}\BibitemShut
  {NoStop}%
\bibitem [{\citenamefont {Abbott}\ \emph
  {et~al.}(2018{\natexlab{a}})\citenamefont {Abbott} \emph
  {et~al.}}]{LIGO:posterior}%
  \BibitemOpen
  \bibfield  {author} {\bibinfo {author} {\bibfnamefont {B.~P.}\ \bibnamefont
  {Abbott}} \emph {et~al.} (\bibinfo {collaboration} {LIGO Scientific,
  Virgo}),\ }\href {\doibase 10.1103/PhysRevLett.121.161101} {\bibfield
  {journal} {\bibinfo  {journal} {Phys. Rev. Lett.}\ }\textbf {\bibinfo
  {volume} {121}},\ \bibinfo {pages} {161101} (\bibinfo {year}
  {2018}{\natexlab{a}})},\ \Eprint {http://arxiv.org/abs/1805.11581}
  {arXiv:1805.11581 [gr-qc]} \BibitemShut {NoStop}%
\bibitem [{\citenamefont {Damour}\ and\ \citenamefont
  {Esposito-Farese}(1996)}]{Damour:1996ke}%
  \BibitemOpen
  \bibfield  {author} {\bibinfo {author} {\bibfnamefont {T.}~\bibnamefont
  {Damour}}\ and\ \bibinfo {author} {\bibfnamefont {G.}~\bibnamefont
  {Esposito-Farese}},\ }\href {\doibase 10.1103/PhysRevD.54.1474} {\bibfield
  {journal} {\bibinfo  {journal} {Phys. Rev.}\ }\textbf {\bibinfo {volume}
  {D54}},\ \bibinfo {pages} {1474} (\bibinfo {year} {1996})},\ \Eprint
  {http://arxiv.org/abs/gr-qc/9602056} {arXiv:gr-qc/9602056 [gr-qc]}
  \BibitemShut {NoStop}%
\bibitem [{\citenamefont {Eling}\ \emph {et~al.}(2007)\citenamefont {Eling},
  \citenamefont {Jacobson},\ and\ \citenamefont
  {Coleman~Miller}}]{Eling:2007xh}%
  \BibitemOpen
  \bibfield  {author} {\bibinfo {author} {\bibfnamefont {C.}~\bibnamefont
  {Eling}}, \bibinfo {author} {\bibfnamefont {T.}~\bibnamefont {Jacobson}}, \
  and\ \bibinfo {author} {\bibfnamefont {M.}~\bibnamefont {Coleman~Miller}},\
  }\href {\doibase 10.1103/PhysRevD.76.042003, 10.1103/PhysRevD.80.129906}
  {\bibfield  {journal} {\bibinfo  {journal} {Phys. Rev.}\ }\textbf {\bibinfo
  {volume} {D76}},\ \bibinfo {pages} {042003} (\bibinfo {year} {2007})},\
  \bibinfo {note} {[Erratum: Phys. Rev.D80,129906(2009)]},\ \Eprint
  {http://arxiv.org/abs/0705.1565} {arXiv:0705.1565 [gr-qc]} \BibitemShut
  {NoStop}%
\bibitem [{\citenamefont {Yagi}\ \emph
  {et~al.}(2014{\natexlab{a}})\citenamefont {Yagi}, \citenamefont {Blas},
  \citenamefont {Barausse},\ and\ \citenamefont {Yunes}}]{Yagi:2013ava}%
  \BibitemOpen
  \bibfield  {author} {\bibinfo {author} {\bibfnamefont {K.}~\bibnamefont
  {Yagi}}, \bibinfo {author} {\bibfnamefont {D.}~\bibnamefont {Blas}}, \bibinfo
  {author} {\bibfnamefont {E.}~\bibnamefont {Barausse}}, \ and\ \bibinfo
  {author} {\bibfnamefont {N.}~\bibnamefont {Yunes}},\ }\href {\doibase
  10.1103/PhysRevD.90.069902, 10.1103/PhysRevD.90.069901,
  10.1103/PhysRevD.89.084067} {\bibfield  {journal} {\bibinfo  {journal} {Phys.
  Rev.}\ }\textbf {\bibinfo {volume} {D89}},\ \bibinfo {pages} {084067}
  (\bibinfo {year} {2014}{\natexlab{a}})},\ \bibinfo {note} {[Erratum: Phys.
  Rev.D90,no.6,069902(2014); Erratum: Phys. Rev.D90,no.6,069901(2014)]},\
  \Eprint {http://arxiv.org/abs/1311.7144} {arXiv:1311.7144 [gr-qc]}
  \BibitemShut {NoStop}%
\bibitem [{\citenamefont {Yagi}\ \emph
  {et~al.}(2014{\natexlab{b}})\citenamefont {Yagi}, \citenamefont {Blas},
  \citenamefont {Yunes},\ and\ \citenamefont {Barausse}}]{Yagi:2013qpa}%
  \BibitemOpen
  \bibfield  {author} {\bibinfo {author} {\bibfnamefont {K.}~\bibnamefont
  {Yagi}}, \bibinfo {author} {\bibfnamefont {D.}~\bibnamefont {Blas}}, \bibinfo
  {author} {\bibfnamefont {N.}~\bibnamefont {Yunes}}, \ and\ \bibinfo {author}
  {\bibfnamefont {E.}~\bibnamefont {Barausse}},\ }\href {\doibase
  10.1103/PhysRevLett.112.161101} {\bibfield  {journal} {\bibinfo  {journal}
  {Phys. Rev. Lett.}\ }\textbf {\bibinfo {volume} {112}},\ \bibinfo {pages}
  {161101} (\bibinfo {year} {2014}{\natexlab{b}})},\ \Eprint
  {http://arxiv.org/abs/1307.6219} {arXiv:1307.6219 [gr-qc]} \BibitemShut
  {NoStop}%
\bibitem [{\citenamefont {Yunes}\ \emph {et~al.}(2010)\citenamefont {Yunes},
  \citenamefont {Psaltis}, \citenamefont {Ozel},\ and\ \citenamefont
  {Loeb}}]{Yunes:2009ch}%
  \BibitemOpen
  \bibfield  {author} {\bibinfo {author} {\bibfnamefont {N.}~\bibnamefont
  {Yunes}}, \bibinfo {author} {\bibfnamefont {D.}~\bibnamefont {Psaltis}},
  \bibinfo {author} {\bibfnamefont {F.}~\bibnamefont {Ozel}}, \ and\ \bibinfo
  {author} {\bibfnamefont {A.}~\bibnamefont {Loeb}},\ }\href {\doibase
  10.1103/PhysRevD.81.064020} {\bibfield  {journal} {\bibinfo  {journal} {Phys.
  Rev.}\ }\textbf {\bibinfo {volume} {D81}},\ \bibinfo {pages} {064020}
  (\bibinfo {year} {2010})},\ \Eprint {http://arxiv.org/abs/0912.2736}
  {arXiv:0912.2736 [gr-qc]} \BibitemShut {NoStop}%
\bibitem [{\citenamefont {Yagi}\ \emph {et~al.}(2013)\citenamefont {Yagi},
  \citenamefont {Stein}, \citenamefont {Yunes},\ and\ \citenamefont
  {Tanaka}}]{Yagi:2013mbt}%
  \BibitemOpen
  \bibfield  {author} {\bibinfo {author} {\bibfnamefont {K.}~\bibnamefont
  {Yagi}}, \bibinfo {author} {\bibfnamefont {L.~C.}\ \bibnamefont {Stein}},
  \bibinfo {author} {\bibfnamefont {N.}~\bibnamefont {Yunes}}, \ and\ \bibinfo
  {author} {\bibfnamefont {T.}~\bibnamefont {Tanaka}},\ }\href {\doibase
  10.1103/PhysRevD.87.084058, 10.1103/PhysRevD.93.089909} {\bibfield  {journal}
  {\bibinfo  {journal} {Phys. Rev.}\ }\textbf {\bibinfo {volume} {D87}},\
  \bibinfo {pages} {084058} (\bibinfo {year} {2013})},\ \bibinfo {note}
  {[Erratum: Phys. Rev.D93,no.8,089909(2016)]},\ \Eprint
  {http://arxiv.org/abs/1302.1918} {arXiv:1302.1918 [gr-qc]} \BibitemShut
  {NoStop}%
\bibitem [{\citenamefont {Gupta}\ \emph {et~al.}(2018)\citenamefont {Gupta},
  \citenamefont {Majumder}, \citenamefont {Yagi},\ and\ \citenamefont
  {Yunes}}]{Gupta:2017vsl}%
  \BibitemOpen
  \bibfield  {author} {\bibinfo {author} {\bibfnamefont {T.}~\bibnamefont
  {Gupta}}, \bibinfo {author} {\bibfnamefont {B.}~\bibnamefont {Majumder}},
  \bibinfo {author} {\bibfnamefont {K.}~\bibnamefont {Yagi}}, \ and\ \bibinfo
  {author} {\bibfnamefont {N.}~\bibnamefont {Yunes}},\ }\href {\doibase
  10.1088/1361-6382/aa9c68} {\bibfield  {journal} {\bibinfo  {journal} {Class.
  Quant. Grav.}\ }\textbf {\bibinfo {volume} {35}},\ \bibinfo {pages} {025009}
  (\bibinfo {year} {2018})},\ \Eprint {http://arxiv.org/abs/1710.07862}
  {arXiv:1710.07862 [gr-qc]} \BibitemShut {NoStop}%
\bibitem [{\citenamefont {Babichev}\ \emph {et~al.}(2016)\citenamefont
  {Babichev}, \citenamefont {Koyama}, \citenamefont {Langlois}, \citenamefont
  {Saito},\ and\ \citenamefont {Sakstein}}]{Babichev:2016jom}%
  \BibitemOpen
  \bibfield  {author} {\bibinfo {author} {\bibfnamefont {E.}~\bibnamefont
  {Babichev}}, \bibinfo {author} {\bibfnamefont {K.}~\bibnamefont {Koyama}},
  \bibinfo {author} {\bibfnamefont {D.}~\bibnamefont {Langlois}}, \bibinfo
  {author} {\bibfnamefont {R.}~\bibnamefont {Saito}}, \ and\ \bibinfo {author}
  {\bibfnamefont {J.}~\bibnamefont {Sakstein}},\ }\href {\doibase
  10.1088/0264-9381/33/23/235014} {\bibfield  {journal} {\bibinfo  {journal}
  {Class. Quant. Grav.}\ }\textbf {\bibinfo {volume} {33}},\ \bibinfo {pages}
  {235014} (\bibinfo {year} {2016})},\ \Eprint
  {http://arxiv.org/abs/1606.06627} {arXiv:1606.06627 [gr-qc]} \BibitemShut
  {NoStop}%
\bibitem [{\citenamefont {Sakstein}\ \emph {et~al.}(2017)\citenamefont
  {Sakstein}, \citenamefont {Babichev}, \citenamefont {Koyama}, \citenamefont
  {Langlois},\ and\ \citenamefont {Saito}}]{Sakstein:2016oel}%
  \BibitemOpen
  \bibfield  {author} {\bibinfo {author} {\bibfnamefont {J.}~\bibnamefont
  {Sakstein}}, \bibinfo {author} {\bibfnamefont {E.}~\bibnamefont {Babichev}},
  \bibinfo {author} {\bibfnamefont {K.}~\bibnamefont {Koyama}}, \bibinfo
  {author} {\bibfnamefont {D.}~\bibnamefont {Langlois}}, \ and\ \bibinfo
  {author} {\bibfnamefont {R.}~\bibnamefont {Saito}},\ }\href {\doibase
  10.1103/PhysRevD.95.064013} {\bibfield  {journal} {\bibinfo  {journal} {Phys.
  Rev.}\ }\textbf {\bibinfo {volume} {D95}},\ \bibinfo {pages} {064013}
  (\bibinfo {year} {2017})},\ \Eprint {http://arxiv.org/abs/1612.04263}
  {arXiv:1612.04263 [gr-qc]} \BibitemShut {NoStop}%
\bibitem [{\citenamefont {Pani}\ and\ \citenamefont
  {Berti}(2014)}]{Pani:2014jra}%
  \BibitemOpen
  \bibfield  {author} {\bibinfo {author} {\bibfnamefont {P.}~\bibnamefont
  {Pani}}\ and\ \bibinfo {author} {\bibfnamefont {E.}~\bibnamefont {Berti}},\
  }\href {\doibase 10.1103/PhysRevD.90.024025} {\bibfield  {journal} {\bibinfo
  {journal} {Phys. Rev.}\ }\textbf {\bibinfo {volume} {D90}},\ \bibinfo {pages}
  {024025} (\bibinfo {year} {2014})},\ \Eprint {http://arxiv.org/abs/1405.4547}
  {arXiv:1405.4547 [gr-qc]} \BibitemShut {NoStop}%
\bibitem [{\citenamefont {Minamitsuji}\ and\ \citenamefont
  {Silva}(2016)}]{Minamitsuji:2016hkk}%
  \BibitemOpen
  \bibfield  {author} {\bibinfo {author} {\bibfnamefont {M.}~\bibnamefont
  {Minamitsuji}}\ and\ \bibinfo {author} {\bibfnamefont {H.~O.}\ \bibnamefont
  {Silva}},\ }\href {\doibase 10.1103/PhysRevD.93.124041} {\bibfield  {journal}
  {\bibinfo  {journal} {Phys. Rev.}\ }\textbf {\bibinfo {volume} {D93}},\
  \bibinfo {pages} {124041} (\bibinfo {year} {2016})},\ \Eprint
  {http://arxiv.org/abs/1604.07742} {arXiv:1604.07742 [gr-qc]} \BibitemShut
  {NoStop}%
\bibitem [{\citenamefont {Maselli}\ \emph {et~al.}(2016)\citenamefont
  {Maselli}, \citenamefont {Silva}, \citenamefont {Minamitsuji},\ and\
  \citenamefont {Berti}}]{Maselli:2016gxk}%
  \BibitemOpen
  \bibfield  {author} {\bibinfo {author} {\bibfnamefont {A.}~\bibnamefont
  {Maselli}}, \bibinfo {author} {\bibfnamefont {H.~O.}\ \bibnamefont {Silva}},
  \bibinfo {author} {\bibfnamefont {M.}~\bibnamefont {Minamitsuji}}, \ and\
  \bibinfo {author} {\bibfnamefont {E.}~\bibnamefont {Berti}},\ }\href
  {\doibase 10.1103/PhysRevD.93.124056} {\bibfield  {journal} {\bibinfo
  {journal} {Phys. Rev.}\ }\textbf {\bibinfo {volume} {D93}},\ \bibinfo {pages}
  {124056} (\bibinfo {year} {2016})},\ \Eprint
  {http://arxiv.org/abs/1603.04876} {arXiv:1603.04876 [gr-qc]} \BibitemShut
  {NoStop}%
\bibitem [{\citenamefont {Lattimer}\ and\ \citenamefont
  {Schutz}(2005)}]{Lattimer:2004nj}%
  \BibitemOpen
  \bibfield  {author} {\bibinfo {author} {\bibfnamefont {J.~M.}\ \bibnamefont
  {Lattimer}}\ and\ \bibinfo {author} {\bibfnamefont {B.~F.}\ \bibnamefont
  {Schutz}},\ }\href {\doibase 10.1086/431543} {\bibfield  {journal} {\bibinfo
  {journal} {Astrophys. J.}\ }\textbf {\bibinfo {volume} {629}},\ \bibinfo
  {pages} {979} (\bibinfo {year} {2005})},\ \Eprint
  {http://arxiv.org/abs/astro-ph/0411470} {arXiv:astro-ph/0411470 [astro-ph]}
  \BibitemShut {NoStop}%
\bibitem [{\citenamefont {Ozel}\ \emph {et~al.}(2016)\citenamefont {Ozel},
  \citenamefont {Psaltis}, \citenamefont {Arzoumanian}, \citenamefont
  {Morsink},\ and\ \citenamefont {Baubock}}]{Ozel:2015ykl}%
  \BibitemOpen
  \bibfield  {author} {\bibinfo {author} {\bibfnamefont {F.}~\bibnamefont
  {Ozel}}, \bibinfo {author} {\bibfnamefont {D.}~\bibnamefont {Psaltis}},
  \bibinfo {author} {\bibfnamefont {Z.}~\bibnamefont {Arzoumanian}}, \bibinfo
  {author} {\bibfnamefont {S.}~\bibnamefont {Morsink}}, \ and\ \bibinfo
  {author} {\bibfnamefont {M.}~\bibnamefont {Baubock}},\ }\href {\doibase
  10.3847/0004-637X/832/1/92} {\bibfield  {journal} {\bibinfo  {journal}
  {Astrophys. J.}\ }\textbf {\bibinfo {volume} {832}},\ \bibinfo {pages} {92}
  (\bibinfo {year} {2016})},\ \Eprint {http://arxiv.org/abs/1512.03067}
  {arXiv:1512.03067 [astro-ph.HE]} \BibitemShut {NoStop}%
\bibitem [{\citenamefont {Doneva}\ and\ \citenamefont
  {Pappas}(2018)}]{Doneva:2017jop}%
  \BibitemOpen
  \bibfield  {author} {\bibinfo {author} {\bibfnamefont {D.~D.}\ \bibnamefont
  {Doneva}}\ and\ \bibinfo {author} {\bibfnamefont {G.}~\bibnamefont
  {Pappas}},\ }\href {\doibase 10.1007/978-3-319-97616-7_13} {\bibfield
  {journal} {\bibinfo  {journal} {Astrophys. Space Sci. Libr.}\ }\textbf
  {\bibinfo {volume} {457}},\ \bibinfo {pages} {737} (\bibinfo {year}
  {2018})},\ \Eprint {http://arxiv.org/abs/1709.08046} {arXiv:1709.08046
  [gr-qc]} \BibitemShut {NoStop}%
\bibitem [{\citenamefont {Carney}\ \emph {et~al.}(2018)\citenamefont {Carney},
  \citenamefont {Wade},\ and\ \citenamefont {Irwin}}]{Carney:2018sdv}%
  \BibitemOpen
  \bibfield  {author} {\bibinfo {author} {\bibfnamefont {M.~F.}\ \bibnamefont
  {Carney}}, \bibinfo {author} {\bibfnamefont {L.~E.}\ \bibnamefont {Wade}}, \
  and\ \bibinfo {author} {\bibfnamefont {B.~S.}\ \bibnamefont {Irwin}},\ }\href
  {\doibase 10.1103/PhysRevD.98.063004} {\bibfield  {journal} {\bibinfo
  {journal} {Phys. Rev.}\ }\textbf {\bibinfo {volume} {D98}},\ \bibinfo {pages}
  {063004} (\bibinfo {year} {2018})},\ \Eprint
  {http://arxiv.org/abs/1805.11217} {arXiv:1805.11217 [gr-qc]} \BibitemShut
  {NoStop}%
\bibitem [{\citenamefont {Lindblom}(2018)}]{Lindblom:2018rfr}%
  \BibitemOpen
  \bibfield  {author} {\bibinfo {author} {\bibfnamefont {L.}~\bibnamefont
  {Lindblom}},\ }\href {\doibase 10.1103/PhysRevD.97.123019} {\bibfield
  {journal} {\bibinfo  {journal} {Phys. Rev.}\ }\textbf {\bibinfo {volume}
  {D97}},\ \bibinfo {pages} {123019} (\bibinfo {year} {2018})},\ \Eprint
  {http://arxiv.org/abs/1804.04072} {arXiv:1804.04072 [astro-ph.HE]}
  \BibitemShut {NoStop}%
\bibitem [{\citenamefont {Paschalidis}\ \emph {et~al.}(2018)\citenamefont
  {Paschalidis}, \citenamefont {Yagi}, \citenamefont {Alvarez-Castillo},
  \citenamefont {Blaschke},\ and\ \citenamefont {Sedrakian}}]{Paschalidis2018}%
  \BibitemOpen
  \bibfield  {author} {\bibinfo {author} {\bibfnamefont {V.}~\bibnamefont
  {Paschalidis}}, \bibinfo {author} {\bibfnamefont {K.}~\bibnamefont {Yagi}},
  \bibinfo {author} {\bibfnamefont {D.}~\bibnamefont {Alvarez-Castillo}},
  \bibinfo {author} {\bibfnamefont {D.~B.}\ \bibnamefont {Blaschke}}, \ and\
  \bibinfo {author} {\bibfnamefont {A.}~\bibnamefont {Sedrakian}},\ }\href
  {\doibase 10.1103/physrevd.97.084038} {\bibfield  {journal} {\bibinfo
  {journal} {Physical Review D}\ }\textbf {\bibinfo {volume} {97}} (\bibinfo
  {year} {2018}),\ 10.1103/physrevd.97.084038}\BibitemShut {NoStop}%
\bibitem [{\citenamefont {Most}\ \emph {et~al.}(2019)\citenamefont {Most},
  \citenamefont {Papenfort}, \citenamefont {Dexheimer}, \citenamefont
  {Hanauske}, \citenamefont {Schramm}, \citenamefont {Stï¿œcker},\ and\
  \citenamefont {Rezzolla}}]{Most:2018eaw}%
  \BibitemOpen
  \bibfield  {author} {\bibinfo {author} {\bibfnamefont {E.~R.}\ \bibnamefont
  {Most}}, \bibinfo {author} {\bibfnamefont {L.~J.}\ \bibnamefont {Papenfort}},
  \bibinfo {author} {\bibfnamefont {V.}~\bibnamefont {Dexheimer}}, \bibinfo
  {author} {\bibfnamefont {M.}~\bibnamefont {Hanauske}}, \bibinfo {author}
  {\bibfnamefont {S.}~\bibnamefont {Schramm}}, \bibinfo {author} {\bibfnamefont
  {H.}~\bibnamefont {Stï¿œcker}}, \ and\ \bibinfo {author} {\bibfnamefont
  {L.}~\bibnamefont {Rezzolla}},\ }\href {\doibase
  10.1103/PhysRevLett.122.061101} {\bibfield  {journal} {\bibinfo  {journal}
  {Phys. Rev. Lett.}\ }\textbf {\bibinfo {volume} {122}},\ \bibinfo {pages}
  {061101} (\bibinfo {year} {2019})},\ \Eprint
  {http://arxiv.org/abs/1807.03684} {arXiv:1807.03684 [astro-ph.HE]}
  \BibitemShut {NoStop}%
\bibitem [{\citenamefont {Burgio}\ \emph {et~al.}(2018)\citenamefont {Burgio},
  \citenamefont {Drago}, \citenamefont {Pagliara}, \citenamefont {Schulze},\
  and\ \citenamefont {Wei}}]{Burgio:2018yix}%
  \BibitemOpen
  \bibfield  {author} {\bibinfo {author} {\bibfnamefont {G.~F.}\ \bibnamefont
  {Burgio}}, \bibinfo {author} {\bibfnamefont {A.}~\bibnamefont {Drago}},
  \bibinfo {author} {\bibfnamefont {G.}~\bibnamefont {Pagliara}}, \bibinfo
  {author} {\bibfnamefont {H.~J.}\ \bibnamefont {Schulze}}, \ and\ \bibinfo
  {author} {\bibfnamefont {J.~B.}\ \bibnamefont {Wei}},\ }\href {\doibase
  10.3847/1538-4357/aac6ee} {\bibfield  {journal} {\bibinfo  {journal}
  {Astrophys. J.}\ }\textbf {\bibinfo {volume} {860}},\ \bibinfo {pages} {139}
  (\bibinfo {year} {2018})},\ \Eprint {http://arxiv.org/abs/1803.09696}
  {arXiv:1803.09696 [astro-ph.HE]} \BibitemShut {NoStop}%
\bibitem [{\citenamefont {Montana}\ \emph {et~al.}(2019)\citenamefont
  {Montana}, \citenamefont {Tolos}, \citenamefont {Hanauske},\ and\
  \citenamefont {Rezzolla}}]{Montana:2018bkb}%
  \BibitemOpen
  \bibfield  {author} {\bibinfo {author} {\bibfnamefont {G.}~\bibnamefont
  {Montana}}, \bibinfo {author} {\bibfnamefont {L.}~\bibnamefont {Tolos}},
  \bibinfo {author} {\bibfnamefont {M.}~\bibnamefont {Hanauske}}, \ and\
  \bibinfo {author} {\bibfnamefont {L.}~\bibnamefont {Rezzolla}},\ }\href
  {\doibase 10.1103/PhysRevD.99.103009} {\bibfield  {journal} {\bibinfo
  {journal} {Phys. Rev.}\ }\textbf {\bibinfo {volume} {D99}},\ \bibinfo {pages}
  {103009} (\bibinfo {year} {2019})},\ \Eprint
  {http://arxiv.org/abs/1811.10929} {arXiv:1811.10929 [astro-ph.HE]}
  \BibitemShut {NoStop}%
\bibitem [{aLI()}]{aLIGO}%
  \BibitemOpen
  \href@noop {} {\enquote {\bibinfo {title} {Advanced {LIGO}},}\ }\bibinfo
  {howpublished} {\url{https://www.advancedligo.mit.edu/}}\BibitemShut
  {NoStop}%
\bibitem [{Ap_()}]{Ap_Voyager_CE}%
  \BibitemOpen
  \href@noop {} {\enquote {\bibinfo {title} {Ligo-t1400316-v4: Instrument
  science white paper},}\ }\bibinfo {howpublished}
  {\url{https://dcc.ligo.org/ligo-T1400316/public}}\BibitemShut {NoStop}%
\bibitem [{ET()}]{ET}%
  \BibitemOpen
  \href@noop {} {\enquote {\bibinfo {title} {The {ET} project website},}\
  }\bibinfo {howpublished} {\url{http://www.et-gw.eu/}}\BibitemShut {NoStop}%
\bibitem [{\citenamefont {Abbott}\ \emph
  {et~al.}(2017{\natexlab{b}})\citenamefont {Abbott} \emph
  {et~al.}}]{Abbott2017}%
  \BibitemOpen
  \bibfield  {author} {\bibinfo {author} {\bibfnamefont {B.}~\bibnamefont
  {Abbott}} \emph {et~al.},\ }\href {\doibase 10.1103/physrevlett.119.161101}
  {\bibfield  {journal} {\bibinfo  {journal} {Physical Review Letters}\
  }\textbf {\bibinfo {volume} {119}} (\bibinfo {year} {2017}{\natexlab{b}}),\
  10.1103/physrevlett.119.161101}\BibitemShut {NoStop}%
\bibitem [{\citenamefont {Khan}\ \emph {et~al.}(2016)\citenamefont {Khan},
  \citenamefont {Husa}, \citenamefont {Hannam}, \citenamefont {Ohme},
  \citenamefont {P\"urrer}, \citenamefont {Forteza},\ and\ \citenamefont
  {Boh\'e}}]{PhenomDI}%
  \BibitemOpen
  \bibfield  {author} {\bibinfo {author} {\bibfnamefont {S.}~\bibnamefont
  {Khan}}, \bibinfo {author} {\bibfnamefont {S.}~\bibnamefont {Husa}}, \bibinfo
  {author} {\bibfnamefont {M.}~\bibnamefont {Hannam}}, \bibinfo {author}
  {\bibfnamefont {F.}~\bibnamefont {Ohme}}, \bibinfo {author} {\bibfnamefont
  {M.}~\bibnamefont {P\"urrer}}, \bibinfo {author} {\bibfnamefont {X.~J.}\
  \bibnamefont {Forteza}}, \ and\ \bibinfo {author} {\bibfnamefont
  {A.}~\bibnamefont {Boh\'e}},\ }\href {\doibase 10.1103/PhysRevD.93.044007}
  {\bibfield  {journal} {\bibinfo  {journal} {Phys. Rev. D}\ }\textbf {\bibinfo
  {volume} {93}},\ \bibinfo {pages} {044007} (\bibinfo {year}
  {2016})}\BibitemShut {NoStop}%
\bibitem [{\citenamefont {Husa}\ \emph {et~al.}(2016)\citenamefont {Husa},
  \citenamefont {Khan}, \citenamefont {Hannam}, \citenamefont {P\"urrer},
  \citenamefont {Ohme}, \citenamefont {Forteza},\ and\ \citenamefont
  {Boh\'e}}]{PhenomDII}%
  \BibitemOpen
  \bibfield  {author} {\bibinfo {author} {\bibfnamefont {S.}~\bibnamefont
  {Husa}}, \bibinfo {author} {\bibfnamefont {S.}~\bibnamefont {Khan}}, \bibinfo
  {author} {\bibfnamefont {M.}~\bibnamefont {Hannam}}, \bibinfo {author}
  {\bibfnamefont {M.}~\bibnamefont {P\"urrer}}, \bibinfo {author}
  {\bibfnamefont {F.}~\bibnamefont {Ohme}}, \bibinfo {author} {\bibfnamefont
  {X.~J.}\ \bibnamefont {Forteza}}, \ and\ \bibinfo {author} {\bibfnamefont
  {A.}~\bibnamefont {Boh\'e}},\ }\href {\doibase 10.1103/PhysRevD.93.044006}
  {\bibfield  {journal} {\bibinfo  {journal} {Phys. Rev. D}\ }\textbf {\bibinfo
  {volume} {93}},\ \bibinfo {pages} {044006} (\bibinfo {year}
  {2016})}\BibitemShut {NoStop}%
\bibitem [{\citenamefont {Dietrich}\ \emph {et~al.}(2017)\citenamefont
  {Dietrich}, \citenamefont {Bernuzzi},\ and\ \citenamefont
  {Tichy}}]{Dietrich:2017aum}%
  \BibitemOpen
  \bibfield  {author} {\bibinfo {author} {\bibfnamefont {T.}~\bibnamefont
  {Dietrich}}, \bibinfo {author} {\bibfnamefont {S.}~\bibnamefont {Bernuzzi}},
  \ and\ \bibinfo {author} {\bibfnamefont {W.}~\bibnamefont {Tichy}},\ }\href
  {\doibase 10.1103/PhysRevD.96.121501} {\bibfield  {journal} {\bibinfo
  {journal} {Phys. Rev.}\ }\textbf {\bibinfo {volume} {D96}},\ \bibinfo {pages}
  {121501} (\bibinfo {year} {2017})},\ \Eprint
  {http://arxiv.org/abs/1706.02969} {arXiv:1706.02969 [gr-qc]} \BibitemShut
  {NoStop}%
\bibitem [{\citenamefont {Samajdar}\ and\ \citenamefont
  {Dietrich}(2018)}]{Samajdar:NRTidal}%
  \BibitemOpen
  \bibfield  {author} {\bibinfo {author} {\bibfnamefont {A.}~\bibnamefont
  {Samajdar}}\ and\ \bibinfo {author} {\bibfnamefont {T.}~\bibnamefont
  {Dietrich}},\ }\href {\doibase 10.1103/PhysRevD.98.124030} {\bibfield
  {journal} {\bibinfo  {journal} {Phys. Rev.}\ }\textbf {\bibinfo {volume}
  {D98}},\ \bibinfo {pages} {124030} (\bibinfo {year} {2018})},\ \Eprint
  {http://arxiv.org/abs/1810.03936} {arXiv:1810.03936 [gr-qc]} \BibitemShut
  {NoStop}%
\bibitem [{\citenamefont {Oertel}\ \emph {et~al.}(2017)\citenamefont {Oertel},
  \citenamefont {Hempel}, \citenamefont {Kl\"ahn},\ and\ \citenamefont
  {Typel}}]{Oertel:Review}%
  \BibitemOpen
  \bibfield  {author} {\bibinfo {author} {\bibfnamefont {M.}~\bibnamefont
  {Oertel}}, \bibinfo {author} {\bibfnamefont {M.}~\bibnamefont {Hempel}},
  \bibinfo {author} {\bibfnamefont {T.}~\bibnamefont {Kl\"ahn}}, \ and\
  \bibinfo {author} {\bibfnamefont {S.}~\bibnamefont {Typel}},\ }\href
  {\doibase 10.1103/RevModPhys.89.015007} {\bibfield  {journal} {\bibinfo
  {journal} {Rev. Mod. Phys.}\ }\textbf {\bibinfo {volume} {89}},\ \bibinfo
  {pages} {015007} (\bibinfo {year} {2017})}\BibitemShut {NoStop}%
\bibitem [{\citenamefont {Baym}\ \emph {et~al.}(2018)\citenamefont {Baym},
  \citenamefont {Hatsuda}, \citenamefont {Kojo}, \citenamefont {Powell},
  \citenamefont {Song},\ and\ \citenamefont {Takatsuka}}]{Baym:Review}%
  \BibitemOpen
  \bibfield  {author} {\bibinfo {author} {\bibfnamefont {G.}~\bibnamefont
  {Baym}}, \bibinfo {author} {\bibfnamefont {T.}~\bibnamefont {Hatsuda}},
  \bibinfo {author} {\bibfnamefont {T.}~\bibnamefont {Kojo}}, \bibinfo {author}
  {\bibfnamefont {P.~D.}\ \bibnamefont {Powell}}, \bibinfo {author}
  {\bibfnamefont {Y.}~\bibnamefont {Song}}, \ and\ \bibinfo {author}
  {\bibfnamefont {T.}~\bibnamefont {Takatsuka}},\ }\href {\doibase
  10.1088/1361-6633/aaae14} {\bibfield  {journal} {\bibinfo  {journal} {Rept.
  Prog. Phys.}\ }\textbf {\bibinfo {volume} {81}},\ \bibinfo {pages} {056902}
  (\bibinfo {year} {2018})},\ \Eprint {http://arxiv.org/abs/1707.04966}
  {arXiv:1707.04966 [astro-ph.HE]} \BibitemShut {NoStop}%
\bibitem [{\citenamefont {Lindblom}(2010)}]{Lindblom:2010bb}%
  \BibitemOpen
  \bibfield  {author} {\bibinfo {author} {\bibfnamefont {L.}~\bibnamefont
  {Lindblom}},\ }\href {\doibase 10.1103/PhysRevD.82.103011} {\bibfield
  {journal} {\bibinfo  {journal} {Phys. Rev.}\ }\textbf {\bibinfo {volume}
  {D82}},\ \bibinfo {pages} {103011} (\bibinfo {year} {2010})},\ \Eprint
  {http://arxiv.org/abs/1009.0738} {arXiv:1009.0738 [astro-ph.HE]} \BibitemShut
  {NoStop}%
\bibitem [{\citenamefont {Lindblom}\ and\ \citenamefont
  {Indik}(2012{\natexlab{a}})}]{Lindblom:2012zi}%
  \BibitemOpen
  \bibfield  {author} {\bibinfo {author} {\bibfnamefont {L.}~\bibnamefont
  {Lindblom}}\ and\ \bibinfo {author} {\bibfnamefont {N.~M.}\ \bibnamefont
  {Indik}},\ }\href {\doibase 10.1103/PhysRevD.86.084003} {\bibfield  {journal}
  {\bibinfo  {journal} {Phys. Rev.}\ }\textbf {\bibinfo {volume} {D86}},\
  \bibinfo {pages} {084003} (\bibinfo {year} {2012}{\natexlab{a}})},\ \Eprint
  {http://arxiv.org/abs/1207.3744} {arXiv:1207.3744 [astro-ph.HE]} \BibitemShut
  {NoStop}%
\bibitem [{\citenamefont {Lindblom}\ and\ \citenamefont
  {Indik}(2014)}]{Lindblom:2013kra}%
  \BibitemOpen
  \bibfield  {author} {\bibinfo {author} {\bibfnamefont {L.}~\bibnamefont
  {Lindblom}}\ and\ \bibinfo {author} {\bibfnamefont {N.~M.}\ \bibnamefont
  {Indik}},\ }\href {\doibase 10.1103/PhysRevD.89.064003,
  10.1103/PhysRevD.93.129903} {\bibfield  {journal} {\bibinfo  {journal} {Phys.
  Rev.}\ }\textbf {\bibinfo {volume} {D89}},\ \bibinfo {pages} {064003}
  (\bibinfo {year} {2014})},\ \bibinfo {note} {[Erratum: Phys.
  Rev.D93,no.12,129903(2016)]},\ \Eprint {http://arxiv.org/abs/1310.0803}
  {arXiv:1310.0803 [astro-ph.HE]} \BibitemShut {NoStop}%
\bibitem [{\citenamefont {Abbott}\ \emph
  {et~al.}(2018{\natexlab{b}})\citenamefont {Abbott} \emph
  {et~al.}}]{Abbott:2018exr}%
  \BibitemOpen
  \bibfield  {author} {\bibinfo {author} {\bibfnamefont {B.~P.}\ \bibnamefont
  {Abbott}} \emph {et~al.} (\bibinfo {collaboration} {LIGO Scientific,
  Virgo}),\ }\href {\doibase 10.1103/PhysRevLett.121.161101} {\bibfield
  {journal} {\bibinfo  {journal} {Phys. Rev. Lett.}\ }\textbf {\bibinfo
  {volume} {121}},\ \bibinfo {pages} {161101} (\bibinfo {year}
  {2018}{\natexlab{b}})},\ \Eprint {http://arxiv.org/abs/1805.11581}
  {arXiv:1805.11581 [gr-qc]} \BibitemShut {NoStop}%
\bibitem [{\citenamefont {Read}\ \emph {et~al.}(2009)\citenamefont {Read},
  \citenamefont {Lackey}, \citenamefont {Owen},\ and\ \citenamefont
  {Friedman}}]{Read2009}%
  \BibitemOpen
  \bibfield  {author} {\bibinfo {author} {\bibfnamefont {J.~S.}\ \bibnamefont
  {Read}}, \bibinfo {author} {\bibfnamefont {B.~D.}\ \bibnamefont {Lackey}},
  \bibinfo {author} {\bibfnamefont {B.~J.}\ \bibnamefont {Owen}}, \ and\
  \bibinfo {author} {\bibfnamefont {J.~L.}\ \bibnamefont {Friedman}},\ }\href
  {\doibase 10.1103/physrevd.79.124032} {\bibfield  {journal} {\bibinfo
  {journal} {Physical Review D}\ }\textbf {\bibinfo {volume} {79}} (\bibinfo
  {year} {2009}),\ 10.1103/physrevd.79.124032}\BibitemShut {NoStop}%
\bibitem [{\citenamefont {Lackey}\ and\ \citenamefont
  {Wade}(2015)}]{Lackey:2014fwa}%
  \BibitemOpen
  \bibfield  {author} {\bibinfo {author} {\bibfnamefont {B.~D.}\ \bibnamefont
  {Lackey}}\ and\ \bibinfo {author} {\bibfnamefont {L.}~\bibnamefont {Wade}},\
  }\href {\doibase 10.1103/PhysRevD.91.043002} {\bibfield  {journal} {\bibinfo
  {journal} {Phys. Rev.}\ }\textbf {\bibinfo {volume} {D91}},\ \bibinfo {pages}
  {043002} (\bibinfo {year} {2015})},\ \Eprint {http://arxiv.org/abs/1410.8866}
  {arXiv:1410.8866 [gr-qc]} \BibitemShut {NoStop}%
\bibitem [{\citenamefont {Douchin}\ and\ \citenamefont
  {Haensel}(2001)}]{Douchin:2001sv}%
  \BibitemOpen
  \bibfield  {author} {\bibinfo {author} {\bibfnamefont {F.}~\bibnamefont
  {Douchin}}\ and\ \bibinfo {author} {\bibfnamefont {P.}~\bibnamefont
  {Haensel}},\ }\href {\doibase 10.1051/0004-6361:20011402} {\bibfield
  {journal} {\bibinfo  {journal} {Astron. Astrophys.}\ }\textbf {\bibinfo
  {volume} {380}},\ \bibinfo {pages} {151} (\bibinfo {year} {2001})},\ \Eprint
  {http://arxiv.org/abs/astro-ph/0111092} {arXiv:astro-ph/0111092 [astro-ph]}
  \BibitemShut {NoStop}%
\bibitem [{\citenamefont {Lindblom}\ and\ \citenamefont
  {Indik}(2012{\natexlab{b}})}]{Lindblom:parameters}%
  \BibitemOpen
  \bibfield  {author} {\bibinfo {author} {\bibfnamefont {L.}~\bibnamefont
  {Lindblom}}\ and\ \bibinfo {author} {\bibfnamefont {N.~M.}\ \bibnamefont
  {Indik}},\ }\href {\doibase 10.1103/PhysRevD.86.084003} {\bibfield  {journal}
  {\bibinfo  {journal} {Phys. Rev. D}\ }\textbf {\bibinfo {volume} {86}},\
  \bibinfo {pages} {084003} (\bibinfo {year} {2012}{\natexlab{b}})}\BibitemShut
  {NoStop}%
\bibitem [{\citenamefont {{Demorest}}\ \emph {et~al.}(2010)\citenamefont
  {{Demorest}}, \citenamefont {{Pennucci}}, \citenamefont {{Ransom}},
  \citenamefont {{Roberts}},\ and\ \citenamefont {{Hessels}}}]{1.97NS}%
  \BibitemOpen
  \bibfield  {author} {\bibinfo {author} {\bibfnamefont {P.~B.}\ \bibnamefont
  {{Demorest}}}, \bibinfo {author} {\bibfnamefont {T.}~\bibnamefont
  {{Pennucci}}}, \bibinfo {author} {\bibfnamefont {S.~M.}\ \bibnamefont
  {{Ransom}}}, \bibinfo {author} {\bibfnamefont {M.~S.~E.}\ \bibnamefont
  {{Roberts}}}, \ and\ \bibinfo {author} {\bibfnamefont {J.~W.~T.}\
  \bibnamefont {{Hessels}}},\ }\href {\doibase 10.1038/nature09466} {\bibfield
  {journal} {\bibinfo  {journal} {Nature}\ }\textbf {\bibinfo {volume} {467}},\
  \bibinfo {pages} {1081} (\bibinfo {year} {2010})},\ \Eprint
  {http://arxiv.org/abs/1010.5788} {arXiv:1010.5788 [astro-ph.HE]} \BibitemShut
  {NoStop}%
\bibitem [{\citenamefont {Antoniadis}\ \emph {et~al.}(2013)\citenamefont
  {Antoniadis}, \citenamefont {Freire}, \citenamefont {Wex}, \citenamefont
  {Tauris}, \citenamefont {Lynch} \emph {et~al.}}]{2.01NS}%
  \BibitemOpen
  \bibfield  {author} {\bibinfo {author} {\bibfnamefont {J.}~\bibnamefont
  {Antoniadis}}, \bibinfo {author} {\bibfnamefont {P.~C.}\ \bibnamefont
  {Freire}}, \bibinfo {author} {\bibfnamefont {N.}~\bibnamefont {Wex}},
  \bibinfo {author} {\bibfnamefont {T.~M.}\ \bibnamefont {Tauris}}, \bibinfo
  {author} {\bibfnamefont {R.~S.}\ \bibnamefont {Lynch}},  \emph {et~al.},\
  }\href {\doibase 10.1126/science.1233232} {\bibfield  {journal} {\bibinfo
  {journal} {Science}\ }\textbf {\bibinfo {volume} {340}},\ \bibinfo {pages}
  {1233232} (\bibinfo {year} {2013})},\ \Eprint
  {http://arxiv.org/abs/1304.6875} {arXiv:1304.6875 [astro-ph.HE]} \BibitemShut
  {NoStop}%
\bibitem [{\citenamefont {Zhao}(2015)}]{Zhao:massiveNS}%
  \BibitemOpen
  \bibfield  {author} {\bibinfo {author} {\bibfnamefont {X.-F.}\ \bibnamefont
  {Zhao}},\ }\href {\doibase 10.1142/S0218271815500583} {\bibfield  {journal}
  {\bibinfo  {journal} {Int. J. Mod. Phys.}\ }\textbf {\bibinfo {volume}
  {D24}},\ \bibinfo {pages} {1550058} (\bibinfo {year} {2015})},\ \Eprint
  {http://arxiv.org/abs/1712.08860} {arXiv:1712.08860 [nucl-th]} \BibitemShut
  {NoStop}%
\bibitem [{\citenamefont {{Gravitational Wave Open Science Center
  (GWOSC)}}()}]{GWOSC}%
  \BibitemOpen
  \bibfield  {author} {\bibinfo {author} {\bibnamefont {{Gravitational Wave
  Open Science Center (GWOSC)}}},\ }\href@noop {} {}\bibinfo {howpublished}
  {\url{https://www.gw-openscience.org/}}\BibitemShut {NoStop}%
\bibitem [{\citenamefont {Vallisneri}\ \emph {et~al.}(2015)\citenamefont
  {Vallisneri}, \citenamefont {Kanner}, \citenamefont {Williams}, \citenamefont
  {Weinstein},\ and\ \citenamefont {Stephens}}]{Vallisneri:2014vxa}%
  \BibitemOpen
  \bibfield  {author} {\bibinfo {author} {\bibfnamefont {M.}~\bibnamefont
  {Vallisneri}}, \bibinfo {author} {\bibfnamefont {J.}~\bibnamefont {Kanner}},
  \bibinfo {author} {\bibfnamefont {R.}~\bibnamefont {Williams}}, \bibinfo
  {author} {\bibfnamefont {A.}~\bibnamefont {Weinstein}}, \ and\ \bibinfo
  {author} {\bibfnamefont {B.}~\bibnamefont {Stephens}},\ }\bibfield
  {booktitle} {\emph {\bibinfo {booktitle} {{Proceedings, 10th International
  LISA Symposium: Gainesville, Florida, USA, May 18-23, 2014}}},\ }\href
  {\doibase 10.1088/1742-6596/610/1/012021} {\bibfield  {journal} {\bibinfo
  {journal} {J. Phys. Conf. Ser.}\ }\textbf {\bibinfo {volume} {610}},\
  \bibinfo {pages} {012021} (\bibinfo {year} {2015})},\ \Eprint
  {http://arxiv.org/abs/1410.4839} {arXiv:1410.4839 [gr-qc]} \BibitemShut
  {NoStop}%
\bibitem [{\citenamefont {Veitch}\ \emph {et~al.}(2015)\citenamefont {Veitch}
  \emph {et~al.}}]{Veitch:2014wba}%
  \BibitemOpen
  \bibfield  {author} {\bibinfo {author} {\bibfnamefont {J.}~\bibnamefont
  {Veitch}} \emph {et~al.},\ }\href {\doibase 10.1103/PhysRevD.91.042003}
  {\bibfield  {journal} {\bibinfo  {journal} {Phys. Rev. D}\ }\textbf {\bibinfo
  {volume} {91}},\ \bibinfo {pages} {042003} (\bibinfo {year} {2015})},\
  \Eprint {http://arxiv.org/abs/1409.7215} {arXiv:1409.7215 [gr-qc]}
  \BibitemShut {NoStop}%
\bibitem [{\citenamefont {{LIGO Scientific Collaboration and Virgo
  Collaboration}}(2017)}]{lalinference_o2}%
  \BibitemOpen
  \bibfield  {author} {\bibinfo {author} {\bibnamefont {{LIGO Scientific
  Collaboration and Virgo Collaboration}}},\ }\href
  {https://git.ligo.org/lscsoft/lalsuite/tree/lalinference_o2} {\enquote
  {\bibinfo {title} {{LALSuite},
  https://git.ligo.org/lscsoft/lalsuite/tree/lalinference\_o2},}\ } (\bibinfo
  {year} {2017})\BibitemShut {NoStop}%
\bibitem [{\citenamefont {Alford}\ and\ \citenamefont
  {Sedrakian}(2017)}]{Alford:2017qgh}%
  \BibitemOpen
  \bibfield  {author} {\bibinfo {author} {\bibfnamefont {M.~G.}\ \bibnamefont
  {Alford}}\ and\ \bibinfo {author} {\bibfnamefont {A.}~\bibnamefont
  {Sedrakian}},\ }\href {\doibase 10.1103/PhysRevLett.119.161104} {\bibfield
  {journal} {\bibinfo  {journal} {Phys. Rev. Lett.}\ }\textbf {\bibinfo
  {volume} {119}},\ \bibinfo {pages} {161104} (\bibinfo {year} {2017})},\
  \Eprint {http://arxiv.org/abs/1706.01592} {arXiv:1706.01592 [astro-ph.HE]}
  \BibitemShut {NoStop}%
\bibitem [{\citenamefont {{Seidov}}(1971)}]{1971SvA....15..347S}%
  \BibitemOpen
  \bibfield  {author} {\bibinfo {author} {\bibfnamefont {Z.~F.}\ \bibnamefont
  {{Seidov}}},\ }\href@noop {} {\bibfield  {journal} {\bibinfo  {journal} {Sov.
  Ast.}\ }\textbf {\bibinfo {volume} {15}},\ \bibinfo {pages} {347} (\bibinfo
  {year} {1971})}\BibitemShut {NoStop}%
\bibitem [{\citenamefont {Zdunik}\ and\ \citenamefont
  {Haensel}(2013)}]{Zdunik:2012dj}%
  \BibitemOpen
  \bibfield  {author} {\bibinfo {author} {\bibfnamefont {J.~L.}\ \bibnamefont
  {Zdunik}}\ and\ \bibinfo {author} {\bibfnamefont {P.}~\bibnamefont
  {Haensel}},\ }\href {\doibase 10.1051/0004-6361/201220697} {\bibfield
  {journal} {\bibinfo  {journal} {Astron. Astrophys.}\ }\textbf {\bibinfo
  {volume} {551}},\ \bibinfo {pages} {A61} (\bibinfo {year} {2013})},\ \Eprint
  {http://arxiv.org/abs/1211.1231} {arXiv:1211.1231 [astro-ph.SR]} \BibitemShut
  {NoStop}%
\bibitem [{\citenamefont {Alford}\ \emph {et~al.}(2013)\citenamefont {Alford},
  \citenamefont {Han},\ and\ \citenamefont {Prakash}}]{Alford:2013aca}%
  \BibitemOpen
  \bibfield  {author} {\bibinfo {author} {\bibfnamefont {M.~G.}\ \bibnamefont
  {Alford}}, \bibinfo {author} {\bibfnamefont {S.}~\bibnamefont {Han}}, \ and\
  \bibinfo {author} {\bibfnamefont {M.}~\bibnamefont {Prakash}},\ }\href
  {\doibase 10.1103/PhysRevD.88.083013} {\bibfield  {journal} {\bibinfo
  {journal} {Phys. Rev.}\ }\textbf {\bibinfo {volume} {D88}},\ \bibinfo {pages}
  {083013} (\bibinfo {year} {2013})},\ \Eprint {http://arxiv.org/abs/1302.4732}
  {arXiv:1302.4732 [astro-ph.SR]} \BibitemShut {NoStop}%
\bibitem [{\citenamefont {Char}\ and\ \citenamefont
  {Datta}(2018)}]{PhysRevD.98.084010}%
  \BibitemOpen
  \bibfield  {author} {\bibinfo {author} {\bibfnamefont {P.}~\bibnamefont
  {Char}}\ and\ \bibinfo {author} {\bibfnamefont {S.}~\bibnamefont {Datta}},\
  }\href {\doibase 10.1103/PhysRevD.98.084010} {\bibfield  {journal} {\bibinfo
  {journal} {Phys. Rev. D}\ }\textbf {\bibinfo {volume} {98}},\ \bibinfo
  {pages} {084010} (\bibinfo {year} {2018})}\BibitemShut {NoStop}%
\bibitem [{\citenamefont {Kumar}\ and\ \citenamefont
  {Landry}(2019)}]{Kumar:2019xgp}%
  \BibitemOpen
  \bibfield  {author} {\bibinfo {author} {\bibfnamefont {B.}~\bibnamefont
  {Kumar}}\ and\ \bibinfo {author} {\bibfnamefont {P.}~\bibnamefont {Landry}},\
  }\href@noop {} {\  (\bibinfo {year} {2019})},\ \Eprint
  {http://arxiv.org/abs/1902.04557} {arXiv:1902.04557 [gr-qc]} \BibitemShut
  {NoStop}%
\bibitem [{\citenamefont {Bandyopadhyay}\ \emph {et~al.}(2018)\citenamefont
  {Bandyopadhyay}, \citenamefont {Bhat}, \citenamefont {Char},\ and\
  \citenamefont {Chatterjee}}]{Bandyopadhyay2018}%
  \BibitemOpen
  \bibfield  {author} {\bibinfo {author} {\bibfnamefont {D.}~\bibnamefont
  {Bandyopadhyay}}, \bibinfo {author} {\bibfnamefont {S.~A.}\ \bibnamefont
  {Bhat}}, \bibinfo {author} {\bibfnamefont {P.}~\bibnamefont {Char}}, \ and\
  \bibinfo {author} {\bibfnamefont {D.}~\bibnamefont {Chatterjee}},\ }\href
  {\doibase 10.1140/epja/i2018-12456-y} {\bibfield  {journal} {\bibinfo
  {journal} {The European Physical Journal A}\ }\textbf {\bibinfo {volume}
  {54}},\ \bibinfo {pages} {26} (\bibinfo {year} {2018})}\BibitemShut {NoStop}%
\bibitem [{\citenamefont {Lau}\ \emph {et~al.}(2017)\citenamefont {Lau},
  \citenamefont {Leung},\ and\ \citenamefont {Lin}}]{PhysRevD.95.101302}%
  \BibitemOpen
  \bibfield  {author} {\bibinfo {author} {\bibfnamefont {S.~Y.}\ \bibnamefont
  {Lau}}, \bibinfo {author} {\bibfnamefont {P.~T.}\ \bibnamefont {Leung}}, \
  and\ \bibinfo {author} {\bibfnamefont {L.-M.}\ \bibnamefont {Lin}},\ }\href
  {\doibase 10.1103/PhysRevD.95.101302} {\bibfield  {journal} {\bibinfo
  {journal} {Phys. Rev. D}\ }\textbf {\bibinfo {volume} {95}},\ \bibinfo
  {pages} {101302} (\bibinfo {year} {2017})}\BibitemShut {NoStop}%
\bibitem [{\citenamefont {Han}\ and\ \citenamefont
  {Steiner}(2018)}]{Han:2018mtj}%
  \BibitemOpen
  \bibfield  {author} {\bibinfo {author} {\bibfnamefont {S.}~\bibnamefont
  {Han}}\ and\ \bibinfo {author} {\bibfnamefont {A.~W.}\ \bibnamefont
  {Steiner}},\ }\href@noop {} {\  (\bibinfo {year} {2018})},\ \Eprint
  {http://arxiv.org/abs/1810.10967} {arXiv:1810.10967 [nucl-th]} \BibitemShut
  {NoStop}%
\bibitem [{\citenamefont {Gamba}\ \emph {et~al.}(2019)\citenamefont {Gamba},
  \citenamefont {Read},\ and\ \citenamefont {Wade}}]{Gamba:2019kwu}%
  \BibitemOpen
  \bibfield  {author} {\bibinfo {author} {\bibfnamefont {R.}~\bibnamefont
  {Gamba}}, \bibinfo {author} {\bibfnamefont {J.~S.}\ \bibnamefont {Read}}, \
  and\ \bibinfo {author} {\bibfnamefont {L.~E.}\ \bibnamefont {Wade}},\
  }\href@noop {} {\  (\bibinfo {year} {2019})},\ \Eprint
  {http://arxiv.org/abs/1902.04616} {arXiv:1902.04616 [gr-qc]} \BibitemShut
  {NoStop}%
\bibitem [{\citenamefont {Annala}\ \emph {et~al.}(2018)\citenamefont {Annala},
  \citenamefont {Gorda}, \citenamefont {Kurkela},\ and\ \citenamefont
  {Vuorinen}}]{Annala:2017llu}%
  \BibitemOpen
  \bibfield  {author} {\bibinfo {author} {\bibfnamefont {E.}~\bibnamefont
  {Annala}}, \bibinfo {author} {\bibfnamefont {T.}~\bibnamefont {Gorda}},
  \bibinfo {author} {\bibfnamefont {A.}~\bibnamefont {Kurkela}}, \ and\
  \bibinfo {author} {\bibfnamefont {A.}~\bibnamefont {Vuorinen}},\ }\href
  {\doibase 10.1103/PhysRevLett.120.172703} {\bibfield  {journal} {\bibinfo
  {journal} {Phys. Rev. Lett.}\ }\textbf {\bibinfo {volume} {120}},\ \bibinfo
  {pages} {172703} (\bibinfo {year} {2018})},\ \Eprint
  {http://arxiv.org/abs/1711.02644} {arXiv:1711.02644 [astro-ph.HE]}
  \BibitemShut {NoStop}%
\bibitem [{\citenamefont {Raithel}\ \emph {et~al.}(2018)\citenamefont
  {Raithel}, \citenamefont {ï¿œzel},\ and\ \citenamefont
  {Psaltis}}]{Raithel:2018ncd}%
  \BibitemOpen
  \bibfield  {author} {\bibinfo {author} {\bibfnamefont {C.}~\bibnamefont
  {Raithel}}, \bibinfo {author} {\bibfnamefont {F.}~\bibnamefont {ï¿œzel}},
  \ and\ \bibinfo {author} {\bibfnamefont {D.}~\bibnamefont {Psaltis}},\ }\href
  {\doibase 10.3847/2041-8213/aabcbf} {\bibfield  {journal} {\bibinfo
  {journal} {Astrophys. J.}\ }\textbf {\bibinfo {volume} {857}},\ \bibinfo
  {pages} {L23} (\bibinfo {year} {2018})},\ \Eprint
  {http://arxiv.org/abs/1803.07687} {arXiv:1803.07687 [astro-ph.HE]}
  \BibitemShut {NoStop}%
\bibitem [{\citenamefont {Acernese}\ \emph {et~al.}(2015)\citenamefont
  {Acernese} \emph {et~al.}}]{TheVirgo:2014hva}%
  \BibitemOpen
  \bibfield  {author} {\bibinfo {author} {\bibfnamefont {F.}~\bibnamefont
  {Acernese}} \emph {et~al.} (\bibinfo {collaboration} {VIRGO}),\ }\href
  {\doibase 10.1088/0264-9381/32/2/024001} {\bibfield  {journal} {\bibinfo
  {journal} {Class. Quant. Grav.}\ }\textbf {\bibinfo {volume} {32}},\ \bibinfo
  {pages} {024001} (\bibinfo {year} {2015})},\ \Eprint
  {http://arxiv.org/abs/1408.3978} {arXiv:1408.3978 [gr-qc]} \BibitemShut
  {NoStop}%
\bibitem [{\citenamefont {Abbott}\ \emph {et~al.}(2019)\citenamefont {Abbott}
  \emph {et~al.}}]{Abbott2018}%
  \BibitemOpen
  \bibfield  {author} {\bibinfo {author} {\bibfnamefont {B.~P.}\ \bibnamefont
  {Abbott}} \emph {et~al.} (\bibinfo {collaboration} {LIGO Scientific,
  Virgo}),\ }\href {\doibase 10.1103/PhysRevX.9.011001} {\bibfield  {journal}
  {\bibinfo  {journal} {Phys. Rev.}\ }\textbf {\bibinfo {volume} {X9}},\
  \bibinfo {pages} {011001} (\bibinfo {year} {2019})},\ \Eprint
  {http://arxiv.org/abs/1805.11579} {arXiv:1805.11579 [gr-qc]} \BibitemShut
  {NoStop}%
\bibitem [{\citenamefont {Finn}(1992)}]{Finn:Fisher}%
  \BibitemOpen
  \bibfield  {author} {\bibinfo {author} {\bibfnamefont {L.~S.}\ \bibnamefont
  {Finn}},\ }\href {\doibase 10.1103/PhysRevD.46.5236} {\bibfield  {journal}
  {\bibinfo  {journal} {Phys. Rev. D}\ }\textbf {\bibinfo {volume} {46}},\
  \bibinfo {pages} {5236} (\bibinfo {year} {1992})}\BibitemShut {NoStop}%
\bibitem [{\citenamefont {Cutler}\ and\ \citenamefont
  {Flanagan}(1994)}]{Cutler:Fisher}%
  \BibitemOpen
  \bibfield  {author} {\bibinfo {author} {\bibfnamefont {C.}~\bibnamefont
  {Cutler}}\ and\ \bibinfo {author} {\bibfnamefont {E.~E.}\ \bibnamefont
  {Flanagan}},\ }\href {\doibase 10.1103/PhysRevD.49.2658} {\bibfield
  {journal} {\bibinfo  {journal} {Phys. Rev. D}\ }\textbf {\bibinfo {volume}
  {49}},\ \bibinfo {pages} {2658} (\bibinfo {year} {1994})}\BibitemShut
  {NoStop}%
\bibitem [{\citenamefont {Berti}\ \emph {et~al.}(2005)\citenamefont {Berti},
  \citenamefont {Buonanno},\ and\ \citenamefont {Will}}]{Berti:Fisher}%
  \BibitemOpen
  \bibfield  {author} {\bibinfo {author} {\bibfnamefont {E.}~\bibnamefont
  {Berti}}, \bibinfo {author} {\bibfnamefont {A.}~\bibnamefont {Buonanno}}, \
  and\ \bibinfo {author} {\bibfnamefont {C.~M.}\ \bibnamefont {Will}},\ }\href
  {\doibase 10.1103/PhysRevD.71.084025} {\bibfield  {journal} {\bibinfo
  {journal} {Phys. Rev.}\ }\textbf {\bibinfo {volume} {D71}},\ \bibinfo {pages}
  {084025} (\bibinfo {year} {2005})},\ \Eprint
  {http://arxiv.org/abs/gr-qc/0411129} {arXiv:gr-qc/0411129 [gr-qc]}
  \BibitemShut {NoStop}%
\bibitem [{\citenamefont {Poisson}\ and\ \citenamefont
  {Will}(1995)}]{Poisson:Fisher}%
  \BibitemOpen
  \bibfield  {author} {\bibinfo {author} {\bibfnamefont {E.}~\bibnamefont
  {Poisson}}\ and\ \bibinfo {author} {\bibfnamefont {C.~M.}\ \bibnamefont
  {Will}},\ }\href {\doibase 10.1103/PhysRevD.52.848} {\bibfield  {journal}
  {\bibinfo  {journal} {Phys. Rev. D}\ }\textbf {\bibinfo {volume} {52}},\
  \bibinfo {pages} {848} (\bibinfo {year} {1995})}\BibitemShut {NoStop}%
\bibitem [{\citenamefont {Wade}\ \emph
  {et~al.}(2014{\natexlab{b}})\citenamefont {Wade}, \citenamefont {Creighton},
  \citenamefont {Ochsner}, \citenamefont {Lackey}, \citenamefont {Farr},
  \citenamefont {Littenberg},\ and\ \citenamefont {Raymond}}]{Wade:2014vqa}%
  \BibitemOpen
  \bibfield  {author} {\bibinfo {author} {\bibfnamefont {L.}~\bibnamefont
  {Wade}}, \bibinfo {author} {\bibfnamefont {J.~D.~E.}\ \bibnamefont
  {Creighton}}, \bibinfo {author} {\bibfnamefont {E.}~\bibnamefont {Ochsner}},
  \bibinfo {author} {\bibfnamefont {B.~D.}\ \bibnamefont {Lackey}}, \bibinfo
  {author} {\bibfnamefont {B.~F.}\ \bibnamefont {Farr}}, \bibinfo {author}
  {\bibfnamefont {T.~B.}\ \bibnamefont {Littenberg}}, \ and\ \bibinfo {author}
  {\bibfnamefont {V.}~\bibnamefont {Raymond}},\ }\href {\doibase
  10.1103/PhysRevD.89.103012} {\bibfield  {journal} {\bibinfo  {journal} {Phys.
  Rev.}\ }\textbf {\bibinfo {volume} {D89}},\ \bibinfo {pages} {103012}
  (\bibinfo {year} {2014}{\natexlab{b}})},\ \Eprint
  {http://arxiv.org/abs/1402.5156} {arXiv:1402.5156 [gr-qc]} \BibitemShut
  {NoStop}%
\bibitem [{\citenamefont {Yagi}(2014)}]{Yagi:Multipole}%
  \BibitemOpen
  \bibfield  {author} {\bibinfo {author} {\bibfnamefont {K.}~\bibnamefont
  {Yagi}},\ }\href {\doibase 10.1103/PhysRevD.97.129901,
  10.1103/PhysRevD.96.129904, 10.1103/PhysRevD.89.043011} {\bibfield  {journal}
  {\bibinfo  {journal} {Phys. Rev.}\ }\textbf {\bibinfo {volume} {D89}},\
  \bibinfo {pages} {043011} (\bibinfo {year} {2014})},\ \bibinfo {note}
  {[Erratum: Phys. Rev.D97,no.12,129901(2018)]},\ \Eprint
  {http://arxiv.org/abs/1311.0872} {arXiv:1311.0872 [gr-qc]} \BibitemShut
  {NoStop}%
\bibitem [{\citenamefont {Lackey}\ \emph {et~al.}(2018)\citenamefont {Lackey},
  \citenamefont {PÃŒrrer}, \citenamefont {Taracchini},\ and\ \citenamefont
  {Marsat}}]{Lackey:Surrogate}%
  \BibitemOpen
  \bibfield  {author} {\bibinfo {author} {\bibfnamefont {B.~D.}\ \bibnamefont
  {Lackey}}, \bibinfo {author} {\bibfnamefont {M.}~\bibnamefont {PÃŒrrer}},
  \bibinfo {author} {\bibfnamefont {A.}~\bibnamefont {Taracchini}}, \ and\
  \bibinfo {author} {\bibfnamefont {S.}~\bibnamefont {Marsat}},\ }\href@noop {}
  {\  (\bibinfo {year} {2018})},\ \Eprint {http://arxiv.org/abs/1812.08643}
  {arXiv:1812.08643 [gr-qc]} \BibitemShut {NoStop}%
\bibitem [{\citenamefont {Lackey}\ \emph {et~al.}(2017)\citenamefont {Lackey},
  \citenamefont {Bernuzzi}, \citenamefont {Galley}, \citenamefont {Meidam},\
  and\ \citenamefont {Van Den~Broeck}}]{Lackey:EOB}%
  \BibitemOpen
  \bibfield  {author} {\bibinfo {author} {\bibfnamefont {B.~D.}\ \bibnamefont
  {Lackey}}, \bibinfo {author} {\bibfnamefont {S.}~\bibnamefont {Bernuzzi}},
  \bibinfo {author} {\bibfnamefont {C.~R.}\ \bibnamefont {Galley}}, \bibinfo
  {author} {\bibfnamefont {J.}~\bibnamefont {Meidam}}, \ and\ \bibinfo {author}
  {\bibfnamefont {C.}~\bibnamefont {Van Den~Broeck}},\ }\href {\doibase
  10.1103/PhysRevD.95.104036} {\bibfield  {journal} {\bibinfo  {journal} {Phys.
  Rev.}\ }\textbf {\bibinfo {volume} {D95}},\ \bibinfo {pages} {104036}
  (\bibinfo {year} {2017})},\ \Eprint {http://arxiv.org/abs/1610.04742}
  {arXiv:1610.04742 [gr-qc]} \BibitemShut {NoStop}%
\bibitem [{\citenamefont {Schutz}(2011)}]{Shutz:SNR}%
  \BibitemOpen
  \bibfield  {author} {\bibinfo {author} {\bibfnamefont {B.~F.}\ \bibnamefont
  {Schutz}},\ }\href {\doibase 10.1088/0264-9381/28/12/125023} {\bibfield
  {journal} {\bibinfo  {journal} {Classical and Quantum Gravity}\ }\textbf
  {\bibinfo {volume} {28}},\ \bibinfo {pages} {125023} (\bibinfo {year}
  {2011})}\BibitemShut {NoStop}%
\bibitem [{\citenamefont {Chen}\ and\ \citenamefont {Holz}(2014)}]{Chen:SNR}%
  \BibitemOpen
  \bibfield  {author} {\bibinfo {author} {\bibfnamefont {H.-Y.}\ \bibnamefont
  {Chen}}\ and\ \bibinfo {author} {\bibfnamefont {D.~E.}\ \bibnamefont
  {Holz}},\ }\href@noop {} {\  (\bibinfo {year} {2014})},\ \Eprint
  {http://arxiv.org/abs/1409.0522} {arXiv:1409.0522 [gr-qc]} \BibitemShut
  {NoStop}%
\bibitem [{\citenamefont {Cutler}\ and\ \citenamefont
  {Harms}(2006)}]{Cutler:BNSmerger}%
  \BibitemOpen
  \bibfield  {author} {\bibinfo {author} {\bibfnamefont {C.}~\bibnamefont
  {Cutler}}\ and\ \bibinfo {author} {\bibfnamefont {J.}~\bibnamefont {Harms}},\
  }\href {\doibase 10.1103/PhysRevD.73.042001} {\bibfield  {journal} {\bibinfo
  {journal} {Phys. Rev.}\ }\textbf {\bibinfo {volume} {D73}},\ \bibinfo {pages}
  {042001} (\bibinfo {year} {2006})},\ \Eprint
  {http://arxiv.org/abs/gr-qc/0511092} {arXiv:gr-qc/0511092 [gr-qc]}
  \BibitemShut {NoStop}%
\bibitem [{\citenamefont {Yagi}\ and\ \citenamefont {Tanaka}(2010)}]{Takahiro}%
  \BibitemOpen
  \bibfield  {author} {\bibinfo {author} {\bibfnamefont {K.}~\bibnamefont
  {Yagi}}\ and\ \bibinfo {author} {\bibfnamefont {T.}~\bibnamefont {Tanaka}},\
  }\href {\doibase 10.1143/PTP.123.1069} {\bibfield  {journal} {\bibinfo
  {journal} {Prog. Theor. Phys.}\ }\textbf {\bibinfo {volume} {123}},\ \bibinfo
  {pages} {1069} (\bibinfo {year} {2010})},\ \Eprint
  {http://arxiv.org/abs/0908.3283} {arXiv:0908.3283 [gr-qc]} \BibitemShut
  {NoStop}%
\end{thebibliography}%
\end{document}